\documentclass[10pt,twocolumn,letterpaper]{article}

\usepackage[pagenumbers]{cvpr}

\usepackage{caption}
\usepackage{float}
\usepackage{array}
\usepackage{fontawesome5}
\usepackage{tikz}
\usetikzlibrary{arrows.meta}
\usepackage{amsthm}
\newtheorem{proposition}{Proposition}
\ifdefined\pdfsuppressptexinfo\pdfsuppressptexinfo=-1 \fi
\newcommand{\DesN}{1024}
\newcommand{\DesKbits}{1024}
\newcommand{\DesNu}{48}
\newcommand{\DesLambda}{532}
\newcommand{\DesPix}{8.0}
\newcommand{\DesWidth}{8.192}
\newcommand{\DesBinPitch}{14.96}
\newcommand{\DesSlmPitch}{3.74}
\newcommand{\DesFocal}{230.4}
\newcommand{\DesRingLo}{16}
\newcommand{\DesRingMid}{80}
\newcommand{\DesRingHi}{144}
\newcommand{\DesLoneBins}{9{,}638}
\newcommand{\DesLtwoBins}{22{,}514}
\newcommand{\DesBinsPerBit}{9}
\newcommand{\DesZTarget}{12}
\newcommand{\DesThr}{25}
\newcommand{\DesLtwoPsnrLo}{42}
\newcommand{\DesLtwoPsnrHi}{54}
\newcommand{\DesStep}{79.6}
\newcommand{\DesNEval}{877}
\newcommand{\DesNAttacks}{8}
\newcommand{\DesPayloadBytes}{128}
\newcommand{\DesDataBytes}{98}
\newcommand{\DesRsParity}{30}
\newcommand{\DesBchN}{255}
\newcommand{\DesBchK}{179}
\newcommand{\DesBchT}{10}
\newcommand{\DesSweepPoints}{50}
\newcommand{\DesPupilWide}{724}
\newcommand{\DesNCalib}{8}
\newcommand{\DesEnrolIter}{6}
\newcommand{\DesRefinePasses}{2}
\newcommand{\CrHzero}{10^{-6}}
\newcommand{\CrAgree}{99}
\newcommand{\CrGap}{2}
\newcommand{\CrMin}{95}
\newcommand{\CrMax}{5}
\newcommand{\CrScaleTest}{1}
\newcommand{\CrShiftTest}{1}
\newcommand{\CrCpLevel}{95}
\newcommand{\DesBIn}{10}
\newcommand{\DesBSlm}{8}
\newcommand{\DesAf}{0.25}
\newcommand{\DesPupil}{512}
\newcommand{\DesFcNe}{3\times10^{4}}
\newcommand{\DesFcRead}{3}
\newcommand{\DesFcBits}{12}
\newcommand{\DesFcNexp}{3}
\newcommand{\DesSNe}{1.2\times10^{5}}
\newcommand{\DesSRead}{6}
\newcommand{\DesSBits}{12}
\newcommand{\DesBPhi}{8}
\newcommand{\DesGamma}{0.95}
\newcommand{\SetNValid}{100}
\newcommand{\SetNTrain}{777}
\newcommand{\SetNSkipped}{15}
\newcommand{\VerHz}{PASS}
\newcommand{\VerHo}{PASS}
\newcommand{\VerHtwo}{PASS}
\newcommand{\VerHf}{PASS}
\newcommand{\VerHv}{PASS}
\newcommand{\VerHs}{PASS}
\newcommand{\VerHe}{PASS}
\newcommand{\SetN}{877}
\newcommand{\SetNCoco}{40{,}670}
\newcommand{\HzN}{877}
\newcommand{\HzMaxRel}{1.7\times10^{-15}}
\newcommand{\HzIdentical}{348}
\newcommand{\HzNegPix}{3{,}249{,}594}
\newcommand{\HoBitAgree}{99.997}
\newcommand{\HoOptDigAgree}{99.988}
\newcommand{\HoPoolOpt}{99.86}
\newcommand{\HoPoolDig}{99.91}
\newcommand{\HoPoolVopt}{99.70}
\newcommand{\HtwoMin}{99.5}
\newcommand{\RDigVdNone}{100}
\newcommand{\RDigVoNone}{99.9}
\newcommand{\RDigBerNone}{0}
\newcommand{\RDigVdJpegA}{100}
\newcommand{\RDigVoJpegA}{99.9}
\newcommand{\RDigBerJpegA}{2.2\times10^{-6}}
\newcommand{\RDigVdJpegB}{100}
\newcommand{\RDigVoJpegB}{99.8}
\newcommand{\RDigBerJpegB}{4.5\times10^{-6}}
\newcommand{\RDigVdJpegC}{100}
\newcommand{\RDigVoJpegC}{99.8}
\newcommand{\RDigBerJpegC}{3.0\times10^{-5}}
\newcommand{\RDigVdWebp}{99.9}
\newcommand{\RDigVoWebp}{99.5}
\newcommand{\RDigBerWebp}{1.6\times10^{-4}}
\newcommand{\RDigVdScaleA}{100}
\newcommand{\RDigVoScaleA}{99.8}
\newcommand{\RDigBerScaleA}{1.6\times10^{-5}}
\newcommand{\RDigVdScaleB}{100}
\newcommand{\RDigVoScaleB}{99.9}
\newcommand{\RDigBerScaleB}{0}
\newcommand{\RDigVdCombo}{99.4}
\newcommand{\RDigVoCombo}{99.3}
\newcommand{\RDigBerCombo}{2.4\times10^{-4}}
\newcommand{\RIdealVdNone}{100}
\newcommand{\RIdealVoNone}{99.9}
\newcommand{\RIdealBerNone}{0}
\newcommand{\RIdealLtNone}{100}
\newcommand{\RIdealZcNone}{12.4}
\newcommand{\RIdealVdJpegA}{100}
\newcommand{\RIdealVoJpegA}{99.9}
\newcommand{\RIdealBerJpegA}{2.2\times10^{-6}}
\newcommand{\RIdealLtJpegA}{99.9}
\newcommand{\RIdealZcJpegA}{12.3}
\newcommand{\RIdealVdJpegB}{100}
\newcommand{\RIdealVoJpegB}{99.9}
\newcommand{\RIdealBerJpegB}{4.5\times10^{-6}}
\newcommand{\RIdealLtJpegB}{99.9}
\newcommand{\RIdealZcJpegB}{11.9}
\newcommand{\RIdealVdJpegC}{100}
\newcommand{\RIdealVoJpegC}{99.8}
\newcommand{\RIdealBerJpegC}{1.4\times10^{-5}}
\newcommand{\RIdealLtJpegC}{99.8}
\newcommand{\RIdealZcJpegC}{11.1}
\newcommand{\RIdealVdWebp}{100}
\newcommand{\RIdealVoWebp}{99.8}
\newcommand{\RIdealBerWebp}{1.3\times10^{-4}}
\newcommand{\RIdealLtWebp}{97.3}
\newcommand{\RIdealZcWebp}{9.1}
\newcommand{\RIdealVdScaleA}{100}
\newcommand{\RIdealVoScaleA}{99.9}
\newcommand{\RIdealBerScaleA}{1.7\times10^{-5}}
\newcommand{\RIdealLtScaleA}{99.9}
\newcommand{\RIdealZcScaleA}{12.4}
\newcommand{\RIdealVdScaleB}{100}
\newcommand{\RIdealVoScaleB}{99.9}
\newcommand{\RIdealBerScaleB}{0}
\newcommand{\RIdealLtScaleB}{99.9}
\newcommand{\RIdealZcScaleB}{12.4}
\newcommand{\RIdealVdCombo}{99.7}
\newcommand{\RIdealVoCombo}{99.5}
\newcommand{\RIdealBerCombo}{2.1\times10^{-4}}
\newcommand{\RIdealLtCombo}{99.2}
\newcommand{\RIdealZcCombo}{10.1}
\newcommand{\ROptDVdNone}{99.9}
\newcommand{\ROptDVoNone}{99.8}
\newcommand{\ROptDBerNone}{2.6\times10^{-5}}
\newcommand{\ROptDLtNone}{99.8}
\newcommand{\ROptDZcNone}{11.8}
\newcommand{\ROptDVdJpegA}{100}
\newcommand{\ROptDVoJpegA}{99.9}
\newcommand{\ROptDBerJpegA}{2.6\times10^{-5}}
\newcommand{\ROptDLtJpegA}{99.9}
\newcommand{\ROptDZcJpegA}{11.8}
\newcommand{\ROptDVdJpegB}{100}
\newcommand{\ROptDVoJpegB}{99.8}
\newcommand{\ROptDBerJpegB}{3.5\times10^{-5}}
\newcommand{\ROptDLtJpegB}{99.9}
\newcommand{\ROptDZcJpegB}{11.4}
\newcommand{\ROptDVdJpegC}{99.9}
\newcommand{\ROptDVoJpegC}{99.8}
\newcommand{\ROptDBerJpegC}{6.1\times10^{-5}}
\newcommand{\ROptDLtJpegC}{99.7}
\newcommand{\ROptDZcJpegC}{10.6}
\newcommand{\ROptDVdWebp}{99.7}
\newcommand{\ROptDVoWebp}{99.4}
\newcommand{\ROptDBerWebp}{2.8\times10^{-4}}
\newcommand{\ROptDLtWebp}{95.9}
\newcommand{\ROptDZcWebp}{8.7}
\newcommand{\ROptDVdScaleA}{99.9}
\newcommand{\ROptDVoScaleA}{99.8}
\newcommand{\ROptDBerScaleA}{5.9\times10^{-5}}
\newcommand{\ROptDLtScaleA}{99.8}
\newcommand{\ROptDZcScaleA}{11.8}
\newcommand{\ROptDVdScaleB}{100}
\newcommand{\ROptDVoScaleB}{99.9}
\newcommand{\ROptDBerScaleB}{2.4\times10^{-5}}
\newcommand{\ROptDLtScaleB}{99.9}
\newcommand{\ROptDZcScaleB}{11.9}
\newcommand{\ROptDVdCombo}{99.5}
\newcommand{\ROptDVoCombo}{99.3}
\newcommand{\ROptDBerCombo}{2.4\times10^{-4}}
\newcommand{\ROptDLtCombo}{98.9}
\newcommand{\ROptDZcCombo}{9.6}
\newcommand{\ROptSVdNone}{100}
\newcommand{\ROptSVoNone}{99.9}
\newcommand{\ROptSBerNone}{2.3\times10^{-5}}
\newcommand{\ROptSLtNone}{100}
\newcommand{\ROptSZcNone}{11.9}
\newcommand{\ROptSVdJpegA}{100}
\newcommand{\ROptSVoJpegA}{99.9}
\newcommand{\ROptSBerJpegA}{2.6\times10^{-5}}
\newcommand{\ROptSLtJpegA}{100}
\newcommand{\ROptSZcJpegA}{11.9}
\newcommand{\ROptSVdJpegB}{99.9}
\newcommand{\ROptSVoJpegB}{99.8}
\newcommand{\ROptSBerJpegB}{3.1\times10^{-5}}
\newcommand{\ROptSLtJpegB}{100}
\newcommand{\ROptSZcJpegB}{11.4}
\newcommand{\ROptSVdJpegC}{99.9}
\newcommand{\ROptSVoJpegC}{99.8}
\newcommand{\ROptSBerJpegC}{5.7\times10^{-5}}
\newcommand{\ROptSLtJpegC}{100}
\newcommand{\ROptSZcJpegC}{10.6}
\newcommand{\ROptSVdWebp}{99.7}
\newcommand{\ROptSVoWebp}{99.3}
\newcommand{\ROptSBerWebp}{2.9\times10^{-4}}
\newcommand{\ROptSLtWebp}{96.5}
\newcommand{\ROptSZcWebp}{8.7}
\newcommand{\ROptSVdScaleA}{99.9}
\newcommand{\ROptSVoScaleA}{99.8}
\newcommand{\ROptSBerScaleA}{4.1\times10^{-5}}
\newcommand{\ROptSLtScaleA}{100}
\newcommand{\ROptSZcScaleA}{11.9}
\newcommand{\ROptSVdScaleB}{100}
\newcommand{\ROptSVoScaleB}{99.9}
\newcommand{\ROptSBerScaleB}{2.2\times10^{-5}}
\newcommand{\ROptSLtScaleB}{100}
\newcommand{\ROptSZcScaleB}{11.9}
\newcommand{\ROptSVdCombo}{99.5}
\newcommand{\ROptSVoCombo}{99.4}
\newcommand{\ROptSBerCombo}{2.3\times10^{-4}}
\newcommand{\ROptSLtCombo}{99.4}
\newcommand{\ROptSZcCombo}{9.6}
\newcommand{\ROptWVdNone}{100}
\newcommand{\ROptWVoNone}{99.9}
\newcommand{\ROptWBerNone}{2.3\times10^{-5}}
\newcommand{\ROptWLtNone}{99.9}
\newcommand{\ROptWZcNone}{11.9}
\newcommand{\ROptWVdJpegA}{100}
\newcommand{\ROptWVoJpegA}{99.9}
\newcommand{\ROptWBerJpegA}{2.7\times10^{-5}}
\newcommand{\ROptWLtJpegA}{99.9}
\newcommand{\ROptWZcJpegA}{11.9}
\newcommand{\ROptWVdJpegB}{100}
\newcommand{\ROptWVoJpegB}{99.9}
\newcommand{\ROptWBerJpegB}{3.2\times10^{-5}}
\newcommand{\ROptWLtJpegB}{99.9}
\newcommand{\ROptWZcJpegB}{11.4}
\newcommand{\ROptWVdJpegC}{99.9}
\newcommand{\ROptWVoJpegC}{99.8}
\newcommand{\ROptWBerJpegC}{7.1\times10^{-5}}
\newcommand{\ROptWLtJpegC}{99.7}
\newcommand{\ROptWZcJpegC}{10.6}
\newcommand{\ROptWVdWebp}{99.7}
\newcommand{\ROptWVoWebp}{99.5}
\newcommand{\ROptWBerWebp}{2.8\times10^{-4}}
\newcommand{\ROptWLtWebp}{96.1}
\newcommand{\ROptWZcWebp}{8.7}
\newcommand{\ROptWVdScaleA}{99.9}
\newcommand{\ROptWVoScaleA}{99.8}
\newcommand{\ROptWBerScaleA}{5.8\times10^{-5}}
\newcommand{\ROptWLtScaleA}{99.8}
\newcommand{\ROptWZcScaleA}{11.8}
\newcommand{\ROptWVdScaleB}{99.9}
\newcommand{\ROptWVoScaleB}{99.9}
\newcommand{\ROptWBerScaleB}{3.1\times10^{-5}}
\newcommand{\ROptWLtScaleB}{99.8}
\newcommand{\ROptWZcScaleB}{11.9}
\newcommand{\ROptWVdCombo}{99.5}
\newcommand{\ROptWVoCombo}{99.3}
\newcommand{\ROptWBerCombo}{3.1\times10^{-4}}
\newcommand{\ROptWLtCombo}{99.1}
\newcommand{\ROptWZcCombo}{9.6}
\newcommand{\LtMinDetD}{95.9}
\newcommand{\LtMinDetS}{96.5}
\newcommand{\QDigPfive}{47.5}
\newcommand{\QDigMed}{47.7}
\newcommand{\QDigMin}{47.1}
\newcommand{\QDigSsimPfive}{0.992}
\newcommand{\QDigConv}{92.1}
\newcommand{\QIdealPfive}{44.7}
\newcommand{\QIdealMed}{46.5}
\newcommand{\QIdealMin}{42.4}
\newcommand{\QIdealSsimPfive}{0.988}
\newcommand{\QIdealConv}{92.5}
\newcommand{\QOptDPfive}{37.1}
\newcommand{\QOptDMed}{42.6}
\newcommand{\QOptDMin}{31.0}
\newcommand{\QOptDSsimPfive}{0.978}
\newcommand{\QOptDConv}{96.4}
\newcommand{\QOptSPfive}{37.1}
\newcommand{\QOptSMed}{42.6}
\newcommand{\QOptSMin}{31.0}
\newcommand{\QOptSSsimPfive}{0.978}
\newcommand{\QOptSConv}{96.1}
\newcommand{\QOptWPfive}{44.4}
\newcommand{\QOptWMed}{46.0}
\newcommand{\QOptWMin}{42.9}
\newcommand{\QOptWSsimPfive}{0.986}
\newcommand{\QOptWConv}{96.8}
\newcommand{\HfCocoN}{40{,}670}
\newcommand{\HfCocoVd}{0}
\newcommand{\HfCocoVo}{0}
\newcommand{\HfDivN}{877}
\newcommand{\HfDivVd}{0}
\newcommand{\HfWrongPk}{0}
\newcommand{\HfWrongPkN}{7{,}016}
\newcommand{\HfLtStatic}{0}
\newcommand{\HfWrongSeed}{0}
\newcommand{\HfWrongSeedN}{17{,}540}
\newcommand{\HfCpUpper}{7.4\times10^{-5}}
\newcommand{\HfNullMax}{19.9}
\newcommand{\HvWone}{0}
\newcommand{\HvWtwo}{0}
\newcommand{\HvWthree}{0}
\newcommand{\HsMatchedMed}{140}
\newcommand{\HsMatchedDet}{99.9}
\newcommand{\HsDetOptDNone}{99.8}
\newcommand{\HsDetOptDJpegB}{99.9}
\newcommand{\HsDetOptSNone}{100}
\newcommand{\HsDetOptSJpegB}{100}
\newcommand{\HsScalePlus}{0}
\newcommand{\HsScaleMinus}{0}
\newcommand{\HsShiftOne}{0}
\newcommand{\HsTolRel}{0.70}
\newcommand{\HsTolNm}{3.7}
\newcommand{\HsTolFmm}{1.6}
\newcommand{\HsTolShift}{0.77}
\newcommand{\HsTolShiftUm}{11.6}
\newcommand{\KsSeed}{256}
\newcommand{\HcScPZpZZZOMed}{139.4}
\newcommand{\HcScPZpZZZODet}{99.9}
\newcommand{\HcScMZpZZZOMed}{140.1}
\newcommand{\HcScMZpZZZODet}{99.9}
\newcommand{\HcScPZpZZZTMed}{139.0}
\newcommand{\HcScPZpZZZTDet}{99.9}
\newcommand{\HcScMZpZZZTMed}{139.7}
\newcommand{\HcScMZpZZZTDet}{99.9}
\newcommand{\HcScPZpZZZVMed}{137.4}
\newcommand{\HcScPZpZZZVDet}{99.9}
\newcommand{\HcScMZpZZZVMed}{138.3}
\newcommand{\HcScMZpZZZVDet}{99.9}
\newcommand{\HcScPZpZZOMed}{133.3}
\newcommand{\HcScPZpZZODet}{99.9}
\newcommand{\HcScMZpZZOMed}{133.6}
\newcommand{\HcScMZpZZODet}{99.9}
\newcommand{\HcScPZpZZTMed}{116.8}
\newcommand{\HcScPZpZZTDet}{99.9}
\newcommand{\HcScMZpZZTMed}{116.6}
\newcommand{\HcScMZpZZTDet}{99.9}
\newcommand{\HcScPZpZZHMed}{91.6}
\newcommand{\HcScPZpZZHDet}{99.8}
\newcommand{\HcScMZpZZHMed}{91.6}
\newcommand{\HcScMZpZZHDet}{99.8}
\newcommand{\HcScPZpZZVMed}{40.9}
\newcommand{\HcScPZpZZVDet}{88.8}
\newcommand{\HcScMZpZZVMed}{41.1}
\newcommand{\HcScMZpZZVDet}{87.0}
\newcommand{\HcScPZpZOMed}{0.4}
\newcommand{\HcScPZpZODet}{0}
\newcommand{\HcScMZpZOMed}{0.5}
\newcommand{\HcScMZpZODet}{0}
\newcommand{\HcScPZpZTMed}{0.5}
\newcommand{\HcScPZpZTDet}{0}
\newcommand{\HcScMZpZTMed}{0.4}
\newcommand{\HcScMZpZTDet}{0}
\newcommand{\HcShZpOMed}{136.2}
\newcommand{\HcShZpODet}{99.9}
\newcommand{\HcShZpTVMed}{122.9}
\newcommand{\HcShZpTVDet}{99.9}
\newcommand{\HcShZpVMed}{92.6}
\newcommand{\HcShZpVDet}{99.8}
\newcommand{\HcShZpNVMed}{27.5}
\newcommand{\HcShZpNVDet}{57.8}
\newcommand{\HcShOMed}{0.4}
\newcommand{\HcShODet}{0}
\newcommand{\HcShTMed}{0.5}
\newcommand{\HcShTDet}{0}
\newcommand{\HcPbOMed}{57.7}
\newcommand{\HcPbODet}{98.4}
\newcommand{\HcPbTMed}{114.9}
\newcommand{\HcPbTDet}{99.9}
\newcommand{\HcPbHMed}{132.4}
\newcommand{\HcPbHDet}{99.9}
\newcommand{\HcPbFMed}{137.8}
\newcommand{\HcPbFDet}{99.9}
\newcommand{\HcSwZpVMed}{138.7}
\newcommand{\HcSwZpVDet}{99.9}
\newcommand{\HcSwOMed}{135.2}
\newcommand{\HcSwODet}{99.9}
\newcommand{\HcSwTMed}{125.1}
\newcommand{\HcSwTDet}{99.9}
\newcommand{\HcSwVMed}{87.5}
\newcommand{\HcSwVDet}{99.9}
\newcommand{\HcSwOZMed}{65.0}
\newcommand{\HcSwOZDet}{99.7}
\newcommand{\HsMatchedQlo}{122}
\newcommand{\HsMatchedQhi}{170}
\newcommand{\HeN}{877}
\newcommand{\HeRead}{876}
\newcommand{\HeVerified}{0}
\newcommand{\HeSketch}{876}
\newcommand{\HeCpUpper}{0.003}
\newcommand{\RcCompared}{996}
\newcommand{\RcMismatch}{0}
\newcommand{\HtN}{877}
\newcommand{\HtVnBslmOZ}{100}
\newcommand{\HtVjBslmOZ}{100}
\newcommand{\HtLtBslmOZ}{99.9}
\newcommand{\HtBerBslmOZ}{2.0\times10^{-5}}
\newcommand{\HtPsBslmOZ}{42.6}
\newcommand{\HtVnBslmH}{13.1}
\newcommand{\HtVjBslmH}{12.3}
\newcommand{\HtLtBslmH}{13.0}
\newcommand{\HtBerBslmH}{4.5\times10^{-2}}
\newcommand{\HtPsBslmH}{41.7}
\newcommand{\HtVnBslmF}{73.9}
\newcommand{\HtVjBslmF}{70.6}
\newcommand{\HtLtBslmF}{73.8}
\newcommand{\HtBerBslmF}{1.2\times10^{-2}}
\newcommand{\HtPsBslmF}{42.2}
\newcommand{\HtVnBslmV}{99.9}
\newcommand{\HtVjBslmV}{99.9}
\newcommand{\HtLtBslmV}{99.8}
\newcommand{\HtBerBslmV}{8.3\times10^{-4}}
\newcommand{\HtPsBslmV}{42.5}
\newcommand{\HtVnBslmS}{99.9}
\newcommand{\HtVjBslmS}{99.9}
\newcommand{\HtLtBslmS}{99.8}
\newcommand{\HtBerBslmS}{6.9\times10^{-5}}
\newcommand{\HtPsBslmS}{42.6}
\newcommand{\HtVnBslmN}{100}
\newcommand{\HtVjBslmN}{100}
\newcommand{\HtLtBslmN}{99.9}
\newcommand{\HtBerBslmN}{3.6\times10^{-5}}
\newcommand{\HtPsBslmN}{42.6}
\newcommand{\HtVnSneHZZZZpZ}{100}
\newcommand{\HtVjSneHZZZZpZ}{100}
\newcommand{\HtLtSneHZZZZpZ}{99.9}
\newcommand{\HtBerSneHZZZZpZ}{2.3\times10^{-5}}
\newcommand{\HtPsSneHZZZZpZ}{42.1}
\newcommand{\HtVnSneOZZpZ}{0}
\newcommand{\HtVjSneOZZpZ}{0}
\newcommand{\HtLtSneOZZpZ}{0}
\newcommand{\HtBerSneOZZpZ}{3.6\times10^{-1}}
\newcommand{\HtPsSneOZZpZ}{19.7}
\newcommand{\HtVnSneHZZpZ}{0.7}
\newcommand{\HtVjSneHZZpZ}{0.7}
\newcommand{\HtLtSneHZZpZ}{0.7}
\newcommand{\HtBerSneHZZpZ}{9.0\times10^{-2}}
\newcommand{\HtPsSneHZZpZ}{26.1}
\newcommand{\HtVnSneOZZZpZ}{99.1}
\newcommand{\HtVjSneOZZZpZ}{98.7}
\newcommand{\HtLtSneOZZZpZ}{99.0}
\newcommand{\HtBerSneOZZZpZ}{1.3\times10^{-3}}
\newcommand{\HtPsSneOZZZpZ}{32.6}
\newcommand{\HtVnSneHZZZpZ}{100}
\newcommand{\HtVjSneHZZZpZ}{99.9}
\newcommand{\HtLtSneHZZZpZ}{99.9}
\newcommand{\HtBerSneHZZZpZ}{2.7\times10^{-5}}
\newcommand{\HtPsSneHZZZpZ}{37.3}
\newcommand{\HtVnSneOZZZZpZ}{99.9}
\newcommand{\HtVjSneOZZZZpZ}{100}
\newcommand{\HtLtSneOZZZZpZ}{99.8}
\newcommand{\HtBerSneOZZZZpZ}{3.2\times10^{-5}}
\newcommand{\HtPsSneOZZZZpZ}{40.6}
\newcommand{\HtVnFcamTxOT}{99.2}
\newcommand{\HtVjFcamTxOT}{99.3}
\newcommand{\HtLtFcamTxOT}{99.2}
\newcommand{\HtBerFcamTxOT}{1.4\times10^{-3}}
\newcommand{\HtPsFcamTxOT}{42.4}
\newcommand{\HtVnFcamHxE}{99.8}
\newcommand{\HtVjFcamHxE}{99.8}
\newcommand{\HtLtFcamHxE}{99.8}
\newcommand{\HtBerFcamHxE}{1.0\times10^{-3}}
\newcommand{\HtPsFcamHxE}{42.4}
\newcommand{\HtVnFcamHxOZ}{99.8}
\newcommand{\HtVjFcamHxOZ}{99.8}
\newcommand{\HtLtFcamHxOZ}{99.7}
\newcommand{\HtBerFcamHxOZ}{1.8\times10^{-4}}
\newcommand{\HtPsFcamHxOZ}{42.6}
\newcommand{\HtVnFcamOxOT}{19.5}
\newcommand{\HtVjFcamOxOT}{18.9}
\newcommand{\HtLtFcamOxOT}{19.5}
\newcommand{\HtBerFcamOxOT}{1.2\times10^{-1}}
\newcommand{\HtPsFcamOxOT}{40.9}
\newcommand{\HtVnAfOpZ}{99.2}
\newcommand{\HtVjAfOpZ}{99.4}
\newcommand{\HtLtAfOpZ}{98.9}
\newcommand{\HtBerAfOpZ}{4.9\times10^{-4}}
\newcommand{\HtPsAfOpZ}{42.0}
\newcommand{\HtVnAfZpZ}{99.9}
\newcommand{\HtVjAfZpZ}{99.9}
\newcommand{\HtLtAfZpZ}{99.8}
\newcommand{\HtBerAfZpZ}{3.7\times10^{-5}}
\newcommand{\HtPsAfZpZ}{42.6}
\newcommand{\HtVnAfZpOTV}{100}
\newcommand{\HtVjAfZpOTV}{100}
\newcommand{\HtLtAfZpOTV}{99.9}
\newcommand{\HtBerAfZpOTV}{2.4\times10^{-5}}
\newcommand{\HtPsAfZpOTV}{42.6}
\newcommand{\HtVnAfZpV}{100}
\newcommand{\HtVjAfZpV}{100}
\newcommand{\HtLtAfZpV}{99.9}
\newcommand{\HtBerAfZpV}{2.0\times10^{-5}}
\newcommand{\HtPsAfZpV}{42.6}
\newcommand{\HtVnAfZpNV}{99.9}
\newcommand{\HtVjAfZpNV}{99.9}
\newcommand{\HtLtAfZpNV}{99.8}
\newcommand{\HtBerAfZpNV}{9.5\times10^{-5}}
\newcommand{\HtPsAfZpNV}{42.5}
\newcommand{\HtVnPupilISpZ}{100}
\newcommand{\HtVjPupilISpZ}{100}
\newcommand{\HtLtPupilISpZ}{100}
\newcommand{\HtBerPupilISpZ}{1.3\times10^{-5}}
\newcommand{\HtPsPupilISpZ}{24.6}
\newcommand{\HtVnPupilOTEpZ}{100}
\newcommand{\HtVjPupilOTEpZ}{100}
\newcommand{\HtLtPupilOTEpZ}{100}
\newcommand{\HtBerPupilOTEpZ}{1.7\times10^{-5}}
\newcommand{\HtPsPupilOTEpZ}{25.9}
\newcommand{\HtVnPupilOSZpZ}{100}
\newcommand{\HtVjPupilOSZpZ}{100}
\newcommand{\HtLtPupilOSZpZ}{99.9}
\newcommand{\HtBerPupilOSZpZ}{1.3\times10^{-5}}
\newcommand{\HtPsPupilOSZpZ}{27.1}
\newcommand{\HtVnPupilTVSpZ}{100}
\newcommand{\HtVjPupilTVSpZ}{100}
\newcommand{\HtLtPupilTVSpZ}{99.9}
\newcommand{\HtBerPupilTVSpZ}{2.2\times10^{-5}}
\newcommand{\HtPsPupilTVSpZ}{30.5}
\newcommand{\HtVnPupilNTFpZ}{100}
\newcommand{\HtVjPupilNTFpZ}{100}
\newcommand{\HtLtPupilNTFpZ}{99.9}
\newcommand{\HtBerPupilNTFpZ}{2.6\times10^{-5}}
\newcommand{\HtPsPupilNTFpZ}{46.0}
\newcommand{\HtVnDzZpV}{100}
\newcommand{\HtVjDzZpV}{100}
\newcommand{\HtLtDzZpV}{100}
\newcommand{\HtBerDzZpV}{2.7\times10^{-5}}
\newcommand{\HtPsDzZpV}{29.8}
\newcommand{\HtVnDzOpZ}{100}
\newcommand{\HtVjDzOpZ}{100}
\newcommand{\HtLtDzOpZ}{100}
\newcommand{\HtBerDzOpZ}{2.8\times10^{-5}}
\newcommand{\HtPsDzOpZ}{26.1}
\newcommand{\HtVnDzTpZ}{99.8}
\newcommand{\HtVjDzTpZ}{99.8}
\newcommand{\HtLtDzTpZ}{99.8}
\newcommand{\HtBerDzTpZ}{1.2\times10^{-4}}
\newcommand{\HtPsDzTpZ}{24.1}
\newcommand{\HtVnDzZpO}{99.9}
\newcommand{\HtVjDzZpO}{100}
\newcommand{\HtLtDzZpO}{99.8}
\newcommand{\HtBerDzZpO}{3.3\times10^{-5}}
\newcommand{\HtPsDzZpO}{42.2}
\newcommand{\HtVnDzFpZ}{98.7}
\newcommand{\HtVjDzFpZ}{98.6}
\newcommand{\HtLtDzFpZ}{98.7}
\newcommand{\HtBerDzFpZ}{1.2\times10^{-3}}
\newcommand{\HtPsDzFpZ}{22.6}
\newcommand{\HtVnDzZpTV}{99.9}
\newcommand{\HtVjDzZpTV}{99.9}
\newcommand{\HtLtDzZpTV}{99.8}
\newcommand{\HtBerDzZpTV}{3.2\times10^{-5}}
\newcommand{\HtPsDzZpTV}{37.0}
\newcommand{\HtVnDzEpZ}{54.0}
\newcommand{\HtVjDzEpZ}{51.7}
\newcommand{\HtLtDzEpZ}{54.0}
\newcommand{\HtBerDzEpZ}{3.7\times10^{-2}}
\newcommand{\HtPsDzEpZ}{21.2}
\newcommand{\HtVnTiltVpZ}{99.0}
\newcommand{\HtVjTiltVpZ}{99.0}
\newcommand{\HtLtTiltVpZ}{98.9}
\newcommand{\HtBerTiltVpZ}{2.0\times10^{-3}}
\newcommand{\HtPsTiltVpZ}{41.0}
\newcommand{\HtVnTiltOZpZ}{0}
\newcommand{\HtVjTiltOZpZ}{0}
\newcommand{\HtLtTiltOZpZ}{0}
\newcommand{\HtBerTiltOZpZ}{1.3\times10^{-1}}
\newcommand{\HtPsTiltOZpZ}{35.1}
\newcommand{\HtVnTiltOpZ}{99.9}
\newcommand{\HtVjTiltOpZ}{99.9}
\newcommand{\HtLtTiltOpZ}{99.8}
\newcommand{\HtBerTiltOpZ}{2.9\times10^{-5}}
\newcommand{\HtPsTiltOpZ}{42.2}
\newcommand{\HtVnTiltTpZ}{100}
\newcommand{\HtVjTiltTpZ}{100}
\newcommand{\HtLtTiltTpZ}{99.9}
\newcommand{\HtBerTiltTpZ}{3.5\times10^{-5}}
\newcommand{\HtPsTiltTpZ}{41.8}
\newcommand{\HtVnShiftZpHV}{96.0}
\newcommand{\HtVjShiftZpHV}{95.7}
\newcommand{\HtLtShiftZpHV}{96.0}
\newcommand{\HtBerShiftZpHV}{2.3\times10^{-3}}
\newcommand{\HtPsShiftZpHV}{40.8}
\newcommand{\HtVnShiftZpZT}{100}
\newcommand{\HtVjShiftZpZT}{100}
\newcommand{\HtLtShiftZpZT}{99.9}
\newcommand{\HtBerShiftZpZT}{2.1\times10^{-5}}
\newcommand{\HtPsShiftZpZT}{42.6}
\newcommand{\HtVnShiftZpV}{8.2}
\newcommand{\HtVjShiftZpV}{7.3}
\newcommand{\HtLtShiftZpV}{8.2}
\newcommand{\HtBerShiftZpV}{5.6\times10^{-2}}
\newcommand{\HtPsShiftZpV}{38.3}
\newcommand{\HtVnShiftZpZV}{100}
\newcommand{\HtVjShiftZpZV}{100}
\newcommand{\HtLtShiftZpZV}{99.9}
\newcommand{\HtBerShiftZpZV}{1.6\times10^{-5}}
\newcommand{\HtPsShiftZpZV}{42.6}
\newcommand{\HtVnShiftOpZ}{0}
\newcommand{\HtVjShiftOpZ}{0}
\newcommand{\HtLtShiftOpZ}{0}
\newcommand{\HtBerShiftOpZ}{5.1\times10^{-1}}
\newcommand{\HtPsShiftOpZ}{35.4}
\newcommand{\HtVnShiftZpO}{99.9}
\newcommand{\HtVjShiftZpO}{100}
\newcommand{\HtLtShiftZpO}{99.8}
\newcommand{\HtBerShiftZpO}{3.2\times10^{-5}}
\newcommand{\HtPsShiftZpO}{42.6}
\newcommand{\HtVnShiftZpT}{100}
\newcommand{\HtVjShiftZpT}{100}
\newcommand{\HtLtShiftZpT}{99.9}
\newcommand{\HtBerShiftZpT}{6.2\times10^{-5}}
\newcommand{\HtPsShiftZpT}{42.4}
\newcommand{\HtVnSbitsS}{98.3}
\newcommand{\HtVjSbitsS}{98.3}
\newcommand{\HtLtSbitsS}{98.3}
\newcommand{\HtBerSbitsS}{3.0\times10^{-3}}
\newcommand{\HtPsSbitsS}{35.5}
\newcommand{\HtVnSbitsE}{100}
\newcommand{\HtVjSbitsE}{100}
\newcommand{\HtLtSbitsE}{99.9}
\newcommand{\HtBerSbitsE}{1.6\times10^{-5}}
\newcommand{\HtPsSbitsE}{40.9}
\newcommand{\HtVnSbitsOZ}{100}
\newcommand{\HtVjSbitsOZ}{100}
\newcommand{\HtLtSbitsOZ}{99.9}
\newcommand{\HtBerSbitsOZ}{2.3\times10^{-5}}
\newcommand{\HtPsSbitsOZ}{42.4}
\newcommand{\HtVnGammaZpTV}{99.9}
\newcommand{\HtVjGammaZpTV}{99.9}
\newcommand{\HtLtGammaZpTV}{8.2}
\newcommand{\HtBerGammaZpTV}{3.5\times10^{-5}}
\newcommand{\HtPsGammaZpTV}{43.0}
\newcommand{\HtVnGammaZpV}{99.9}
\newcommand{\HtVjGammaZpV}{99.9}
\newcommand{\HtLtGammaZpV}{88.6}
\newcommand{\HtBerGammaZpV}{2.4\times10^{-5}}
\newcommand{\HtPsGammaZpV}{43.0}
\newcommand{\HtVnGammaZpNV}{99.9}
\newcommand{\HtVjGammaZpNV}{99.9}
\newcommand{\HtLtGammaZpNV}{99.5}
\newcommand{\HtBerGammaZpNV}{3.1\times10^{-5}}
\newcommand{\HtPsGammaZpNV}{42.8}
\newcommand{\BudBslm}{5}
\newcommand{\BudShift}{0.35}
\newcommand{\BudDz}{4.0}
\newcommand{\BudSne}{1\times10^{3}}
\newcommand{\BudTilt}{5}
\newcommand{\NxtBslm}{4}
\newcommand{\NxtVnBslm}{73.9}
\newcommand{\NxtShift}{0.5}
\newcommand{\NxtVnShift}{8.2}
\newcommand{\NxtSne}{3\times10^{2}}
\newcommand{\NxtVnSne}{0.7}
\newcommand{\NxtTilt}{10}
\newcommand{\NxtVnTilt}{0}
\newcommand{\SwDzA}{0.1}
\newcommand{\SwDzB}{0.25}
\newcommand{\FChanN}{180}
\newcommand{\FChanPsnr}{47.5}
\newcommand{\FEeDiv}{180/180}
\newcommand{\FEeCoco}{997/1{,}000}
\newcommand{\FEeFaN}{40{,}410}
\newcommand{\FEeCopyN}{1{,}000}
\newcommand{\FBindSearchBar}{100/100}
\newcommand{\FBarPsnr}{38}
\newcommand{\FBarLpips}{0.02}
\newcommand{\FBaseTmBits}{61}
\newcommand{\FBaseSealBits}{256}
\newcommand{\FBaseWamBits}{32}
\newcommand{\FBaseChunkyBits}{1024}
\newcommand{\FBaseChunkyPsnr}{41.6}
\newcommand{\FBaseChunkyLpips}{0.021}
\newcommand{\FBaseOursPsnr}{47.5}
\newcommand{\FBaseOursLpips}{0.014}
\newcommand{\FBaseTmqPsnr}{38.3}
\newcommand{\FBaseTmqLpips}{0.004}
\newcommand{\FBaseTmpPsnr}{44.6}
\newcommand{\FBaseTmpLpips}{0.008}
\newcommand{\FBaseVsPsnr}{41.4}
\newcommand{\FBaseVsLpips}{0.004}
\newcommand{\FBasePsPsnr}{42.3}
\newcommand{\FBasePsLpips}{0.004}
\newcommand{\FBaseWamPsnr}{35.8}
\newcommand{\FBaseWamLpips}{0.080}
\newcommand{\FBaseGainAcc}{0.978}
\newcommand{\FBaseOursGainHi}{0.81}
\newcommand{\FBaseOursGeom}{0.50}
\newcommand{\FRtChunkyHundred}{0.85}
\newcommand{\FRtWamHundred}{0.39}
\newcommand{\FRtTmLo}{0.31}
\newcommand{\FRtTmHi}{0.76}
\newcommand{\FRtSeal}{0.04}
\newcommand{\FRtOurs}{0.00}
\newcommand{\FGeomCrop}{0.006}
\newcommand{\FGeomRot}{0.07}
\newcommand{\FGeomRotDeg}{3}
\newcommand{\FRtS}{100}
\newcommand{\FNewBenign}{0.98}
\newcommand{\FNewSteer}{0/100}
\newcommand{\FNewLpipsOnly}{87/100}
\newcommand{\FNewSameScene}{20/34}
\newcommand{\FNewRegrade}{21}
\newcommand{\OldVerHz}{PASS}
\newcommand{\OldVerHo}{PASS}
\newcommand{\OldVerHtwo}{PASS}
\newcommand{\OldVerHf}{PASS}
\newcommand{\OldVerHv}{PASS}
\newcommand{\OldVerHs}{PASS}
\newcommand{\OldVerHe}{PASS}
\newcommand{\OldSetN}{100}
\newcommand{\OldSetNCoco}{1{,}000}
\newcommand{\OldHzN}{100}
\newcommand{\OldHzMaxRel}{1.6\times10^{-15}}
\newcommand{\OldHzIdentical}{37}
\newcommand{\OldHoBitAgree}{99.999}
\newcommand{\OldHoOptDigAgree}{99.994}
\newcommand{\OldHoPoolOpt}{99.75}
\newcommand{\OldHoPoolDig}{99.88}
\newcommand{\OldHoPoolVopt}{99.62}
\newcommand{\OldHtwoMin}{99.0}
\newcommand{\OldRDigVdNone}{100}
\newcommand{\OldRDigVoNone}{100}
\newcommand{\OldRDigBerNone}{0}
\newcommand{\OldRDigVdJpegA}{100}
\newcommand{\OldRDigVoJpegA}{100}
\newcommand{\OldRDigBerJpegA}{0}
\newcommand{\OldRDigVdJpegB}{100}
\newcommand{\OldRDigVoJpegB}{100}
\newcommand{\OldRDigBerJpegB}{0}
\newcommand{\OldRDigVdJpegC}{100}
\newcommand{\OldRDigVoJpegC}{100}
\newcommand{\OldRDigBerJpegC}{9.8\times10^{-6}}
\newcommand{\OldRDigVdWebp}{100}
\newcommand{\OldRDigVoWebp}{99.0}
\newcommand{\OldRDigBerWebp}{2.7\times10^{-4}}
\newcommand{\OldRDigVdScaleA}{100}
\newcommand{\OldRDigVoScaleA}{100}
\newcommand{\OldRDigBerScaleA}{0}
\newcommand{\OldRDigVdScaleB}{100}
\newcommand{\OldRDigVoScaleB}{100}
\newcommand{\OldRDigBerScaleB}{0}
\newcommand{\OldRDigVdCombo}{99.0}
\newcommand{\OldRDigVoCombo}{99.0}
\newcommand{\OldRDigBerCombo}{4.7\times10^{-4}}
\newcommand{\OldRIdealVdNone}{100}
\newcommand{\OldRIdealVoNone}{100}
\newcommand{\OldRIdealBerNone}{0}
\newcommand{\OldRIdealLtNone}{100}
\newcommand{\OldRIdealZcNone}{12.0}
\newcommand{\OldRIdealVdJpegA}{100}
\newcommand{\OldRIdealVoJpegA}{100}
\newcommand{\OldRIdealBerJpegA}{0}
\newcommand{\OldRIdealLtJpegA}{100}
\newcommand{\OldRIdealZcJpegA}{12.0}
\newcommand{\OldRIdealVdJpegB}{100}
\newcommand{\OldRIdealVoJpegB}{100}
\newcommand{\OldRIdealBerJpegB}{0}
\newcommand{\OldRIdealLtJpegB}{100}
\newcommand{\OldRIdealZcJpegB}{11.6}
\newcommand{\OldRIdealVdJpegC}{100}
\newcommand{\OldRIdealVoJpegC}{100}
\newcommand{\OldRIdealBerJpegC}{0}
\newcommand{\OldRIdealLtJpegC}{99.0}
\newcommand{\OldRIdealZcJpegC}{10.9}
\newcommand{\OldRIdealVdWebp}{100}
\newcommand{\OldRIdealVoWebp}{100}
\newcommand{\OldRIdealBerWebp}{2.0\times10^{-4}}
\newcommand{\OldRIdealLtWebp}{95.0}
\newcommand{\OldRIdealZcWebp}{8.8}
\newcommand{\OldRIdealVdScaleA}{100}
\newcommand{\OldRIdealVoScaleA}{100}
\newcommand{\OldRIdealBerScaleA}{0}
\newcommand{\OldRIdealLtScaleA}{100}
\newcommand{\OldRIdealZcScaleA}{12.0}
\newcommand{\OldRIdealVdScaleB}{100}
\newcommand{\OldRIdealVoScaleB}{100}
\newcommand{\OldRIdealBerScaleB}{0}
\newcommand{\OldRIdealLtScaleB}{100}
\newcommand{\OldRIdealZcScaleB}{12.0}
\newcommand{\OldRIdealVdCombo}{99.0}
\newcommand{\OldRIdealVoCombo}{99.0}
\newcommand{\OldRIdealBerCombo}{5.5\times10^{-4}}
\newcommand{\OldRIdealLtCombo}{97.0}
\newcommand{\OldRIdealZcCombo}{9.9}
\newcommand{\OldROptDVdNone}{100}
\newcommand{\OldROptDVoNone}{100}
\newcommand{\OldROptDBerNone}{9.8\times10^{-6}}
\newcommand{\OldROptDLtNone}{100}
\newcommand{\OldROptDZcNone}{11.4}
\newcommand{\OldROptDVdJpegA}{100}
\newcommand{\OldROptDVoJpegA}{100}
\newcommand{\OldROptDBerJpegA}{2.0\times10^{-5}}
\newcommand{\OldROptDLtJpegA}{100}
\newcommand{\OldROptDZcJpegA}{11.4}
\newcommand{\OldROptDVdJpegB}{100}
\newcommand{\OldROptDVoJpegB}{100}
\newcommand{\OldROptDBerJpegB}{3.9\times10^{-5}}
\newcommand{\OldROptDLtJpegB}{100}
\newcommand{\OldROptDZcJpegB}{11.1}
\newcommand{\OldROptDVdJpegC}{100}
\newcommand{\OldROptDVoJpegC}{100}
\newcommand{\OldROptDBerJpegC}{3.9\times10^{-5}}
\newcommand{\OldROptDLtJpegC}{99.0}
\newcommand{\OldROptDZcJpegC}{10.4}
\newcommand{\OldROptDVdWebp}{99.0}
\newcommand{\OldROptDVoWebp}{99.0}
\newcommand{\OldROptDBerWebp}{5.5\times10^{-4}}
\newcommand{\OldROptDLtWebp}{93.0}
\newcommand{\OldROptDZcWebp}{8.4}
\newcommand{\OldROptDVdScaleA}{100}
\newcommand{\OldROptDVoScaleA}{100}
\newcommand{\OldROptDBerScaleA}{2.9\times10^{-5}}
\newcommand{\OldROptDLtScaleA}{100}
\newcommand{\OldROptDZcScaleA}{11.4}
\newcommand{\OldROptDVdScaleB}{100}
\newcommand{\OldROptDVoScaleB}{100}
\newcommand{\OldROptDBerScaleB}{2.0\times10^{-5}}
\newcommand{\OldROptDLtScaleB}{100}
\newcommand{\OldROptDZcScaleB}{11.4}
\newcommand{\OldROptDVdCombo}{99.0}
\newcommand{\OldROptDVoCombo}{98.0}
\newcommand{\OldROptDBerCombo}{4.6\times10^{-4}}
\newcommand{\OldROptDLtCombo}{96.0}
\newcommand{\OldROptDZcCombo}{9.3}
\newcommand{\OldROptSVdNone}{100}
\newcommand{\OldROptSVoNone}{100}
\newcommand{\OldROptSBerNone}{9.8\times10^{-6}}
\newcommand{\OldROptSLtNone}{100}
\newcommand{\OldROptSZcNone}{11.7}
\newcommand{\OldROptSVdJpegA}{100}
\newcommand{\OldROptSVoJpegA}{100}
\newcommand{\OldROptSBerJpegA}{0}
\newcommand{\OldROptSLtJpegA}{100}
\newcommand{\OldROptSZcJpegA}{11.6}
\newcommand{\OldROptSVdJpegB}{100}
\newcommand{\OldROptSVoJpegB}{100}
\newcommand{\OldROptSBerJpegB}{9.8\times10^{-6}}
\newcommand{\OldROptSLtJpegB}{100}
\newcommand{\OldROptSZcJpegB}{11.3}
\newcommand{\OldROptSVdJpegC}{100}
\newcommand{\OldROptSVoJpegC}{100}
\newcommand{\OldROptSBerJpegC}{5.9\times10^{-5}}
\newcommand{\OldROptSLtJpegC}{100}
\newcommand{\OldROptSZcJpegC}{10.4}
\newcommand{\OldROptSVdWebp}{100}
\newcommand{\OldROptSVoWebp}{98.0}
\newcommand{\OldROptSBerWebp}{3.8\times10^{-4}}
\newcommand{\OldROptSLtWebp}{95.0}
\newcommand{\OldROptSZcWebp}{8.5}
\newcommand{\OldROptSVdScaleA}{100}
\newcommand{\OldROptSVoScaleA}{100}
\newcommand{\OldROptSBerScaleA}{0}
\newcommand{\OldROptSLtScaleA}{100}
\newcommand{\OldROptSZcScaleA}{11.6}
\newcommand{\OldROptSVdScaleB}{100}
\newcommand{\OldROptSVoScaleB}{100}
\newcommand{\OldROptSBerScaleB}{9.8\times10^{-6}}
\newcommand{\OldROptSLtScaleB}{100}
\newcommand{\OldROptSZcScaleB}{11.7}
\newcommand{\OldROptSVdCombo}{99.0}
\newcommand{\OldROptSVoCombo}{99.0}
\newcommand{\OldROptSBerCombo}{3.5\times10^{-4}}
\newcommand{\OldROptSLtCombo}{98.0}
\newcommand{\OldROptSZcCombo}{9.5}
\newcommand{\OldROptWVdNone}{100}
\newcommand{\OldROptWVoNone}{100}
\newcommand{\OldROptWBerNone}{2.0\times10^{-5}}
\newcommand{\OldROptWLtNone}{100}
\newcommand{\OldROptWZcNone}{11.5}
\newcommand{\OldROptWVdJpegA}{100}
\newcommand{\OldROptWVoJpegA}{100}
\newcommand{\OldROptWBerJpegA}{2.0\times10^{-5}}
\newcommand{\OldROptWLtJpegA}{100}
\newcommand{\OldROptWZcJpegA}{11.5}
\newcommand{\OldROptWVdJpegB}{100}
\newcommand{\OldROptWVoJpegB}{100}
\newcommand{\OldROptWBerJpegB}{2.9\times10^{-5}}
\newcommand{\OldROptWLtJpegB}{100}
\newcommand{\OldROptWZcJpegB}{11.1}
\newcommand{\OldROptWVdJpegC}{100}
\newcommand{\OldROptWVoJpegC}{100}
\newcommand{\OldROptWBerJpegC}{4.9\times10^{-5}}
\newcommand{\OldROptWLtJpegC}{99.0}
\newcommand{\OldROptWZcJpegC}{10.5}
\newcommand{\OldROptWVdWebp}{99.0}
\newcommand{\OldROptWVoWebp}{99.0}
\newcommand{\OldROptWBerWebp}{5.6\times10^{-4}}
\newcommand{\OldROptWLtWebp}{94.0}
\newcommand{\OldROptWZcWebp}{8.4}
\newcommand{\OldROptWVdScaleA}{100}
\newcommand{\OldROptWVoScaleA}{100}
\newcommand{\OldROptWBerScaleA}{2.9\times10^{-5}}
\newcommand{\OldROptWLtScaleA}{100}
\newcommand{\OldROptWZcScaleA}{11.4}
\newcommand{\OldROptWVdScaleB}{100}
\newcommand{\OldROptWVoScaleB}{100}
\newcommand{\OldROptWBerScaleB}{2.0\times10^{-5}}
\newcommand{\OldROptWLtScaleB}{100}
\newcommand{\OldROptWZcScaleB}{11.5}
\newcommand{\OldROptWVdCombo}{99.0}
\newcommand{\OldROptWVoCombo}{99.0}
\newcommand{\OldROptWBerCombo}{4.7\times10^{-4}}
\newcommand{\OldROptWLtCombo}{97.0}
\newcommand{\OldROptWZcCombo}{9.3}
\newcommand{\OldQDigPfive}{47.5}
\newcommand{\OldQDigMed}{47.7}
\newcommand{\OldQDigMin}{47.5}
\newcommand{\OldQDigSsimPfive}{0.993}
\newcommand{\OldQDigConv}{92.0}
\newcommand{\OldQIdealPfive}{44.7}
\newcommand{\OldQIdealMed}{46.3}
\newcommand{\OldQIdealMin}{43.3}
\newcommand{\OldQIdealSsimPfive}{0.988}
\newcommand{\OldQIdealConv}{91.0}
\newcommand{\OldQOptDPfive}{37.2}
\newcommand{\OldQOptDMed}{42.9}
\newcommand{\OldQOptDMin}{33.5}
\newcommand{\OldQOptDSsimPfive}{0.978}
\newcommand{\OldQOptDConv}{97.0}
\newcommand{\OldQOptSPfive}{37.2}
\newcommand{\OldQOptSMed}{42.9}
\newcommand{\OldQOptSMin}{33.5}
\newcommand{\OldQOptSSsimPfive}{0.978}
\newcommand{\OldQOptSConv}{97.0}
\newcommand{\OldQOptWPfive}{44.3}
\newcommand{\OldQOptWMed}{46.0}
\newcommand{\OldQOptWMin}{43.0}
\newcommand{\OldQOptWSsimPfive}{0.987}
\newcommand{\OldQOptWConv}{97.0}
\newcommand{\OldHfCocoN}{1{,}000}
\newcommand{\OldHfCocoVd}{0}
\newcommand{\OldHfCocoVo}{0}
\newcommand{\OldHfDivN}{100}
\newcommand{\OldHfDivVd}{0}
\newcommand{\OldHfWrongPk}{0}
\newcommand{\OldHfWrongPkN}{800}
\newcommand{\OldHfWrongSeed}{0}
\newcommand{\OldHfWrongSeedN}{2{,}000}
\newcommand{\OldHfCpUpper}{0.0030}
\newcommand{\OldHfNullMax}{18.1}
\newcommand{\OldHvWone}{0}
\newcommand{\OldHvWtwo}{0}
\newcommand{\OldHvWthree}{0}
\newcommand{\OldHsMatchedDet}{100}
\newcommand{\OldHsScalePlus}{0}
\newcommand{\OldHsScaleMinus}{0}
\newcommand{\OldHsShiftOne}{0}
\newcommand{\OldHsTolRel}{0.68}
\newcommand{\OldHsTolShift}{0.78}
\newcommand{\OldHcScPZpZZZOMed}{129.7}
\newcommand{\OldHcScPZpZZZODet}{100}
\newcommand{\OldHcScMZpZZZOMed}{129.8}
\newcommand{\OldHcScMZpZZZODet}{100}
\newcommand{\OldHcScPZpZZZTMed}{130.2}
\newcommand{\OldHcScPZpZZZTDet}{100}
\newcommand{\OldHcScMZpZZZTMed}{129.7}
\newcommand{\OldHcScMZpZZZTDet}{100}
\newcommand{\OldHcScPZpZZZVMed}{128.2}
\newcommand{\OldHcScPZpZZZVDet}{100}
\newcommand{\OldHcScMZpZZZVMed}{128.7}
\newcommand{\OldHcScMZpZZZVDet}{100}
\newcommand{\OldHcScPZpZZOMed}{124.3}
\newcommand{\OldHcScPZpZZODet}{100}
\newcommand{\OldHcScMZpZZOMed}{124.3}
\newcommand{\OldHcScMZpZZODet}{100}
\newcommand{\OldHcScPZpZZTMed}{106.0}
\newcommand{\OldHcScPZpZZTDet}{100}
\newcommand{\OldHcScMZpZZTMed}{107.5}
\newcommand{\OldHcScMZpZZTDet}{100}
\newcommand{\OldHcScPZpZZHMed}{86.6}
\newcommand{\OldHcScPZpZZHDet}{99.0}
\newcommand{\OldHcScMZpZZHMed}{87.2}
\newcommand{\OldHcScMZpZZHDet}{100}
\newcommand{\OldHcScPZpZZVMed}{39.8}
\newcommand{\OldHcScPZpZZVDet}{86.0}
\newcommand{\OldHcScMZpZZVMed}{37.1}
\newcommand{\OldHcScMZpZZVDet}{88.0}
\newcommand{\OldHcScPZpZOMed}{0.5}
\newcommand{\OldHcScPZpZODet}{0}
\newcommand{\OldHcScMZpZOMed}{0.4}
\newcommand{\OldHcScMZpZODet}{0}
\newcommand{\OldHcScPZpZTMed}{0.4}
\newcommand{\OldHcScPZpZTDet}{0}
\newcommand{\OldHcScMZpZTMed}{0.7}
\newcommand{\OldHcScMZpZTDet}{0}
\newcommand{\OldHcShZpOMed}{125.8}
\newcommand{\OldHcShZpODet}{100}
\newcommand{\OldHcShZpTVMed}{113.9}
\newcommand{\OldHcShZpTVDet}{100}
\newcommand{\OldHcShZpVMed}{86.5}
\newcommand{\OldHcShZpVDet}{100}
\newcommand{\OldHcShZpNVMed}{28.0}
\newcommand{\OldHcShZpNVDet}{52.0}
\newcommand{\OldHcShOMed}{0.5}
\newcommand{\OldHcShODet}{0}
\newcommand{\OldHcShTMed}{0.4}
\newcommand{\OldHcShTDet}{0}
\newcommand{\OldHcPbOMed}{53.6}
\newcommand{\OldHcPbODet}{96.0}
\newcommand{\OldHcPbTMed}{107.6}
\newcommand{\OldHcPbTDet}{100}
\newcommand{\OldHcPbHMed}{125.6}
\newcommand{\OldHcPbHDet}{100}
\newcommand{\OldHcPbFMed}{129.3}
\newcommand{\OldHcPbFDet}{100}
\newcommand{\OldHcSwZpVMed}{128.6}
\newcommand{\OldHcSwZpVDet}{100}
\newcommand{\OldHcSwOMed}{126.2}
\newcommand{\OldHcSwODet}{100}
\newcommand{\OldHcSwTMed}{115.5}
\newcommand{\OldHcSwTDet}{100}
\newcommand{\OldHcSwVMed}{82.0}
\newcommand{\OldHcSwVDet}{100}
\newcommand{\OldHcSwOZMed}{59.4}
\newcommand{\OldHcSwOZDet}{100}
\newcommand{\OldHeN}{100}
\newcommand{\OldHeVerified}{0}
\newcommand{\OldHeCpUpper}{0.030}
\newcommand{\OldRcCompared}{966}
\newcommand{\OldRcMismatch}{0}
\newcommand{\OldHtN}{50}
\newcommand{\OldHtVnBslmOZ}{100}
\newcommand{\OldHtVjBslmOZ}{100}
\newcommand{\OldHtLtBslmOZ}{100}
\newcommand{\OldHtBerBslmOZ}{0}
\newcommand{\OldHtPsBslmOZ}{42.8}
\newcommand{\OldHtVnBslmF}{78.0}
\newcommand{\OldHtVjBslmF}{74.0}
\newcommand{\OldHtLtBslmF}{78.0}
\newcommand{\OldHtBerBslmF}{1.2\times10^{-2}}
\newcommand{\OldHtPsBslmF}{42.3}
\newcommand{\OldHtVnBslmH}{10.0}
\newcommand{\OldHtVjBslmH}{8.0}
\newcommand{\OldHtLtBslmH}{10.0}
\newcommand{\OldHtBerBslmH}{5.0\times10^{-2}}
\newcommand{\OldHtPsBslmH}{41.7}
\newcommand{\OldHtVnBslmN}{100}
\newcommand{\OldHtVjBslmN}{100}
\newcommand{\OldHtLtBslmN}{100}
\newcommand{\OldHtBerBslmN}{5.9\times10^{-5}}
\newcommand{\OldHtPsBslmN}{42.8}
\newcommand{\OldHtVnBslmS}{100}
\newcommand{\OldHtVjBslmS}{100}
\newcommand{\OldHtLtBslmS}{100}
\newcommand{\OldHtBerBslmS}{3.9\times10^{-5}}
\newcommand{\OldHtPsBslmS}{42.7}
\newcommand{\OldHtVnBslmV}{100}
\newcommand{\OldHtVjBslmV}{100}
\newcommand{\OldHtLtBslmV}{100}
\newcommand{\OldHtBerBslmV}{9.6\times10^{-4}}
\newcommand{\OldHtPsBslmV}{42.6}
\newcommand{\OldHtVnAfZpOTV}{100}
\newcommand{\OldHtVjAfZpOTV}{100}
\newcommand{\OldHtLtAfZpOTV}{100}
\newcommand{\OldHtBerAfZpOTV}{2.0\times10^{-5}}
\newcommand{\OldHtPsAfZpOTV}{42.7}
\newcommand{\OldHtVnAfZpZ}{100}
\newcommand{\OldHtVjAfZpZ}{100}
\newcommand{\OldHtLtAfZpZ}{100}
\newcommand{\OldHtBerAfZpZ}{0}
\newcommand{\OldHtPsAfZpZ}{42.7}
\newcommand{\OldHtVnAfZpV}{100}
\newcommand{\OldHtVjAfZpV}{100}
\newcommand{\OldHtLtAfZpV}{100}
\newcommand{\OldHtBerAfZpV}{0}
\newcommand{\OldHtPsAfZpV}{42.7}
\newcommand{\OldHtVnAfOpZ}{100}
\newcommand{\OldHtVjAfOpZ}{100}
\newcommand{\OldHtLtAfOpZ}{100}
\newcommand{\OldHtBerAfOpZ}{2.5\times10^{-4}}
\newcommand{\OldHtPsAfOpZ}{42.0}
\newcommand{\OldHtVnAfZpNV}{100}
\newcommand{\OldHtVjAfZpNV}{100}
\newcommand{\OldHtLtAfZpNV}{100}
\newcommand{\OldHtBerAfZpNV}{0}
\newcommand{\OldHtPsAfZpNV}{42.6}
\newcommand{\OldHtVnShiftZpZT}{100}
\newcommand{\OldHtVjShiftZpZT}{100}
\newcommand{\OldHtLtShiftZpZT}{100}
\newcommand{\OldHtBerShiftZpZT}{0}
\newcommand{\OldHtPsShiftZpZT}{42.7}
\newcommand{\OldHtVnShiftZpZV}{100}
\newcommand{\OldHtVjShiftZpZV}{100}
\newcommand{\OldHtLtShiftZpZV}{100}
\newcommand{\OldHtBerShiftZpZV}{0}
\newcommand{\OldHtPsShiftZpZV}{42.8}
\newcommand{\OldHtVnShiftZpO}{100}
\newcommand{\OldHtVjShiftZpO}{100}
\newcommand{\OldHtLtShiftZpO}{100}
\newcommand{\OldHtBerShiftZpO}{0}
\newcommand{\OldHtPsShiftZpO}{42.8}
\newcommand{\OldHtVnShiftZpT}{100}
\newcommand{\OldHtVjShiftZpT}{100}
\newcommand{\OldHtLtShiftZpT}{100}
\newcommand{\OldHtBerShiftZpT}{0}
\newcommand{\OldHtPsShiftZpT}{42.6}
\newcommand{\OldHtVnShiftZpHV}{98.0}
\newcommand{\OldHtVjShiftZpHV}{98.0}
\newcommand{\OldHtLtShiftZpHV}{98.0}
\newcommand{\OldHtBerShiftZpHV}{1.9\times10^{-3}}
\newcommand{\OldHtPsShiftZpHV}{41.0}
\newcommand{\OldHtVnShiftOpZ}{0}
\newcommand{\OldHtVjShiftOpZ}{0}
\newcommand{\OldHtLtShiftOpZ}{0}
\newcommand{\OldHtBerShiftOpZ}{5.1\times10^{-1}}
\newcommand{\OldHtPsShiftOpZ}{35.2}
\newcommand{\OldHtVnShiftZpV}{2.0}
\newcommand{\OldHtVjShiftZpV}{4.0}
\newcommand{\OldHtLtShiftZpV}{2.0}
\newcommand{\OldHtBerShiftZpV}{5.8\times10^{-2}}
\newcommand{\OldHtPsShiftZpV}{38.1}
\newcommand{\OldHtVnDzZpO}{100}
\newcommand{\OldHtVjDzZpO}{100}
\newcommand{\OldHtLtDzZpO}{100}
\newcommand{\OldHtBerDzZpO}{0}
\newcommand{\OldHtPsDzZpO}{42.4}
\newcommand{\OldHtVnDzZpTV}{100}
\newcommand{\OldHtVjDzZpTV}{100}
\newcommand{\OldHtLtDzZpTV}{100}
\newcommand{\OldHtBerDzZpTV}{0}
\newcommand{\OldHtPsDzZpTV}{37.3}
\newcommand{\OldHtVnDzZpV}{100}
\newcommand{\OldHtVjDzZpV}{100}
\newcommand{\OldHtLtDzZpV}{100}
\newcommand{\OldHtBerDzZpV}{0}
\newcommand{\OldHtPsDzZpV}{30.0}
\newcommand{\OldHtVnDzOpZ}{100}
\newcommand{\OldHtVjDzOpZ}{100}
\newcommand{\OldHtLtDzOpZ}{100}
\newcommand{\OldHtBerDzOpZ}{2.0\times10^{-5}}
\newcommand{\OldHtPsDzOpZ}{26.1}
\newcommand{\OldHtVnDzTpZ}{98.0}
\newcommand{\OldHtVjDzTpZ}{98.0}
\newcommand{\OldHtLtDzTpZ}{98.0}
\newcommand{\OldHtBerDzTpZ}{8.2\times10^{-4}}
\newcommand{\OldHtPsDzTpZ}{23.8}
\newcommand{\OldHtVnDzFpZ}{98.0}
\newcommand{\OldHtVjDzFpZ}{98.0}
\newcommand{\OldHtLtDzFpZ}{98.0}
\newcommand{\OldHtBerDzFpZ}{7.6\times10^{-3}}
\newcommand{\OldHtPsDzFpZ}{22.3}
\newcommand{\OldHtVnDzEpZ}{50.0}
\newcommand{\OldHtVjDzEpZ}{50.0}
\newcommand{\OldHtLtDzEpZ}{50.0}
\newcommand{\OldHtBerDzEpZ}{4.8\times10^{-2}}
\newcommand{\OldHtPsDzEpZ}{21.0}
\newcommand{\OldHtVnSneHZZZpZ}{100}
\newcommand{\OldHtVjSneHZZZpZ}{100}
\newcommand{\OldHtLtSneHZZZpZ}{98.0}
\newcommand{\OldHtBerSneHZZZpZ}{0}
\newcommand{\OldHtPsSneHZZZpZ}{37.4}
\newcommand{\OldHtVnSneOZZZpZ}{98.0}
\newcommand{\OldHtVjSneOZZZpZ}{98.0}
\newcommand{\OldHtLtSneOZZZpZ}{96.0}
\newcommand{\OldHtBerSneOZZZpZ}{1.8\times10^{-3}}
\newcommand{\OldHtPsSneOZZZpZ}{32.5}
\newcommand{\OldHtVnSneHZZpZ}{0}
\newcommand{\OldHtVjSneHZZpZ}{0}
\newcommand{\OldHtLtSneHZZpZ}{0}
\newcommand{\OldHtBerSneHZZpZ}{9.2\times10^{-2}}
\newcommand{\OldHtPsSneHZZpZ}{25.9}
\newcommand{\OldHtVnSneHZZZZpZ}{100}
\newcommand{\OldHtVjSneHZZZZpZ}{100}
\newcommand{\OldHtLtSneHZZZZpZ}{100}
\newcommand{\OldHtBerSneHZZZZpZ}{2.0\times10^{-5}}
\newcommand{\OldHtPsSneHZZZZpZ}{42.2}
\newcommand{\OldHtVnSneOZZZZpZ}{100}
\newcommand{\OldHtVjSneOZZZZpZ}{100}
\newcommand{\OldHtLtSneOZZZZpZ}{98.0}
\newcommand{\OldHtBerSneOZZZZpZ}{0}
\newcommand{\OldHtPsSneOZZZZpZ}{40.8}
\newcommand{\OldHtVnSneOZZpZ}{0}
\newcommand{\OldHtVjSneOZZpZ}{0}
\newcommand{\OldHtLtSneOZZpZ}{0}
\newcommand{\OldHtBerSneOZZpZ}{3.6\times10^{-1}}
\newcommand{\OldHtPsSneOZZpZ}{19.6}
\newcommand{\OldHtVnSbitsE}{100}
\newcommand{\OldHtVjSbitsE}{100}
\newcommand{\OldHtLtSbitsE}{100}
\newcommand{\OldHtBerSbitsE}{0}
\newcommand{\OldHtPsSbitsE}{40.9}
\newcommand{\OldHtVnSbitsOZ}{100}
\newcommand{\OldHtVjSbitsOZ}{100}
\newcommand{\OldHtLtSbitsOZ}{100}
\newcommand{\OldHtBerSbitsOZ}{0}
\newcommand{\OldHtPsSbitsOZ}{42.7}
\newcommand{\OldHtVnSbitsS}{96.0}
\newcommand{\OldHtVjSbitsS}{96.0}
\newcommand{\OldHtLtSbitsS}{96.0}
\newcommand{\OldHtBerSbitsS}{2.5\times10^{-3}}
\newcommand{\OldHtPsSbitsS}{34.9}
\newcommand{\OldHtVnFcamOxOT}{26.0}
\newcommand{\OldHtVjFcamOxOT}{20.0}
\newcommand{\OldHtLtFcamOxOT}{26.0}
\newcommand{\OldHtBerFcamOxOT}{1.2\times10^{-1}}
\newcommand{\OldHtPsFcamOxOT}{40.9}
\newcommand{\OldHtVnFcamTxOT}{100}
\newcommand{\OldHtVjFcamTxOT}{100}
\newcommand{\OldHtLtFcamTxOT}{100}
\newcommand{\OldHtBerFcamTxOT}{0}
\newcommand{\OldHtPsFcamTxOT}{42.5}
\newcommand{\OldHtVnFcamHxE}{100}
\newcommand{\OldHtVjFcamHxE}{100}
\newcommand{\OldHtLtFcamHxE}{100}
\newcommand{\OldHtBerFcamHxE}{0}
\newcommand{\OldHtPsFcamHxE}{42.6}
\newcommand{\OldHtVnFcamHxOZ}{100}
\newcommand{\OldHtVjFcamHxOZ}{100}
\newcommand{\OldHtLtFcamHxOZ}{100}
\newcommand{\OldHtBerFcamHxOZ}{0}
\newcommand{\OldHtPsFcamHxOZ}{42.7}
\newcommand{\OldHtVnTiltTpZ}{100}
\newcommand{\OldHtVjTiltTpZ}{100}
\newcommand{\OldHtLtTiltTpZ}{100}
\newcommand{\OldHtBerTiltTpZ}{0}
\newcommand{\OldHtPsTiltTpZ}{41.9}
\newcommand{\OldHtVnTiltOpZ}{100}
\newcommand{\OldHtVjTiltOpZ}{100}
\newcommand{\OldHtLtTiltOpZ}{100}
\newcommand{\OldHtBerTiltOpZ}{0}
\newcommand{\OldHtPsTiltOpZ}{42.4}
\newcommand{\OldHtVnTiltVpZ}{98.0}
\newcommand{\OldHtVjTiltVpZ}{98.0}
\newcommand{\OldHtLtTiltVpZ}{98.0}
\newcommand{\OldHtBerTiltVpZ}{1.6\times10^{-3}}
\newcommand{\OldHtPsTiltVpZ}{41.1}
\newcommand{\OldHtVnTiltOZpZ}{0}
\newcommand{\OldHtVjTiltOZpZ}{0}
\newcommand{\OldHtLtTiltOZpZ}{0}
\newcommand{\OldHtBerTiltOZpZ}{1.3\times10^{-1}}
\newcommand{\OldHtPsTiltOZpZ}{34.9}
\newcommand{\OldHtVnGammaZpTV}{100}
\newcommand{\OldHtVjGammaZpTV}{100}
\newcommand{\OldHtLtGammaZpTV}{6.0}
\newcommand{\OldHtBerGammaZpTV}{0}
\newcommand{\OldHtPsGammaZpTV}{43.3}
\newcommand{\OldHtVnGammaZpV}{100}
\newcommand{\OldHtVjGammaZpV}{100}
\newcommand{\OldHtLtGammaZpV}{82.0}
\newcommand{\OldHtBerGammaZpV}{0}
\newcommand{\OldHtPsGammaZpV}{43.2}
\newcommand{\OldHtVnGammaZpNV}{100}
\newcommand{\OldHtVjGammaZpNV}{100}
\newcommand{\OldHtLtGammaZpNV}{100}
\newcommand{\OldHtBerGammaZpNV}{0}
\newcommand{\OldHtPsGammaZpNV}{42.9}
\newcommand{\OldHtVnPupilISpZ}{100}
\newcommand{\OldHtVjPupilISpZ}{100}
\newcommand{\OldHtLtPupilISpZ}{100}
\newcommand{\OldHtBerPupilISpZ}{0}
\newcommand{\OldHtPsPupilISpZ}{24.3}
\newcommand{\OldHtVnPupilOTEpZ}{100}
\newcommand{\OldHtVjPupilOTEpZ}{100}
\newcommand{\OldHtLtPupilOTEpZ}{100}
\newcommand{\OldHtBerPupilOTEpZ}{2.0\times10^{-5}}
\newcommand{\OldHtPsPupilOTEpZ}{25.7}
\newcommand{\OldHtVnPupilOSZpZ}{100}
\newcommand{\OldHtVjPupilOSZpZ}{100}
\newcommand{\OldHtLtPupilOSZpZ}{100}
\newcommand{\OldHtBerPupilOSZpZ}{0}
\newcommand{\OldHtPsPupilOSZpZ}{26.9}
\newcommand{\OldHtVnPupilNTFpZ}{100}
\newcommand{\OldHtVjPupilNTFpZ}{100}
\newcommand{\OldHtLtPupilNTFpZ}{100}
\newcommand{\OldHtBerPupilNTFpZ}{2.0\times10^{-5}}
\newcommand{\OldHtPsPupilNTFpZ}{45.9}
\newcommand{\OldHtVnPupilTVSpZ}{100}
\newcommand{\OldHtVjPupilTVSpZ}{100}
\newcommand{\OldHtLtPupilTVSpZ}{100}
\newcommand{\OldHtBerPupilTVSpZ}{0}
\newcommand{\OldHtPsPupilTVSpZ}{30.6}
\newcommand{\OldBudBslm}{5}
\newcommand{\OldBudShift}{0.35}
\newcommand{\OldBudDz}{4.0}
\newcommand{\OldBudSne}{1\times10^{3}}
\newcommand{\OldBudTilt}{5}
\newcommand{\XvDigOptNone}{99.9}
\newcommand{\XvDigOptJpegA}{99.9}
\newcommand{\XvDigOptJpegB}{99.8}
\newcommand{\XvDigOptJpegC}{99.8}
\newcommand{\XvDigOptWebp}{99.5}
\newcommand{\XvDigOptScaleA}{99.8}
\newcommand{\XvDigOptScaleB}{99.9}
\newcommand{\XvDigOptCombo}{99.3}
\newcommand{\XvMinRate}{99.3}
\newcommand{\HwBenchBaseW}{668}
\newcommand{\HwBenchBaseD}{747}
\newcommand{\HwBenchF}{230.36}
\newcommand{\HwBenchPartTypes}{25}
\newcommand{\HwPortL}{132.7}
\newcommand{\HwPortW}{74.3}
\newcommand{\HwPortH}{29.6}
\newcommand{\VtLambda}{532}
\newcommand{\VtN}{1024}
\newcommand{\VtObjPitch}{8.0}
\newcommand{\VtObjW}{8.192}
\newcommand{\VtObjDiag}{11.59}
\newcommand{\VtBinPitch}{14.96}
\newcommand{\VtModPitch}{3.74}
\newcommand{\VtPxPerBin}{4}
\newcommand{\VtFzero}{230.36}
\newcommand{\VtAf}{0.25}
\newcommand{\VtActivePx}{3.74}
\newcommand{\VtBinsLoneIn}{16}
\newcommand{\VtRLoneIn}{0.24}
\newcommand{\VtBinsLoneOut}{80}
\newcommand{\VtRLoneOut}{1.20}
\newcommand{\VtBinsLtwoOut}{144}
\newcommand{\VtRLtwoOut}{2.15}
\newcommand{\VtBinsRoi}{160}
\newcommand{\VtBinsPupil}{512}
\newcommand{\VtPupilD}{15.32}
\newcommand{\VtPupilPx}{4096}
\newcommand{\VtLtwoPx}{1152}
\newcommand{\VtRoiD}{4.79}
\newcommand{\VtStopD}{0.24}
\newcommand{\VtPixShift}{0.25}
\newcommand{\VtResidShift}{0.125}
\newcommand{\VtBIn}{10}
\newcommand{\VtBSlm}{8}
\newcommand{\VtBPhi}{8}
\newcommand{\VtFcNe}{3\times10^{4}}
\newcommand{\VtFcRead}{3}
\newcommand{\VtFcBits}{12}
\newcommand{\VtFcNexp}{3}
\newcommand{\VtFcGain}{16}
\newcommand{\VtSNe}{1.2\times10^{5}}
\newcommand{\VtSRead}{6}
\newcommand{\VtSBits}{12}
\newcommand{\VtSFrames}{4}
\newcommand{\VtSWell}{3\times10^{4}}
\newcommand{\VtGamma}{0.95}
\newcommand{\VtCube}{25.4}
\newcommand{\VtGap}{12}
\newcommand{\VtXthree}{75}
\newcommand{\VtOptD}{25.4}
\newcommand{\VtFcol}{100}
\newcommand{\VtRail}{20}
\newcommand{\VtAxisH}{60}
\newcommand{\VtGlassN}{1.5195}
\newcommand{\VtFootW}{668}
\newcommand{\VtFootD}{747}
\newcommand{\VtLegAone}{201.6}
\newcommand{\VtLegAthree}{135.3}
\newcommand{\VtLegGthree}{78.3}
\newcommand{\VtArmMismatch}{106}
\newcommand{\VtRailN}{387}
\newcommand{\VtRailS}{588}
\newcommand{\VtRailW}{369}
\newcommand{\VtRailRef}{369}
\newcommand{\VtRailStub}{121}
\newcommand{\VtCaLens}{26.9}
\newcommand{\VtCaComb}{20.1}
\newcommand{\VtCaCubeHave}{22.9}
\newcommand{\VtCaLensHave}{45.7}
\newcommand{\VtInBitsClass}{8}
\newcommand{\VtPanelPxW}{4160}
\newcommand{\VtPanelPxH}{2464}
\newcommand{\VtPanelHalfW}{520}
\newcommand{\VtPanelHalfH}{308}
\newcommand{\VtPanelFrac}{71.7}
\newcommand{\VtRefRange}{3.98}
\newcommand{\VtRefRangeDb}{12}
\newcommand{\VtRefRelLo}{3.8\times10^{-6}}
\newcommand{\VtRefRelHi}{5.9\times10^{-5}}
\newcommand{\VtThrCombInv}{256}
\newcommand{\VtThrCubeInv}{64}
\newcommand{\VtThrInv}{6.5\times10^{4}}
\newcommand{\VtThrFrameNj}{11.7}
\newcommand{\VtThrFrameMj}{0.76}
\newcommand{\VtThrExp}{0.1}
\newcommand{\VtThrPower}{7.6}
\newcommand{\VtThrPowerFull}{0.03}
\newcommand{\VtBudN}{877}
\newcommand{\VtBudMin}{95}
\newcommand{\VtNomPsnr}{42.6}
\newcommand{\VtShiftBudUm}{5.2}
\newcommand{\VtShiftFailUm}{7.5}
\newcommand{\VtResidShiftUm}{1.87}
\newcommand{\VtTiltFore}{0.38}
\newcommand{\VtTiltForeFail}{1.52}
\newcommand{\VtTiltEdgeBins}{0.30}
\newcommand{\VtKeyN}{877}
\newcommand{\VtKeyScaleTol}{0.70}
\newcommand{\VtKeyShiftTol}{0.77}
\newcommand{\VtKeyShiftTolUm}{11.6}
\newcommand{\VtKeyScaleOk}{0.3}
\newcommand{\VtKeyScaleOkDetP}{99.8}
\newcommand{\VtKeyScaleOkDetM}{99.8}
\newcommand{\VtKeyScaleOkNm}{1.6}
\newcommand{\VtKeyScaleOkFmm}{0.69}
\newcommand{\VtKeyScaleBad}{0.5}
\newcommand{\VtKeyScaleBadDetP}{88.8}
\newcommand{\VtKeyScaleBadDetM}{87.0}
\newcommand{\VtKeyShiftOk}{0.5}
\newcommand{\VtKeyShiftOkDet}{99.8}
\newcommand{\VtKeyShiftOkUm}{7.5}
\newcommand{\VtKeyShiftBad}{0.75}
\newcommand{\VtKeyShiftBadDet}{57.8}
\newcommand{\VtBudShift}{0.35}
\newcommand{\VtBudVnShift}{96.0}
\newcommand{\VtBudVjShift}{95.7}
\newcommand{\VtNxtShift}{0.5}
\newcommand{\VtNxtVnShift}{8.2}
\newcommand{\VtNxtVjShift}{7.3}
\newcommand{\VtBudTilt}{5}
\newcommand{\VtBudVnTilt}{99.0}
\newcommand{\VtBudVjTilt}{99.0}
\newcommand{\VtNxtTilt}{10}
\newcommand{\VtNxtVnTilt}{0}
\newcommand{\VtNxtVjTilt}{0}
\newcommand{\VtBudDz}{4}
\newcommand{\VtBudVnDz}{98.7}
\newcommand{\VtBudVjDz}{98.6}
\newcommand{\VtNxtDz}{8}
\newcommand{\VtNxtVnDz}{54.0}
\newcommand{\VtNxtVjDz}{51.7}
\newcommand{\VtBudBslm}{5}
\newcommand{\VtBudVnBslm}{99.9}
\newcommand{\VtBudVjBslm}{99.9}
\newcommand{\VtNxtBslm}{4}
\newcommand{\VtNxtVnBslm}{73.9}
\newcommand{\VtNxtVjBslm}{70.6}
\newcommand{\VtBudAf}{1}
\newcommand{\VtBudVnAf}{99.2}
\newcommand{\VtBudVjAf}{99.4}
\newcommand{\VtBudSne}{$10^{3}$}
\newcommand{\VtBudVnSne}{99.1}
\newcommand{\VtBudVjSne}{98.7}
\newcommand{\VtNxtSne}{300}
\newcommand{\VtNxtVnSne}{0.7}
\newcommand{\VtNxtVjSne}{0.7}
\newcommand{\VtBudSbits}{6}
\newcommand{\VtBudVnSbits}{98.3}
\newcommand{\VtBudVjSbits}{98.3}
\newcommand{\VtBudPupil}{96}
\newcommand{\VtBudVnPupil}{100}
\newcommand{\VtBudVjPupil}{100}
\newcommand{\VtBudGamma}{0.25}
\newcommand{\VtBudVnGamma}{99.9}
\newcommand{\VtBudVjGamma}{99.9}
\newcommand{\VtBslmMargin}{3}
\newcommand{\VtBracketLo}{0.1}
\newcommand{\VtBracketHi}{0.2}
\newcommand{\VtVnBracketLo}{99.9}
\newcommand{\VtVnBracketHi}{100}
\newcommand{\VtDzA}{0.1}
\newcommand{\VtPsDzA}{42.2}
\newcommand{\VtDzB}{0.25}
\newcommand{\VtPsDzB}{37.0}
\newcommand{\VtPupilSmall}{256}
\newcommand{\VtPsPupilSmall}{30.5}
\newcommand{\VtVnnPupilSmall}{100}
\newcommand{\VtGamA}{0.75}
\newcommand{\VtLttGamA}{99.5}
\newcommand{\VtGamB}{0.5}
\newcommand{\VtLttGamB}{88.6}
\newcommand{\VtGamC}{0.25}
\newcommand{\VtLttGamC}{8.2}
\newcommand{\VtSneA}{$10^{4}$}
\newcommand{\VtPsSneA}{40.6}
\newcommand{\VtSneB}{$10^{3}$}
\newcommand{\VtPsSneB}{32.6}
\newcommand{\VtSbitsA}{8}
\newcommand{\VtPsSbitsA}{40.9}
\newcommand{\VtSbitsB}{6}
\newcommand{\VtPsSbitsB}{35.5}
\newcommand{\VtLttAfFull}{98.9}
\newcommand{\VtPhaseA}{38}
\newcommand{\VtPhasePathA}{56}
\newcommand{\VtPhaseB}{58}
\newcommand{\VtXN}{877}
\newcommand{\VtXNbits}{1024}
\newcommand{\VtXNatt}{8}
\newcommand{\VtXNtrials}{7016}
\newcommand{\VtXkDpNone}{877}
\newcommand{\VtXloDpNone}{99.58}
\newcommand{\VtXhiDpNone}{100}
\newcommand{\VtXkDpJpegA}{877}
\newcommand{\VtXloDpJpegA}{99.58}
\newcommand{\VtXhiDpJpegA}{100}
\newcommand{\VtXkDpJpegB}{877}
\newcommand{\VtXloDpJpegB}{99.58}
\newcommand{\VtXhiDpJpegB}{100}
\newcommand{\VtXkDpJpegC}{877}
\newcommand{\VtXloDpJpegC}{99.58}
\newcommand{\VtXhiDpJpegC}{100}
\newcommand{\VtXkDpWebp}{876}
\newcommand{\VtXloDpWebp}{99.36}
\newcommand{\VtXhiDpWebp}{100.00}
\newcommand{\VtXkDpScaleA}{877}
\newcommand{\VtXloDpScaleA}{99.58}
\newcommand{\VtXhiDpScaleA}{100}
\newcommand{\VtXkDpScaleB}{877}
\newcommand{\VtXloDpScaleB}{99.58}
\newcommand{\VtXhiDpScaleB}{100}
\newcommand{\VtXkDpCombo}{872}
\newcommand{\VtXloDpCombo}{98.67}
\newcommand{\VtXhiDpCombo}{99.82}
\newcommand{\VtXkDpAll}{871}
\newcommand{\VtXloDpAll}{98.51}
\newcommand{\VtXhiDpAll}{99.75}
\newcommand{\VtXkDpPool}{7010}
\newcommand{\VtXrDpPool}{99.9}
\newcommand{\VtXkDoNone}{876}
\newcommand{\VtXloDoNone}{99.36}
\newcommand{\VtXhiDoNone}{100.00}
\newcommand{\VtXkDoJpegA}{876}
\newcommand{\VtXloDoJpegA}{99.36}
\newcommand{\VtXhiDoJpegA}{100.00}
\newcommand{\VtXkDoJpegB}{875}
\newcommand{\VtXloDoJpegB}{99.17}
\newcommand{\VtXhiDoJpegB}{99.98}
\newcommand{\VtXkDoJpegC}{875}
\newcommand{\VtXloDoJpegC}{99.17}
\newcommand{\VtXhiDoJpegC}{99.98}
\newcommand{\VtXkDoWebp}{873}
\newcommand{\VtXloDoWebp}{98.83}
\newcommand{\VtXhiDoWebp}{99.88}
\newcommand{\VtXkDoScaleA}{875}
\newcommand{\VtXloDoScaleA}{99.17}
\newcommand{\VtXhiDoScaleA}{99.98}
\newcommand{\VtXkDoScaleB}{876}
\newcommand{\VtXloDoScaleB}{99.36}
\newcommand{\VtXhiDoScaleB}{100.00}
\newcommand{\VtXkDoCombo}{871}
\newcommand{\VtXloDoCombo}{98.51}
\newcommand{\VtXhiDoCombo}{99.75}
\newcommand{\VtXkDoAll}{869}
\newcommand{\VtXloDoAll}{98.21}
\newcommand{\VtXhiDoAll}{99.61}
\newcommand{\VtXkDoPool}{6997}
\newcommand{\VtXrDoPool}{99.7}
\newcommand{\VtXkOpNone}{876}
\newcommand{\VtXloOpNone}{99.36}
\newcommand{\VtXhiOpNone}{100.00}
\newcommand{\VtXkOpJpegA}{877}
\newcommand{\VtXloOpJpegA}{99.58}
\newcommand{\VtXhiOpJpegA}{100}
\newcommand{\VtXkOpJpegB}{877}
\newcommand{\VtXloOpJpegB}{99.58}
\newcommand{\VtXhiOpJpegB}{100}
\newcommand{\VtXkOpJpegC}{876}
\newcommand{\VtXloOpJpegC}{99.36}
\newcommand{\VtXhiOpJpegC}{100.00}
\newcommand{\VtXkOpWebp}{874}
\newcommand{\VtXloOpWebp}{99.00}
\newcommand{\VtXhiOpWebp}{99.93}
\newcommand{\VtXkOpScaleA}{876}
\newcommand{\VtXloOpScaleA}{99.36}
\newcommand{\VtXhiOpScaleA}{100.00}
\newcommand{\VtXkOpScaleB}{877}
\newcommand{\VtXloOpScaleB}{99.58}
\newcommand{\VtXhiOpScaleB}{100}
\newcommand{\VtXkOpCombo}{873}
\newcommand{\VtXloOpCombo}{98.83}
\newcommand{\VtXhiOpCombo}{99.88}
\newcommand{\VtXkOpAll}{870}
\newcommand{\VtXloOpAll}{98.36}
\newcommand{\VtXhiOpAll}{99.68}
\newcommand{\VtXkOpPool}{7006}
\newcommand{\VtXrOpPool}{99.9}
\newcommand{\VtXkOoNone}{875}
\newcommand{\VtXloOoNone}{99.17}
\newcommand{\VtXhiOoNone}{99.98}
\newcommand{\VtXkOoJpegA}{876}
\newcommand{\VtXloOoJpegA}{99.36}
\newcommand{\VtXhiOoJpegA}{100.00}
\newcommand{\VtXkOoJpegB}{875}
\newcommand{\VtXloOoJpegB}{99.17}
\newcommand{\VtXhiOoJpegB}{99.98}
\newcommand{\VtXkOoJpegC}{875}
\newcommand{\VtXloOoJpegC}{99.17}
\newcommand{\VtXhiOoJpegC}{99.98}
\newcommand{\VtXkOoWebp}{872}
\newcommand{\VtXloOoWebp}{98.67}
\newcommand{\VtXhiOoWebp}{99.82}
\newcommand{\VtXkOoScaleA}{875}
\newcommand{\VtXloOoScaleA}{99.17}
\newcommand{\VtXhiOoScaleA}{99.98}
\newcommand{\VtXkOoScaleB}{876}
\newcommand{\VtXloOoScaleB}{99.36}
\newcommand{\VtXhiOoScaleB}{100.00}
\newcommand{\VtXkOoCombo}{871}
\newcommand{\VtXloOoCombo}{98.51}
\newcommand{\VtXhiOoCombo}{99.75}
\newcommand{\VtXkOoAll}{868}
\newcommand{\VtXloOoAll}{98.06}
\newcommand{\VtXhiOoAll}{99.53}
\newcommand{\VtXkOoPool}{6995}
\newcommand{\VtXrOoPool}{99.7}
\newcommand{\VtXcDBoth}{6997}
\newcommand{\VtXcDPubOnly}{13}
\newcommand{\VtXcDOptOnly}{0}
\newcommand{\VtXcDNeither}{6}
\newcommand{\VtXcDSame}{7003}
\newcommand{\VtXcDBothNone}{876}
\newcommand{\VtXcDPubNone}{1}
\newcommand{\VtXcDOptNone}{0}
\newcommand{\VtXcDNeiNone}{0}
\newcommand{\VtXbDMeanNone}{0.040}
\newcommand{\VtXbDMaxNone}{26}
\newcommand{\VtXcDBothJpegA}{876}
\newcommand{\VtXcDPubJpegA}{1}
\newcommand{\VtXcDOptJpegA}{0}
\newcommand{\VtXcDNeiJpegA}{0}
\newcommand{\VtXbDMeanJpegA}{0.049}
\newcommand{\VtXbDMaxJpegA}{29}
\newcommand{\VtXcDBothJpegB}{875}
\newcommand{\VtXcDPubJpegB}{2}
\newcommand{\VtXcDOptJpegB}{0}
\newcommand{\VtXcDNeiJpegB}{0}
\newcommand{\VtXbDMeanJpegB}{0.060}
\newcommand{\VtXbDMaxJpegB}{25}
\newcommand{\VtXcDBothJpegC}{875}
\newcommand{\VtXcDPubJpegC}{2}
\newcommand{\VtXcDOptJpegC}{0}
\newcommand{\VtXcDNeiJpegC}{0}
\newcommand{\VtXbDMeanJpegC}{0.151}
\newcommand{\VtXbDMaxJpegC}{44}
\newcommand{\VtXcDBothWebp}{873}
\newcommand{\VtXcDPubWebp}{3}
\newcommand{\VtXcDOptWebp}{0}
\newcommand{\VtXcDNeiWebp}{1}
\newcommand{\VtXbDMeanWebp}{0.300}
\newcommand{\VtXbDMaxWebp}{64}
\newcommand{\VtXcDBothScaleA}{875}
\newcommand{\VtXcDPubScaleA}{2}
\newcommand{\VtXcDOptScaleA}{0}
\newcommand{\VtXcDNeiScaleA}{0}
\newcommand{\VtXbDMeanScaleA}{0.091}
\newcommand{\VtXbDMaxScaleA}{41}
\newcommand{\VtXcDBothScaleB}{876}
\newcommand{\VtXcDPubScaleB}{1}
\newcommand{\VtXcDOptScaleB}{0}
\newcommand{\VtXcDNeiScaleB}{0}
\newcommand{\VtXbDMeanScaleB}{0.040}
\newcommand{\VtXbDMaxScaleB}{25}
\newcommand{\VtXcDBothCombo}{871}
\newcommand{\VtXcDPubCombo}{1}
\newcommand{\VtXcDOptCombo}{0}
\newcommand{\VtXcDNeiCombo}{5}
\newcommand{\VtXbDMeanCombo}{0.308}
\newcommand{\VtXbDMaxCombo}{77}
\newcommand{\VtXbDMeanAll}{0.130}
\newcommand{\VtXbDMaxAll}{77}
\newcommand{\VtXcOBoth}{6995}
\newcommand{\VtXcOPubOnly}{11}
\newcommand{\VtXcOOptOnly}{0}
\newcommand{\VtXcONeither}{10}
\newcommand{\VtXcOSame}{7005}
\newcommand{\VtXcOBothNone}{875}
\newcommand{\VtXcOPubNone}{1}
\newcommand{\VtXcOOptNone}{0}
\newcommand{\VtXcONeiNone}{1}
\newcommand{\VtXbOMeanNone}{0.124}
\newcommand{\VtXbOMaxNone}{52}
\newcommand{\VtXcOBothJpegA}{876}
\newcommand{\VtXcOPubJpegA}{1}
\newcommand{\VtXcOOptJpegA}{0}
\newcommand{\VtXcONeiJpegA}{0}
\newcommand{\VtXbOMeanJpegA}{0.114}
\newcommand{\VtXbOMaxJpegA}{46}
\newcommand{\VtXcOBothJpegB}{875}
\newcommand{\VtXcOPubJpegB}{2}
\newcommand{\VtXcOOptJpegB}{0}
\newcommand{\VtXcONeiJpegB}{0}
\newcommand{\VtXbOMeanJpegB}{0.151}
\newcommand{\VtXbOMaxJpegB}{67}
\newcommand{\VtXcOBothJpegC}{875}
\newcommand{\VtXcOPubJpegC}{1}
\newcommand{\VtXcOOptJpegC}{0}
\newcommand{\VtXcONeiJpegC}{1}
\newcommand{\VtXbOMeanJpegC}{0.202}
\newcommand{\VtXbOMaxJpegC}{62}
\newcommand{\VtXcOBothWebp}{872}
\newcommand{\VtXcOPubWebp}{2}
\newcommand{\VtXcOOptWebp}{0}
\newcommand{\VtXcONeiWebp}{3}
\newcommand{\VtXbOMeanWebp}{0.384}
\newcommand{\VtXbOMaxWebp}{67}
\newcommand{\VtXcOBothScaleA}{875}
\newcommand{\VtXcOPubScaleA}{1}
\newcommand{\VtXcOOptScaleA}{0}
\newcommand{\VtXcONeiScaleA}{1}
\newcommand{\VtXbOMeanScaleA}{0.170}
\newcommand{\VtXbOMaxScaleA}{62}
\newcommand{\VtXcOBothScaleB}{876}
\newcommand{\VtXcOPubScaleB}{1}
\newcommand{\VtXcOOptScaleB}{0}
\newcommand{\VtXcONeiScaleB}{0}
\newcommand{\VtXbOMeanScaleB}{0.117}
\newcommand{\VtXbOMaxScaleB}{54}
\newcommand{\VtXcOBothCombo}{871}
\newcommand{\VtXcOPubCombo}{2}
\newcommand{\VtXcOOptCombo}{0}
\newcommand{\VtXcONeiCombo}{4}
\newcommand{\VtXbOMeanCombo}{0.387}
\newcommand{\VtXbOMaxCombo}{84}
\newcommand{\VtXbOMeanAll}{0.206}
\newcommand{\VtXbOMaxAll}{84}
\newcommand{\VtXminK}{871}
\newcommand{\VtXmaxK}{877}
\newcommand{\VtXminRate}{99.3}
\newcommand{\VtXminLo}{98.51}
\newcommand{\VtXminAllK}{868}
\newcommand{\VtXmaxGap}{0.46}
\newcommand{\VtXminAgree}{99.96}
\newcommand{\VtXmaxDis}{0.39}
\newcommand{\VtXminDis}{0.04}
\newcommand{\VtXoptOnly}{0}
\newcommand{\VtXmaxAllK}{871}
\newcommand{\VtXmaxDisBits}{84}
\newcommand{\VtXnotVerified}{0}
\newcommand{\VtXfailUnion}{12}
\newcommand{\VtXfailOptOnly}{2}
\newcommand{\VtXnoPayload}{56}
\newcommand{\VtXallTrials}{28064}
\newcommand{\VtCadParts}{25}
\newcommand{\VtCadUsed}{24}
\newcommand{\VtCadInstances}{84}
\newcommand{\VtCadVolume}{1569}
\newcommand{\VtCadUnionDiff}{0.001}
\newcommand{\VtCadBuildX}{220}
\newcommand{\VtCadBuildZ}{250}
\newcommand{\VtCadRails}{5}
\newcommand{\VtCarL}{30}
\newcommand{\VtCarW}{40}
\newcommand{\VtPanShift}{1.5}
\newcommand{\VtSlideTravel}{3}
\newcommand{\VtCadMargin}{1}
\newcommand{\VbPLambda}{520}
\newcommand{\VbPObjPitch}{3.74}
\newcommand{\VbPObjW}{3.83}
\newcommand{\VbPBin}{7.48}
\newcommand{\VbPPxPerBin}{2}
\newcommand{\VbPF}{55.09}
\newcommand{\VbPTwoF}{110.2}
\newcommand{\VbPAf}{0.5}
\newcommand{\VbPPupil}{7.66}
\newcommand{\VbPPupilPx}{2048}
\newcommand{\VbPHalfField}{2.8}
\newcommand{\VbPRegRes}{0.25}
\newcommand{\VbPRingLL}{1.08}
\newcommand{\VbPStop}{0.12}
\newcommand{\VbPThroughGain}{16}
\newcommand{\VbPFramesWell}{12}
\newcommand{\VbPBitsNeeded}{9216}
\newcommand{\VbPBitsHave}{9638}
\newcommand{\VbEnvTarget}{150$\times$75$\times$30}
\newcommand{\VbPAltLambda}{638}
\newcommand{\VbPAltF}{44.9}
\newcommand{\VbLabF}{230.36}
\newcommand{\VbLabFourF}{921}
\newcommand{\VbPupilBins}{512}
\newcommand{\VbPanelPx}{4160$\times$2464}
\newcommand{\VbEnvLen}{150}
\newcommand{\VbLensPLen}{14.1}
\newcommand{\VbLensPFfd}{47.9}
\newcommand{\VbLensPBfd}{50.8}
\newcommand{\VbPOutL}{132.7}
\newcommand{\VbPOutW}{74.3}
\newcommand{\VbPOutH}{29.6}
\newcommand{\VbPVol}{292}
\newcommand{\VbPMarginL}{17.3}
\newcommand{\VbPMarginW}{0.7}
\newcommand{\VbPMarginH}{0.4}
\newcommand{\VbPNParts}{seven}
\newcommand{\VbPClear}{1}
\newcommand{\AppPreregDate}{16 November 2026}
\newcommand{\AppDevWhenA}{16 Nov., 21:30}
\newcommand{\AppDevWhenB}{16 Nov., 21:30}
\newcommand{\AppDevWhenC}{16 Nov., 21:35}
\newcommand{\AppDevWhenD}{16 Nov., 22:08}
\newcommand{\AppDevWhenE}{16 Nov., 22:08}
\newcommand{\AppDevWhenF}{16 Nov., 22:28}
\newcommand{\AppDevWhenG}{16 Nov., 23:08}
\newcommand{\AppDevWhenH}{16 Nov., 23:45}
\newcommand{\AppDevWhenI}{17 Nov., 00:05}
\newcommand{\AppDevWhenJ}{17 Nov., 00:31}
\newcommand{\AppDevN}{9}
\newcommand{\SpLabLensPubK}{875}
\newcommand{\SpLabLensPubPct}{99.8}
\newcommand{\SpLabLensOptK}{875}
\newcommand{\SpLabLensOptPct}{99.8}
\newcommand{\SpLabTolPubK}{877}
\newcommand{\SpLabTolPubPct}{100}
\newcommand{\SpLabTolOptK}{874}
\newcommand{\SpLabTolOptPct}{99.7}
\newcommand{\SpPortTolPubK}{850}
\newcommand{\SpPortTolPubPct}{96.9}
\newcommand{\SpPortTolOptK}{837}
\newcommand{\SpPortTolOptPct}{95.4}
\newcommand{\SpN}{877}
\newcommand{\SpNSettings}{8}
\newcommand{\SpFrames}{16}
\newcommand{\SpNegN}{1{,}877}
\newcommand{\SpNegWideN}{40{,}670}
\newcommand{\SpNegOverlap}{1{,}000}
\newcommand{\SpTolCamReg}{0.0025}
\newcommand{\SpTolCamRegPx}{0.01}
\newcommand{\SpTolShift}{0.125}
\newcommand{\SpTolScalePct}{0.02}
\newcommand{\SpTolDTLab}{0.75}
\newcommand{\SpTolDTPort}{2.55}

\definecolor{cvprblue}{rgb}{0.21,0.49,0.74}
\usepackage[breaklinks,colorlinks,allcolors=cvprblue]{hyperref}
\hypersetup{pdftitle={Asymmetric Optical Image Watermarking},pdfauthor={Guang Yang}}

\def\confName{CVPR}
\def\confYear{2027}

\title{Asymmetric Optical Image Watermarking}

\author{Guang Yang\\
University of California, Los Angeles\\[2pt]
{\normalsize\href{https://github.com/GY19A/asymmetric-optical-image-watermarking}{\faGithub\ \texttt{github.com/GY19A/asymmetric-optical-image-watermarking}}}}

\begin{document}
\maketitle
\begin{abstract}
Watermark detectors held by a provider cannot be checked by third parties, and a party that can detect a mark can
usually also forge it. We study a two-layer image watermark in which the Fourier transform at the core of the pipeline
is performed by a coherent optical 4F unit. Layer~1 carries an Ed25519 signature bound to the image content in Fourier
magnitudes, so anyone with the public key verifies offline. Layer~2 is a key-defined carrier written by a secret
phase hologram and read by a matched-filter correlator. The optics is a scalar-diffraction simulation with finite
modulator bits, pixel aperture, pupil, defocus, misalignment, and sensor noise. No hardware is built. All criteria
were fixed in a hashed specification before any optical result. The simulated optical signer reproduces the digital pipeline (field error
$\HzMaxRel$, bit agreement \HoBitAgree\%), verifies at least \HtwoMin\% of \DesNEval{} images under each of
\DesNAttacks{} post-sensor settings, one unattacked, and accepts none of \HfCocoN{} unmarked images. The optical carrier is
key-selective: the matched key detects it on \HsDetOptDNone\% of the unattacked images, and detection falls to
\HsScalePlus\% under a scale error of \CrScaleTest\% and to \HsShiftOne\% under a mask shift of \CrShiftTest{} bin.
Security rests on the signature and the content binding,
for which we separate what is provable from what is measured.
\end{abstract}

\begin{figure*}[t]
\centering
\includegraphics{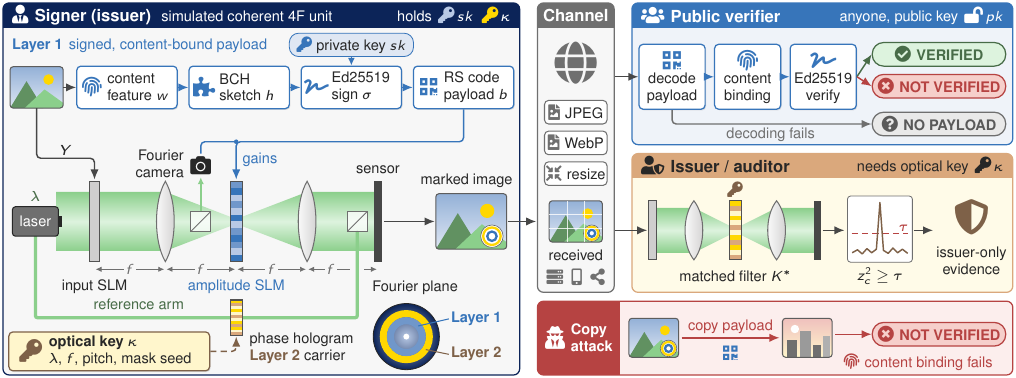}
\caption{Two-layer architecture. Left: the signer is a simulated coherent 4F unit. A Fourier-plane camera reads the
magnitudes of the luminance spectrum, and an amplitude spatial light modulator (SLM) writes the targets that carry the Ed25519-signed,
content-bound payload (Layer 1). A reference arm with a secret phase hologram adds a key-defined carrier (Layer 2).
Right, top: after the channel, anyone with the public key reads Layer 1 and checks the content binding and the
signature digitally. Right, middle: only a holder of the optical key can run the correlator that detects Layer 2.
Bottom: a payload copied to another image fails the content binding. All optics in this paper are numerical simulation.}
\label{fig:pipeline}
\end{figure*}

\section{Introduction}\label{sec:intro}

Invisible watermarks label the output of image generators and attach provenance to photographs~\cite{gowal2025synthid,fernandez2023stablesignature,bui2023trustmark}.
In these systems detection is not a cryptographic proof of origin.
Where one secret key embeds and detects, no third party can check and a key holder can forge~\cite{zhao2024sok}; even with a public decoder, a mark can be copied onto an unrelated image~\cite{kutter2000copyattack,dong2026residual}.
Signed provenance metadata~\cite{c2pa2025spec} is publicly verifiable, but embedded metadata can be stripped, and soft bindings are optional recovery aids.

A public-key design addresses both weaknesses and lets a third party check AI-generated content without trusting the provider's detector.
The issuer signs a message together with a compact description of the image content, and the pixels carry the signature.
Anyone with the public key extracts the payload and checks it offline, and a payload copied to a different image fails the content check.
Only part of this is provable.
A forged VERIFIED image implies a signature forgery or a collision of the content binding (Sec.~\ref{sec:security}).
Survival under compression or resizing is an empirical property of the embedding channel, and removal remains possible~\cite{zhao2024regeneration,zhang2023sand}.
The idea is not new~\cite{zhou2026metaseal,graf2026deepsignature,choi2025readme,yakushev2026cpw}; our public layer follows that pattern with a deterministic Fourier-magnitude channel.

The core transform of such a watermark is a Fourier transform, which a lens performs on a coherent field~\cite{goodman1968fourier}.
Two lenses with a filter in their shared focal plane form the 4F processor, the classical instrument for spatial filtering and matched-filter correlation~\cite{vanderlugt1964}.
Optical encryption by double random phase encoding~\cite{refregier1995} places secret phase masks in the same geometry as a symmetric key, and fell to chosen-ciphertext and plaintext attacks~\cite{carnicer2005,peng2006kpa,frauel2007}.
We therefore ask a two-part question.
Can an optical 4F unit serve as the transform core of a public-key-verifiable watermark, and can the same unit carry a key-selective mark for the issuer?

We study a two-layer design (Fig.~\ref{fig:pipeline}).
Layer~1 is public.
An Ed25519 signature~\cite{bernstein2012highspeed} over a message and a content feature is written into Fourier magnitudes of the luminance.
In the signer, a lens forms the spectrum, a camera reads its magnitudes, and an amplitude modulator sets the target values.
Layer~2 is private.
A secret phase hologram in a reference arm of the same unit adds a key-defined carrier, which only a holder of the optical key detects with a matched-filter correlator.
We model the unit by scalar diffraction with finite modulator resolution, pixel aperture, pupil, defocus, misalignment, sensor noise, and quantization.
All criteria (Tab.~\ref{tab:hyp}) were fixed in a hashed specification before any optical result existed.

In this numerical simulation the public layer survives the optics.
With all non-idealities off, the simulated optical write equals the digital pipeline up to a relative field error of $\HzMaxRel$.
With the nominal non-ideal optics, the public verifier reads back \HoBitAgree\% of the embedded bits, and its VERIFIED rate is at least \HtwoMin\% for each of \DesNAttacks{} attack settings (no attack, compression, downscaling).
No unmarked image among \HfCocoN{} is accepted.
The private layer is key-selective.
The matched correlator detects the carrier on \HsDetOptDNone\% of the unattacked images, and detection falls to \HsScalePlus\% under a scale error of \CrScaleTest\% and to \HsShiftOne\% under a mask shift of \CrShiftTest{} bin.
Security rests on Layer~1, and the optical carrier is issuer-side supporting evidence.

Every optical result in this paper comes from a numerical model; no hardware was built, and speed, energy, and cost are outside the scope of this paper.
Our contributions are threefold.
First, a two-layer architecture in which a publicly verifiable signature and an issuer-only optical carrier share one 4F unit.
Second, a scalar-diffraction model of the optical signer and verifier, evaluated under a pre-registered protocol with an independent recount.
Third, a simulated tolerance budget of the optical components and a measurement of the key selectivity of the optical carrier.

\section{Related Work}\label{sec:related}

\paragraph{Learned and generation-time watermarks.}
Encoder and decoder networks trained through differentiable distortions hide a bit string in an existing image~\cite{zhu2018hidden,bui2023trustmark,fernandez2024videoseal,soucek2025pixelseal,petrov2025chunkyseal}, and some also localize the marked region~\cite{sander2025wam}.
Generation-time methods place the mark in the decoder weights or in the initial noise of a diffusion model~\cite{fernandez2023stablesignature,wen2023treering,yang2024gaussianshading}.
A provider-operated system reports deployment at internet scale~\cite{gowal2025synthid}.
Their detection gives neither cryptographic unforgeability nor public verifiability of origin, even with a released decoder.
Stress tests and regeneration attacks show that pixel-level marks can be removed~\cite{an2024waves,zhao2024regeneration}, and strong watermarking is impossible given quality and perturbation oracles and a mixing assumption~\cite{zhang2023sand}.

\paragraph{Public-key and signed watermarking.}
Asymmetric detection was studied early for spread-spectrum marks, with detectors that need no embedding secret~\cite{furon1999asymmetric,eggers2000publickey}.
Formal treatments define watermark security and copy resistance~\cite{hopper2007weak,adelsbach2003provably}, construct publicly detectable marks for language models~\cite{fairoze2023publicly}, and analyze why a robust, publicly detectable image mark is hard to build~\cite{fairoze2025difficulty}.
Recent image systems embed a digital signature that is bound to the image content~\cite{zhou2026metaseal,graf2026deepsignature,choi2025readme,yakushev2026cpw}.
Asymmetric phase coding embeds Ed25519 signatures in audio~\cite{apca2026} and video~\cite{apcv2026}.
Fuzzy commitments~\cite{juels1999fuzzy} and secure sketches~\cite{dodis2008fuzzy} let such a binding tolerate small feature changes.
The binding answers the copy attack~\cite{kutter2000copyattack}, whose success on learned marks has been measured as residual transferability~\cite{dong2026residual}, and a benchmark shows that perceptual fingerprints can themselves be forged or evaded~\cite{kazempourradi2026softbind}.
We do not claim a new public-key watermark concept.
Our difference is the optical transform core, the pre-registered measurements of optical tolerance and key sensitivity, and the two-layer trust model.

\paragraph{Optical correlation and optical encryption.}
A lens forms the Fourier transform of a coherent field in its focal plane, and a 4F system filters that spectrum~\cite{goodman1968fourier}.
A matched spatial filter in the Fourier plane turns the system into a correlator~\cite{vanderlugt1964}.
Double random phase encoding~\cite{refregier1995} uses secret phase masks in a 4F system as a symmetric cipher.
The system is linear, and its masks are recovered by chosen-ciphertext, known-plaintext, and chosen-plaintext attacks~\cite{carnicer2005,peng2006kpa,frauel2007}.
Many later optical security schemes~\cite{javidi2016roadmap} and physical one-way functions~\cite{pappu2002} rely on a secret key or a physical secret.
Here the optics computes the transform of a public-key-verifiable watermark, and no public claim depends on an optical secret.
We measure the key sensitivity of the optical carrier and place no security claim on it.
Passive diffractive layers compute optically~\cite{lin2018d2nn,wetzstein2020}, and hybrid 4F systems add an electronic classifier~\cite{chang2018hybrid}; we borrow their wave-optics modeling practice and make no speed or energy claim.

\section{Method}\label{sec:method}

Figure~\ref{fig:pipeline} shows the architecture.
Every optical component in this paper is a numerical simulation.

\begin{table}[t]
\centering
\footnotesize
\begin{tabular}{@{}ll@{}}
\toprule
Parameter & Value \\
\midrule
Wavelength $\lambda$ & \DesLambda\,nm \\
Object pixel pitch $\Delta x$, width $L$ & \DesPix\,$\mu$m, \DesWidth\,mm \\
Focal length $f$ & \DesFocal\,mm \\
Fourier-plane bin pitch $\lambda f/L$ & \DesBinPitch\,$\mu$m \\
Modulator pixel pitch & \DesSlmPitch\,$\mu$m \\
Input, modulator bits $b_{\mathrm{in}}$, $b_T$ & \DesBIn, \DesBSlm \\
Pixel aperture fraction $a_f$ & \DesAf \\
Pupil radius $R_p$ & \DesPupil\ bins \\
Fourier camera: full well, read noise & $\DesFcNe$\,e$^-$, \DesFcRead\,e$^-$ \\
Fourier camera: ADC bits, exposures & \DesFcBits, \DesFcNexp \\
Sensor: full well $N_s$, read noise $\sigma_r$ & $\DesSNe$\,e$^-$, \DesSRead\,e$^-$ \\
Sensor ADC bits $b_s$, hologram bits $b_\phi$ & \DesSBits, \DesBPhi \\
Arm coherence $\gamma$ & \DesGamma \\
\bottomrule
\end{tabular}
\caption{Nominal configuration of the simulated 4F unit.
The ideal configuration switches off every quantization, noise source, aperture, and pupil, and uses full coherence.}
\label{tab:optics}
\end{table}

\subsection{Overview and trust model}\label{sec:overview}

A signer marks an $N\times N$ RGB frame $X$, $N=\DesN$, in two disjoint rings of the Fourier spectrum of its luminance $Y$.
Layer~1 is public: an Ed25519 signature bound to the image content, written into Fourier magnitudes with a public layout.
Layer~2 is private: a key-defined carrier that a secret phase hologram adds in a reference arm of the same unit.
The signer holds the signing key $sk$ and the optical key $\kappa$.
The public verifier holds only the public key $pk$, and an auditor acting for the issuer holds $\kappa$.
Layer~2 is a symmetric, physical secret: whoever can detect it can also write it, so it gives no public verifiability.
The combination gives verification by anyone plus issuer-side evidence, and security claims rest on Layer~1 (Sec.~\ref{sec:security}).

\subsection{Layer 1: signed, content-bound payload}\label{sec:layer1}

\paragraph{Feature, sketch and signature.}
The content feature $w=c(X)\in\{0,1\}^{\DesBchN}$ holds the signs of the lowest-frequency AC coefficients of the DCT of a small grayscale thumbnail, taken against their median.
Benign edits flip a few bits of $w$, so we store a secure sketch~\cite{dodis2008fuzzy}.
The helper $h=\mathrm{SS}(w)$ is the syndrome of $w$ in a binary BCH[\DesBchN,\DesBchK] code, and $\mathrm{Rec}(w',h)$ returns $w$ for any $w'$ within Hamming distance $t=\DesBchT$.
The signer computes $\sigma=\mathrm{Sign}_{sk}(H(\mathrm{dom}\,\|\,m\,\|\,h\,\|\,H(w)))$ with Ed25519~\cite{bernstein2012highspeed,rfc8032}.
Here $H$ is SHA-256, $m$ a short message, and $\mathrm{dom}$ a fixed domain string; the digest includes $H(w)$ because $h$ alone does not determine $w$.
A Reed--Solomon code with \DesRsParity{} parity bytes extends the \DesDataBytes{} bytes of data $D=(m,\sigma,h)$ to \DesPayloadBytes{} bytes, the \DesKbits{} payload bits $b$.

\paragraph{Magnitude channel.}
Let $F=\mathcal{F}\{Y\}$ be the unitary DFT of the luminance, with bin index $\mathbf{u}$ and radius $\rho=\|\mathbf{u}\|$.
The ring $\mathcal{R}_1=\{\DesRingLo\le\rho<\DesRingMid\}$ holds \DesLoneBins{} bins in one half plane.
A public seeded permutation gives bit $j$ a set of $n=\DesBinsPerBit$ bins, signs $s_i=\pm1/\sqrt{n}$ and a dither $d_j$.
Spread-transform dither modulation moves the projection $p_j=\sum_i s_i\,|F(\mathbf{u}_i)|$ to the nearest point of a shifted lattice:
\begin{equation}
p_j^{\star}=\Delta\left(\left\lfloor \frac{p_j}{\Delta}-d_j-\frac{b_j}{2}\right\rceil+d_j+\frac{b_j}{2}\right),
\label{eq:stdm}
\end{equation}
where $\lfloor\cdot\rceil$ rounds to the nearest integer and $\Delta=\DesStep$ at the nominal strength of \DesNu\,dB.
Each magnitude changes by $(p_j^{\star}-p_j)\,s_i$, is clamped at zero, and keeps its phase.
The decoder recomputes $p_j$ and outputs the bit whose sub-lattice is nearer.
Layout, dither and step are public, so anyone can read or write this channel.

\paragraph{Signing loop and verification.}
Embedding can flip feature bits, so the signer signs the current feature, embeds, re-reads it from the marked image, and repeats until they agree, at most \DesEnrolIter{} times.
The verifier resamples the received image $X'$ to the frame size, decodes the \DesKbits{} bits, and applies Reed--Solomon decoding; a failure gives NO PAYLOAD.
It computes $\hat w=\mathrm{Rec}(c(X'),h)$ and checks $\sigma$ against $H(\mathrm{dom}\,\|\,m\,\|\,h\,\|\,H(\hat w))$ under $pk$.
The outcome is VERIFIED if the check passes, else NOT VERIFIED.

\subsection{The 4F unit as transform core}\label{sec:optics}

We model a coherent 4F unit with scalar, paraxial, thin-lens, monochromatic Fourier optics~\cite{goodman1968fourier}, sampled on the image grid.
Polarization, aberrations other than defocus, pixel crosstalk, and stray light are not modeled.
Table~\ref{tab:optics} lists the nominal configuration.

\paragraph{Fourier plane and camera.}
An input amplitude modulator with $b_{\mathrm{in}}$ bits displays $Y$ as a real, non-negative field in the front focal plane of the first lens, whose back focal plane holds the scaled Fourier transform.
On the grid of DFT bins,
\begin{equation}
F(\mathbf{u})=\frac{1}{N}\sum_{\mathbf{x}}Y(\mathbf{x})\,e^{-2\pi i\,\mathbf{u}\cdot\mathbf{x}/N},
\qquad
\boldsymbol{\xi}=\frac{\lambda f}{L}\,\mathbf{u},
\label{eq:fourier}
\end{equation}
where $\boldsymbol{\xi}$ is the physical position of bin $\mathbf{u}$ and $L=N\Delta x$ is the object width.
A beam splitter taps the Fourier plane, where a camera behind a zero-order stop samples $|F(\mathbf{u})|^2$ at the bin centers.
It combines \DesFcNexp{} exposures of increasing gain, each with shot noise, read noise, saturation, and an ADC, into the magnitude estimate $\hat a(\mathbf{u})$.

\begin{table}[t]
\centering
\footnotesize
\begin{tabular}{@{}p{0.34\linewidth}@{\hspace{6pt}}p{0.39\linewidth}@{\hspace{6pt}}p{0.20\linewidth}@{}}
\toprule
criterion & pre-registered threshold & outcome \\
\midrule
ideal-optics equivalence & ideal optics equals FFT: error $<$ $\CrHzero$, same 8-bit output & \VerHz{} (field level; see text) \\
optical fidelity & bit agreement $\ge$ \CrAgree\%; rates within \CrGap{} points & \VerHo \\
post-sensor robustness & VERIFIED $\ge$ \CrMin\% after each attack & \VerHtwo \\
tolerance sweeps & one parameter at a time & descriptive \\
false acceptance & no false accept and no false detection & \VerHf \\
wrong variants & wrong layout seed, wrong public key, dark-field signer never VERIFIED & \VerHv \\
key sensitivity & matched key $\ge$ \CrMin\%; mismatched $\le$ \CrMax\% & \VerHs \\
payload copy & no copied payload VERIFIED & \VerHe \\
\bottomrule
\end{tabular}
\caption{Pre-registered criteria of the results reported here, abbreviated, and their outcomes in the numerical simulation on \DesNEval{} images.
For key sensitivity, mismatched means \CrScaleTest\% of scale error or \CrShiftTest{} bin of mask shift.
}
\label{tab:hyp}
\end{table}

\paragraph{Modulated pass.}
An amplitude-only modulator in the Fourier plane writes Layer~1.
With $a^{\star}$ the target magnitudes from Eq.~\eqref{eq:stdm}, the gain is $g=a^{\star}/\hat a$ on $\mathcal{R}_1$ and its conjugate bins, and $g=1$ elsewhere.
The simulated pass is
\begin{equation}
\begin{aligned}
T(\mathbf{u})&=Q_{b_T}\!\left[T_0\,g(\mathbf{u})\right],\\
t(\mathbf{d})&=t_T(\mathbf{d})\,\eta(\mathbf{d})\,e^{2\pi i\,\mathbf{s}\cdot\mathbf{d}/N},\\
E&=T_0^{-1}\,\mathcal{F}^{-1}\!\left\{\mathcal{H}\,\mathcal{F}\{Y\ast t\}\right\}.
\end{aligned}
\label{eq:modulator}
\end{equation}
Here $Q_b$ rounds to $b$ bits on $[0,1]$, and the bias $T_0\approx1/2$ allows gains up to $1/T_0$.
The sampled mask has the periodic impulse response $t_T$, with $Y\ast t_T=\mathcal{F}^{-1}\{TF\}$.
A pixel aperture of linear fraction $a_f$ multiplies it by the envelope $\eta(\mathbf{d})=\mathrm{sinc}(a_f d_y/N)\,\mathrm{sinc}(a_f d_x/N)$, and a lateral modulator shift of $\mathbf{s}$ bins adds the linear phase.
A tilt $\theta$ is modeled as foreshortening by $\cos\theta$ along one axis.
The transfer function $\mathcal{H}(\mathbf{k})=P(\mathbf{k})\exp(-i\pi\lambda\,\Delta z\,|\mathbf{k}|^2/L^2)$ holds a circular pupil $P$ of radius $R_p$ and the paraxial phase of a defocus $\Delta z$.
With all non-idealities off, Eq.~\eqref{eq:modulator} reduces to $E=\mathcal{F}^{-1}\{gF\}$, the digital embedding.

\paragraph{Sensor.}
With the Layer-2 reference field $r$ of Sec.~\ref{sec:layer2} and arm coherence $\gamma$, the image sensor is modeled as
\begin{equation}
\begin{aligned}
I&=|E+\gamma r|^{2}+(1-\gamma^{2})\,|r|^{2},\\
n_e&\sim\mathrm{Poisson}\!\left(N_s I/A^{2}\right)+\mathcal{N}\!\left(0,\sigma_r^{2}\right),\\
\hat Y&=A\,\sqrt{Q_{b_s}\!\left[n_e/N_s\right]},
\end{aligned}
\label{eq:sensor}
\end{equation}
with full-scale amplitude $A$.
The marked image is $\hat Y$, recombined digitally with the host chroma.
The square-root readout is intentional.
In the linear model, $E=Y+v$ with a real mark $v$, and the real, positive host field is its own in-line reference.
The intensity keeps the sign of $v$ in its cross term, and the square root returns $Y+v$ wherever $Y+v\ge0$.
At near-black pixels the field can turn negative and is rectified.
The ideal unit therefore matches the digital pipeline at field level, not in every rounded pixel (Sec.~\ref{sec:equiv}).

\paragraph{Optical signer and verifier.}
The signer displays the host, measures $\hat a$ on both rings, and computes the targets of Eq.~\eqref{eq:stdm} and the gains digitally.
The unit writes the gains with the reference arm on, and the sensor records the marked image.
The signer re-displays it, re-measures the projections, and rewrites from the host if they miss the targets, at most \DesRefinePasses{} times.
The optical verifier displays the received image on a second unit and reads the magnitudes on $\mathcal{R}_1$ with its Fourier camera; Layer~1 needs no correlation filter.
Decoding, the sketch, and the signature check stay digital, as in the public verifier, which is software only.

\subsection{Layer 2: optical key and correlator}\label{sec:layer2}

\paragraph{Hologram and carrier.}
A phase-only modulator in the Fourier plane of the reference arm displays a secret mask $K(\mathbf{u})=e^{i\phi_k(\mathbf{u})}$.
Its support is the ring $\mathcal{R}_2=\{\DesRingMid\le\rho<\DesRingHi\}$ with $n_2=\DesLtwoBins$ half-plane bins, disjoint from $\mathcal{R}_1$.
The phases are pseudorandom, uniform and quantized to $b_\phi$ bits.
The carrier at the sensor is
\begin{equation}
r(\mathbf{x})=\beta\,\eta_r(\mathbf{x})\,\mathcal{F}^{-1}\{K\}(\mathbf{x}),
\qquad
K(-\mathbf{u})=K^{*}(\mathbf{u}),
\label{eq:carrier}
\end{equation}
which is real by symmetry; $\eta_r$ is the pixel-aperture envelope.
The amplitude is $\beta=z_t/(\sqrt{2n_2}\,\langle 1/(2\hat a)\rangle)$, with target statistic $z_t=\DesZTarget$ and $\langle\cdot\rangle$ the average over the host reading on $\mathcal{R}_2$.
It is clipped so that the PSNR of Layer~2 alone lies between \DesLtwoPsnrLo{} and \DesLtwoPsnrHi\,dB.
The static mask seeds the phases with the mask key alone, like a fixed phase plate.
The per-image mask seeds them with an HMAC of the signed data $D$ under the mask key, so each image has its own mask.

\paragraph{Correlator.}
The auditor displays the received luminance on a 4F unit with Fourier-plane filter $W$.
This is a whitened Vander Lugt matched-filter correlator~\cite{vanderlugt1964}:
\begin{equation}
\begin{aligned}
W(\mathbf{u})&=K^{*}(\mathbf{u})\,\min\!\left(1,\,a_0/\hat a(\mathbf{u})\right),\\
C(\mathbf{x})&=\mathcal{F}^{-1}\{F'\,W\}(\mathbf{x}),\qquad
z_c=\frac{C(\mathbf{0})}{\mathrm{rms}(C)}.
\end{aligned}
\label{eq:corr}
\end{equation}
Here $F'$ is the received spectrum, $\hat a$ its camera magnitudes, and $a_0$ a low percentile of $\hat a$ over $\mathcal{R}_2$.
The rms excludes a small window at the origin.
An intensity sensor observes $z_c^2$; the carrier is detected if and only if $z_c^2\ge\tau=\DesThr$.

\paragraph{Optical parameters as key.}
The optical key is $\kappa=(\lambda,f,\text{pitch},\text{seed},\text{phase bits},\text{position})$; by Eq.~\eqref{eq:fourier}, the product $\lambda f$ sets the spatial frequency that a mask pixel addresses.
An auditor unit with relative scale mismatch $\delta=\lambda'f'/(\lambda f)-1$ and mask shift $\mathbf{s}$ meets the spectrum at $(\mathbf{u}+\mathbf{s})/(1+\delta)$ instead of $\mathbf{u}$.
A bin at radius $\rho$ moves by about $\delta\rho$ bins, so the peak should vanish once this nears one bin (Sec.~\ref{sec:keysens}).
For the per-image mask the auditor derives the mask from the data $D$ that the public verifier decodes.
If no payload is readable, Layer~2 counts as not detected.

\section{Security Analysis}\label{sec:security}

\begin{table}[t]
\centering
\footnotesize
\begin{tabular}{@{}l@{\hspace{4pt}}c@{\hspace{4pt}}c@{\hspace{4pt}}c@{\hspace{4pt}}c@{\hspace{4pt}}c@{\hspace{4pt}}c@{\hspace{6pt}}c@{\hspace{4pt}}c@{}}
\toprule
 & \multicolumn{6}{c}{Layer 1 VERIFIED (\%)} & \multicolumn{2}{c}{Layer 2 (\%)} \\
\cmidrule(r){2-7}\cmidrule(l){8-9}
signer & \multicolumn{2}{c}{digital} & ideal & \multicolumn{2}{c}{nominal} & static & nominal & static \\
verifier & pub. & opt. & pub. & pub. & opt. & pub. & corr. & corr. \\
\midrule
none & \RDigVdNone & \XvDigOptNone & \RIdealVdNone & \ROptDVdNone & \ROptDVoNone & \ROptSVdNone & \ROptDLtNone & \ROptSLtNone \\
JPEG 95 & \RDigVdJpegA & \XvDigOptJpegA & \RIdealVdJpegA & \ROptDVdJpegA & \ROptDVoJpegA & \ROptSVdJpegA & \ROptDLtJpegA & \ROptSLtJpegA \\
JPEG 75 & \RDigVdJpegB & \XvDigOptJpegB & \RIdealVdJpegB & \ROptDVdJpegB & \ROptDVoJpegB & \ROptSVdJpegB & \ROptDLtJpegB & \ROptSLtJpegB \\
JPEG 50 & \RDigVdJpegC & \XvDigOptJpegC & \RIdealVdJpegC & \ROptDVdJpegC & \ROptDVoJpegC & \ROptSVdJpegC & \ROptDLtJpegC & \ROptSLtJpegC \\
WebP 75 & \RDigVdWebp & \XvDigOptWebp & \RIdealVdWebp & \ROptDVdWebp & \ROptDVoWebp & \ROptSVdWebp & \ROptDLtWebp & \ROptSLtWebp \\
resize 0.5x & \RDigVdScaleA & \XvDigOptScaleA & \RIdealVdScaleA & \ROptDVdScaleA & \ROptDVoScaleA & \ROptSVdScaleA & \ROptDLtScaleA & \ROptSLtScaleA \\
resize 0.75x & \RDigVdScaleB & \XvDigOptScaleB & \RIdealVdScaleB & \ROptDVdScaleB & \ROptDVoScaleB & \ROptSVdScaleB & \ROptDLtScaleB & \ROptSLtScaleB \\
JPEG 75, 0.5x & \RDigVdCombo & \XvDigOptCombo & \RIdealVdCombo & \ROptDVdCombo & \ROptDVoCombo & \ROptSVdCombo & \ROptDLtCombo & \ROptSLtCombo \\
\bottomrule
\end{tabular}

\caption{Numerical simulation, \DesNEval{} images: Layer-1 VERIFIED rate and Layer-2 detection rate with the matched key after each post-sensor attack.
Signers: digital signer, ideal optics, nominal optics with the per-image mask (nominal) and with the static mask (static); verifiers: public verifier (pub.), optical verifier (opt.) and the private correlator (corr.); conditions: Sec.~\ref{sec:protocol}.
For the per-image mask, an unreadable payload counts as no Layer-2 detection.}
\label{tab:attacks}
\end{table}

\paragraph{What is provable.}
Let $H$ be SHA-256.
We say that the signer signed $(m,h,w)$ if it issued $\mathrm{Sign}_{sk}(H(\mathrm{dom}\,\|\,m\,\|\,h\,\|\,H(w)))$.
The adversary knows $pk$ and all algorithms, and may obtain marked images for hosts and messages of its choice.

\begin{proposition}\label{prop:forgery}
Suppose an adversary without $sk$ outputs an image $X'$ that verifies under $pk$.
Let $(m,\sigma,h)$ be the data decoded from $X'$ and $\hat w=\mathrm{Rec}(c(X'),h)$.
If the signer never signed $(m,h,\hat w)$, the adversary has produced an Ed25519 forgery under chosen-message attack (EUF-CMA) or a SHA-256 collision.
Hence, unless such an event occurs, every accepted image carries a signature that the signer issued on some $(m,h,H(w))$, and its feature decodes to that $w$ under the sketch, $\mathrm{Rec}(c(X'),h)=w$.
\end{proposition}

Proof sketch.
Acceptance means that $\sigma$ verifies under $pk$ on the digest $H(\mathrm{dom}\,\|\,m\,\|\,h\,\|\,H(\hat w))$.
If the signer never signed this digest, the pair is an existential forgery.
Otherwise the digest equals one that the signer computed from some $(m_i,h_i,H(w_i))$.
All fields have fixed length, so a difference in any field is a collision of the outer hash, and equal fields with $\hat w\neq w_i$ give a collision of the inner hash.
In the remaining case the triple $(m,h,\hat w)$ was signed.

The proposition does not state that $X'$ is the image that was signed.
It leaves one condition on content: $c(X')$ must lie within Hamming distance $t$ of a signed $w$ with the same helper.
Whether a different image can be steered to meet this condition is the binding strength of the feature, which is empirical and can fail (Sec.~\ref{sec:digital}).
Robustness is not covered either; both are measured by pre-registered tests (Sec.~\ref{sec:protocol}).
The public verifier is deterministic and specified at bit level, with no learned component.

\begin{figure}[t]
\centering
\includegraphics{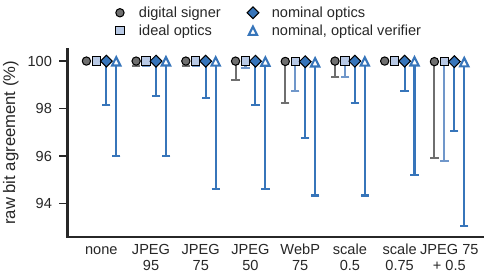}
\caption{Raw bit agreement with the embedded codeword after each attack, in the numerical simulation.
Shown is the mean over the \DesNEval{} evaluation images for the digital signer, the ideal optics, and the nominal optics read by the public verifier and by the optical verifier, in legend order (conditions: Sec.~\ref{sec:protocol}; ``scale'': resize).
Whiskers reach the worst image.}
\label{fig:agreement}
\end{figure}

\begin{figure*}[t]
\centering
\includegraphics{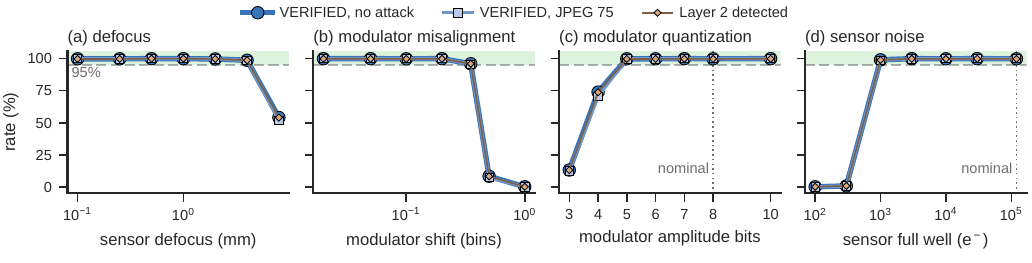}
\caption{Tolerance of the simulated optical signer to one departure from the nominal configuration at a time (descriptive, \HtN{} images): VERIFIED rate without attack and after JPEG~75, and Layer-2 detection rate.}
\label{fig:tolerance}
\end{figure*}

\paragraph{Role of the optical key.}
Layer~2 is a symmetric secret with a digital part, the mask seed of \KsSeed{} bits, and physical parts.
The physical parameters enter the correlator only through the scale $\lambda f$ and the mask position, so wavelength, focal length, and pixel pitch act as one degree of freedom.
Proposition~\ref{prop:forgery} does not depend on Layer~2: the optical layer is issuer-side supporting evidence, counted only on an image that Layer~1 verifies, and security rests on the signature and the content binding.
For Layer~1 we test the copy attack~\cite{kutter2000copyattack} (Sec.~\ref{sec:copy}).
Physical unclonability~\cite{pappu2002} is not claimed.

\section{Simulation Protocol}\label{sec:protocol}

\paragraph{Pre-registration.}
We wrote the specification of the simulation, with a numeric criterion for every test, before any optical result existed, then froze it and recorded its hash.
Later amendments are logged as deviations (Appendix~\ref{sec:app_prereg}).
Those that touch a reported number are named where the number appears, and no numeric threshold changed.
An independent script recomputed \RcCompared{} verdict values from the raw per-image records with \RcMismatch{} mismatches.

\paragraph{Data.}
The evaluation set has \DesNEval{} DIV2K images~\cite{agustsson2017div2k}: the \SetNValid{} validation images and the \SetNTrain{} training images with a short side of at least \DesN{} pixels.
Each image is resized so that its short side is \DesN{} pixels and then center cropped; one validation image is smaller and is upscaled (a logged deviation).
The negatives are all \HfCocoN{} unmarked COCO test images~\cite{lin2014coco}.
The specification named the \SetNValid{} validation images; a logged deviation only enlarges the sample, and both sets give the same outcomes (Sec.~\ref{sec:replicate}).

\paragraph{Conditions.}
The digital signer writes Layer~1 only, without optics.
The ideal optics is the optical path with every non-ideality off.
The nominal optics follows Tab.~\ref{tab:optics}, with the per-image Layer-2 mask in the primary condition and the static mask in a second one.
The wide-pupil optics differs from the nominal optics only in a pupil radius of \DesPupilWide{} bins instead of \DesPupil; it was added after the specification, in a deviation logged before any evaluation result, and supports no verdict.
Three wrong variants must not verify: decoding with a wrong layout seed, checking with a wrong public key, and signing with a reference-free dark-field sensor (the dark-field signer).
Every marked image then passes the \DesNAttacks{} post-sensor attacks of Tab.~\ref{tab:attacks}, one of which is no attack.

\section{Results}\label{sec:results}

All optical numbers below come from the numerical simulation on the \DesNEval{} evaluation images.

\begin{figure*}[t]
\centering
\includegraphics{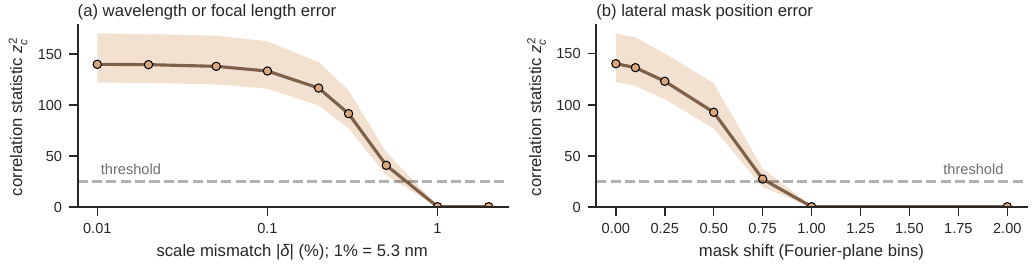}
\caption{Key sensitivity of the optical carrier in the numerical simulation.
Layer-2 statistic $z_c^2$ of the \DesNEval{} unattacked images signed with the nominal optics and the per-image mask, against (a) the scale mismatch and (b) the mask shift of the auditor unit (median and quartiles; (a) pools both signs).}
\label{fig:key}
\end{figure*}

\subsection{Optical versus digital}\label{sec:equiv}

\paragraph{Equivalence and optical fidelity.}
With every non-ideality off, the simulated optical write equals the digital pipeline at field level: the largest relative difference is $\HzMaxRel$.
The 8-bit outputs are identical on only \HzIdentical{} of \HzN{} images, although the specification required identity.
An intensity sensor cannot return the negative field values that occur at \HzNegPix{} near-black pixels.
A deviation logged before the evaluation run moves the comparison to the field before the sensor, so the ideal-optics equivalence criterion is met at field level only.
With the nominal optics, the public verifier reads \HoBitAgree\% of the embedded bits from unattacked images, and the optical verifier agrees with the public verifier on \HoOptDigAgree\% of the raw bits.
Pooled over the \DesNAttacks{} attacks, the VERIFIED rate under the public verifier is \HoPoolOpt\% for the nominal optics and \HoPoolDig\% for the digital signer, and \HoPoolVopt\% for the nominal optics under the optical verifier.
The gaps are within the pre-registered margin, so the optical-fidelity criterion is met.
The optical verifier also accepts images from the digital signer, and the public verifier those from the nominal optics (Tab.~\ref{tab:attacks}), so a server could issue signatures in bulk with the digital signer while a device in the field checks them with the optical verifier, or the other way round.
Over all signer--verifier pairs and attacks, the lowest VERIFIED rate is \XvMinRate\% (full matrix: Appendix~\ref{sec:app_cross}).

\paragraph{Post-sensor robustness and quality.}
Under the public verifier, the nominal optics is VERIFIED on at least \HtwoMin\% of the images after every attack (Tab.~\ref{tab:attacks}, Fig.~\ref{fig:agreement}), so the post-sensor robustness criterion is met.
Layer-2 detection is lowest under WebP, at \LtMinDetD\% for the per-image mask and \LtMinDetS\% for the static mask.
The fifth-percentile PSNR is \QDigPfive{} dB for the digital signer, \QOptDPfive{} dB for the nominal optics, whose pupil has a radius of \DesPupil{} bins, and \QOptWPfive{} dB for the wide-pupil optics, whose public-verifier counts are within one image of the nominal optics.

\subsection{Tolerance budget}\label{sec:tolerance}

Each sweep changes one parameter of the nominal configuration, on \HtN{} evaluation images.
Figure~\ref{fig:tolerance} shows four of the \DesSweepPoints{} sweep points; Appendix~\ref{sec:app_sweeps} lists all of them.
In the simulation, the public layer keeps a VERIFIED rate of at least \CrMin\%, without attack and after JPEG~75, down to \BudBslm{} modulator bits and a sensor full well of $\BudSne$ electrons.
It also tolerates a modulator shift of \BudShift{} bins and a tilt of \BudTilt{} degrees.
One step further it fails: \NxtVnBslm\% at \NxtBslm{} bits, \NxtVnShift\% at \NxtShift{} bins, \NxtVnTilt\% at \NxtTilt{} degrees, and \NxtVnSne\% at $\NxtSne$ electrons.
A single exposure of the Fourier camera is not enough (\HtVnFcamOxOT\%), because weak bins fall below one code.
Verification tolerates defocus up to \BudDz\,mm, because defocus is a pure phase in the Fourier plane and the public mark lives in magnitudes.
Image quality sets the real budget: the median PSNR is \HtPsDzZpO{} dB at \SwDzA\,mm of defocus and \HtPsDzZpTV{} dB at \SwDzB\,mm.
At the lowest arm coherence, Layer~1 is unaffected (\HtVnGammaZpTV\%), but Layer~2 is detected on \HtLtGammaZpTV\% of the images.

\subsection{False acceptance and wrong variants}\label{sec:false}

No unmarked or wrongly keyed input was accepted in the simulation, so the false-acceptance criterion is met.
Layer~1 gave \HfCocoVd{} VERIFIED among \HfCocoN{} unmarked COCO images under the public verifier and \HfCocoVo{} under the optical verifier, \HfDivVd{} among the \HfDivN{} unmarked evaluation frames, and \HfWrongPk{} among \HfWrongPkN{} wrong-key checks.
The one-sided \CrCpLevel\% Clopper--Pearson upper bound for the COCO count is $\HfCpUpper$.
Layer~2 gave \HfLtStatic{} detections on the COCO images and \HfWrongSeed{} in \HfWrongSeedN{} wrong-seed trials; the largest null value of $z_c^2$ is \HfNullMax{} against $\tau=\DesThr$.
The wrong layout seed, the wrong public key, and the dark-field signer give \HvWone, \HvWtwo, and \HvWthree{} VERIFIED images, so the wrong-variant criterion is met.

\subsection{Key sensitivity}\label{sec:keysens}

With the matched key, the correlator detects Layer~2 on \HsDetOptDNone\% of the unattacked images with the per-image mask and on \HsDetOptDJpegB\% after JPEG~75; the static mask gives \HsDetOptSNone\% and \HsDetOptSJpegB\%.
An auditor unit with \CrScaleTest\% of scale error detects \HsScalePlus\% and \HsScaleMinus\% for the two signs, and \CrShiftTest{} bin of mask shift gives \HsShiftOne\% (Fig.~\ref{fig:key}).
The complete sweep, which also varies the phase bits of the auditor and the width of the source, is listed in Appendix~\ref{sec:app_key} (Tab.~\ref{tab:keysweep}).
The key-sensitivity criterion is met: Layer~2 is key-selective in the simulation.
The median statistic crosses the threshold at a scale mismatch of \HsTolRel\%, that is \HsTolNm{} nm of wavelength or \HsTolFmm{} mm of focal length, and at a mask shift of \HsTolShift{} bins, or \HsTolShiftUm{}\,$\mu$m.

\subsection{Copy of the public payload}\label{sec:copy}

Each decoded payload is re-embedded into another unmarked host with the public embedder.
The public decoder reads \HeRead{} of \HeN{} payloads; \HeVerified{} copies are VERIFIED, the sketch rejects \HeSketch, and the one-sided upper bound is \HeCpUpper.
The payload-copy criterion is met.

\begin{figure*}[t]
\centering
\includegraphics[width=\textwidth]{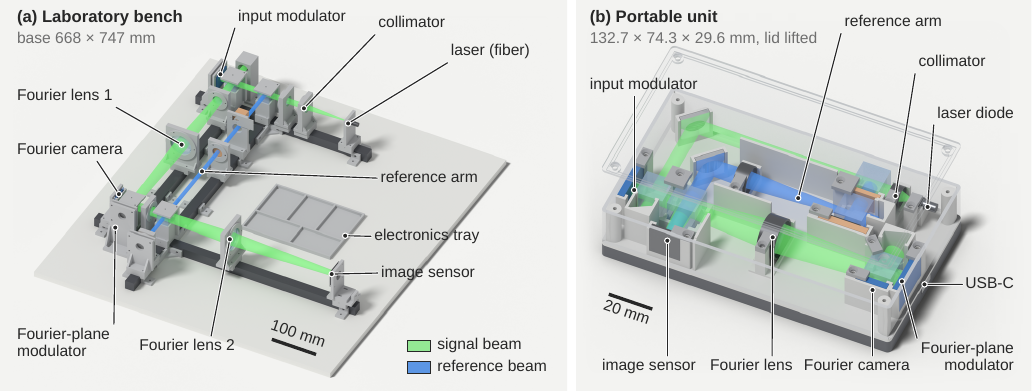}
\caption{Reference hardware designs, rendered from the CAD models; no hardware was built. (a)~Laboratory bench with printed mounts on slotted rails; the enclosure is not shown. (b)~Portable unit with the lid lifted; lid and side walls are drawn transparent. Green volumes are the beam envelopes of the illumination and signal arm, blue those of the reference arm (legend in (a), valid for both panels); the two also differ in lightness. In (a), the electronics tray holds the USB hub, the modulator driver boards, and the microcontroller board. Each scale bar lies in the base plane, along the front edge.}
\label{fig:devices}
\end{figure*}

\subsection{Replication and independent recount}\label{sec:replicate}
The full protocol was first run on an independent replicate of \OldHfDivN{} evaluation images and \OldHfCocoN{} unmarked COCO images, before the enlarged set was evaluated.
Every criterion reported in this section gave the same outcome on both sets (Appendix~\ref{sec:app_sets}): bit agreement \OldHoBitAgree\% against \HoBitAgree\%, a minimum post-sensor VERIFIED rate of \OldHtwoMin\% against \HtwoMin\%, \OldHeVerified{} of \OldHeN{} copied payloads VERIFIED, and no false acceptance.
An independent script that does not share code with the evaluation recomputes every reported value from the raw records; it found \RcMismatch{} mismatches in \RcCompared{} values on the enlarged set and \OldRcMismatch{} in \OldRcCompared{} values on the replicate.

\subsection{Digital evidence for the public layer}\label{sec:digital}

The following results were obtained on the digital pipeline alone, without optics and on other image sets.
On \FChanN{} DIV2K crops, the channel carries \DesKbits{} bits at a fifth-percentile PSNR of \FChanPsnr{} dB.
End to end, \FEeDiv{} DIV2K and at least \FEeCoco{} COCO images are VERIFIED under each codec and resize attack.
No false accept occurred among the \FEeFaN{} unmarked COCO images of that run, and no copied payload verified among \FEeCopyN{} random pairs.

\begin{table}[t]
\centering
\small
\setlength{\tabcolsep}{4pt}
\begin{tabular}{@{}lcccc@{}}
\toprule
method & bits & PSNR & LPIPS & RT \\
\midrule
this channel & \DesKbits & \FBaseOursPsnr & \FBaseOursLpips & \FRtOurs \\
TrustMark-Q~\cite{bui2023trustmark} & \FBaseTmBits & \FBaseTmqPsnr & \FBaseTmqLpips & \FRtTmLo{} to \FRtTmHi$^{\dagger}$ \\
TrustMark-P~\cite{bui2023trustmark} & \FBaseTmBits & \FBaseTmpPsnr & \FBaseTmpLpips & \FRtTmLo{} to \FRtTmHi$^{\dagger}$ \\
Video Seal~\cite{fernandez2024videoseal} & \FBaseSealBits & \FBaseVsPsnr & \FBaseVsLpips & $\le$\,\FRtSeal \\
Pixel Seal~\cite{soucek2025pixelseal} & \FBaseSealBits & \FBasePsPsnr & \FBasePsLpips & $\le$\,\FRtSeal \\
ChunkySeal~\cite{petrov2025chunkyseal} & \FBaseChunkyBits & \FBaseChunkyPsnr & \FBaseChunkyLpips & \FRtChunkyHundred \\
WAM~\cite{sander2025wam} & \FBaseWamBits & \FBaseWamPsnr & \FBaseWamLpips & \FRtWamHundred \\
\bottomrule
\end{tabular}
\caption{Digital pipeline, no optics: the magnitude channel against open learned watermarks at default strength.
Bits: native payload. PSNR: fifth percentile (dB). LPIPS~\cite{zhang2018lpips}: ninety-fifth percentile.
RT: residual transferability~\cite{dong2026residual} with \FRtS{} ground-truth residuals. $\dagger$: range over the TrustMark variants.}
\label{tab:baselines}
\end{table}

\paragraph{Baselines.}
No open learned baseline of Tab.~\ref{tab:baselines} matches the \DesKbits{}-bit operating point, but all are more robust to gain and geometry.
They keep a mean bit accuracy of at least \FBaseGainAcc{} where this channel drops to \FBaseOursGainHi{} or less under gain changes and to \FBaseOursGeom{} under small crops or rotation.
The residual transplant scores \FRtOurs{} here, which is not copy resistance: a public decoder and embedder copy any payload.

\paragraph{Binding.}
The optical simulation uses the content feature $c$ of Sec.~\ref{sec:layer1}, the one of the verification results above.
A white-box attacker can steer it: after a search for a flat target, \FBindSearchBar{} forgeries stay within the quality bar (PSNR of at least \FBarPsnr{} dB, LPIPS of at most \FBarLpips).
An improved binding that also signs the feature energy allows \FNewSteer{} forgeries within the bar, at a benign verification rate of \FNewBenign{} or more.
Its pass depends on the PSNR half of the bar, which is post hoc: with LPIPS alone, \FNewLpipsOnly{} are forged, and \FNewSameScene{} same-scene pairs are not separated.
This binding was verified digitally only; the optical simulation did not use it.

\section{Reference Hardware Designs}\label{sec:hardware}

To make the simulated parameters concrete, we derived two reference hardware designs from them (Fig.~\ref{fig:devices}).
The laboratory bench places printed mounts on slotted rails over a base board of \HwBenchBaseW$\times$\HwBenchBaseD\,mm, with Fourier lenses of focal length $f=\HwBenchF$\,mm and \HwBenchPartTypes{} printed part types.
The portable unit arranges the same optical functions in a USB-connected housing of \HwPortL$\times$\HwPortW$\times$\HwPortH\,mm.
Both are designs only: Appendices~\ref{sec:app_layouts} to~\ref{sec:app_portable} give the optical layouts, the parts lists, and the tolerance budget, and the repository named on the first page holds their CAD sources and STL files (design files only).
No hardware was built; in a simulation of the laboratory bench with its lens prescription and software calibration, the lowest VERIFIED rate over the \SpNSettings{} post-sensor settings is \SpLabLensPubPct\% under the public verifier and \SpLabLensOptPct\% under the optical verifier, and \SpLabTolPubPct\% and \SpLabTolOptPct\% when assembly tolerances at the favorable end of published ranges are added ($n=\SpN$ images; Appendix~\ref{sec:app_simperf}).
Every other simulation reported here uses the ideal thin-lens model of Sec.~\ref{sec:optics}, not these designs.

\section{Limitations}\label{sec:limits}

\paragraph{Simulation only.}
Every optical result is a numerical simulation with a scalar, paraxial, thin-lens, monochromatic model; no hardware was built, and speed, energy, and cost are outside the scope of this paper.
Unmodeled are polarization, pixel crosstalk, speckle, stray light, and partial coherence.

\paragraph{Evaluation size.}
The verdicts rest on \DesNEval{} images from one photographic dataset and on one wavelength.
A count of zero at this size supports only a bound.

\paragraph{Crop and rotation.}
The magnitude channel is a global pattern.
On the digital pipeline it decodes at most a fraction \FGeomCrop{} of the images after any tested crop and \FGeomRot{} after a rotation of \FGeomRotDeg{} degrees; gain changes and blur are also weak.

\paragraph{Regeneration.}
Diffusion regeneration removes pixel-level watermarks~\cite{zhao2024regeneration}, and we expect it to remove ours; we did not run it.
Removal gives NO PAYLOAD or NOT VERIFIED, the safe direction: provenance is lost, not forged.

\paragraph{Content binding.}
The binding feature used in the simulation is steerable on flat targets (Sec.~\ref{sec:digital}).
The improved binding resists that attack, but a visible global re-grade at a PSNR of about \FNewRegrade{} dB still forges it, and same-scene shots are not separated.

\paragraph{Optical layer.}
The optical layer is issuer-side supporting evidence; security rests on the signature and the content binding.

\section{Conclusion}\label{sec:conclusion}

We embed an Ed25519 signature bound to the image content in Fourier magnitudes, with a coherent 4F unit as the transform core of signer and verifier.
In a numerical simulation on \DesNEval{} images, the optical signer matches the digital pipeline (bit agreement \HoBitAgree\%), keeps at least \HtwoMin\% of the images VERIFIED under every tested compression and resizing, and accepts none of \HfCocoN{} unmarked images and no copied payload; an independent replicate gives the same outcomes.
The optical carrier is key-selective and serves the issuer as supporting evidence.
Security rests on the signature and the content binding.
We give two reference hardware designs but built neither; hardware validation is future work.

\section*{Acknowledgments}
We thank the Phi Lab Foundation for providing computing resources and the experimental environment.
\section*{LLM Usage Statement}
The authors take full responsibility for all submitted content, regardless of any use of LLMs or other AI tools. LLM-based assistants (OpenAI ChatGPT/GPT, Anthropic Claude, and Google Gemini) were used only in an assistive role to edit and refine author-produced material, including the manuscript text, the related-work discussion and citations, the code, the description and analysis of experiments, and the figures. For the data, these tools were used to generate the simulated optical measurements and the watermarked and attacked test images derived from the public DIV2K and COCO images, which the authors reviewed, edited, and approved before use. The research ideas, hypotheses, claims, and conclusions were developed by the authors without LLM or AI-tool involvement. The full manuscript was also provided to these services to obtain reviewer-style feedback, and all resulting suggestions were checked and verified by the authors.

{
    \small
    \bibliographystyle{ieeenat_fullname}
    \bibliography{refs}
}

\clearpage
\onecolumn
\appendix
\raggedbottom
\renewcommand{\topfraction}{0.95}\renewcommand{\bottomfraction}{0.95}\renewcommand{\textfraction}{0.05}
\renewcommand{\floatpagefraction}{0.75}\setcounter{topnumber}{3}\setcounter{bottomnumber}{2}\setcounter{totalnumber}{5}
\section{Pre-registration and deviations}
\label{sec:app_prereg}
The specification of the simulation was written on \AppPreregDate, before any optical simulation result existed.
It fixes the optical model with its nominal parameters (Tab.~\ref{tab:optics}), the image sets, the signer conditions and the wrong variants, the post-sensor attacks, the points of the tolerance and key-sensitivity sweeps, and a numeric criterion with a pass rule for each test; Tab.~\ref{tab:hyp} lists the criteria of the results reported in this paper.
The file was then frozen, and its checksum was recorded at that time.
Later amendments were logged with a time stamp and a reason.
Table~\ref{tab:app_dev} lists the \AppDevN{} logged deviations and one addendum; none changes a numeric threshold.
Outcomes are computed from the raw per-image records.
An independent implementation, written from the record format and the specification without access to the evaluation code, recomputed \RcCompared{} outcome values of the enlarged set with \RcMismatch{} mismatches, and \OldRcCompared{} values of replicate~1 with \OldRcMismatch{} mismatches.

\begin{table}[!htbp]
\centering\footnotesize
\begin{tabular}{@{}l@{\hspace{6pt}}p{0.17\linewidth}p{0.53\linewidth}p{0.19\linewidth}@{}}
\toprule
no. & logged & deviation & affects \\
\midrule
1 & \AppDevWhenA; before any optical result & The ideal-optics equivalence is evaluated on the field in front of the sensor. An intensity sensor returns the modulus of the field, so the 8-bit outputs cannot be identical at pixels where the ideal marked field is negative; these pixels are counted separately. The threshold of $\CrHzero$ is unchanged. & wording of the ideal-optics equivalence criterion \\
\addlinespace[2pt]
2 & \AppDevWhenB; before any optical result & A zero reading of the Fourier camera at its longest exposure is taken as half a least-significant bit, so that the gains and the Layer-2 amplitude stay finite. & implementation detail; no criterion \\
\addlinespace[2pt]
3 & \AppDevWhenC; after a trial on calibration images, before any evaluation result & The wide-pupil optics (pupil radius of \DesPupilWide{} bins instead of \DesPupil) is added as an exploratory condition, and the same pupil is added to the pupil sweep. The nominal pupil is not changed. & supports no criterion \\
\addlinespace[2pt]
4 & \AppDevWhenD & Compute limits after an interrupted run: complete record sets were kept, and incomplete ones were discarded and computed again with fewer parallel workers. & none; noise seeds depend on condition and image only \\
\addlinespace[2pt]
5 & \AppDevWhenE & In replicate~1, the tolerance sweeps use the first \OldHtN{} evaluation images, the fallback fixed in the specification. All sweep points are kept. & tolerance sweeps of replicate~1 (descriptive) \\
\addlinespace[2pt]
6 & \AppDevWhenF & The sweep run of replicate~1 was stopped once and restarted from its first task with the same settings; the partial records enter no number. & none \\
\addlinespace[2pt]
7 & \AppDevWhenG; after results, documentation only & Points where the simulation code is more specific than the specification: the modulator gain is clipped only when the modulator is quantized; the Fourier camera reads one half plane and mirrors the reading to the conjugate bins; the host spectrum is measured once per image and reused in the refinement passes and in the content-binding loop; the amplitude of the correlator filter is quantized whenever the modulator is; the sweep list has \DesSweepPoints{} points, including the wide pupil of deviation~3. & no number changed \\
\addlinespace[2pt]
8 & \AppDevWhenH; before any record of the enlarged set & The enlarged set: \SetN{} evaluation images (the \SetNValid{} validation images and \SetNTrain{} training images) and all \SetNCoco{} COCO test images as negatives, computed on a second platform. Specification, parameters, thresholds, and pass rules are unchanged; the counts in the pass rules follow the set size. The records of the \OldSetN{} validation images are kept as replicate~1. & sample size of every test \\
\addlinespace[2pt]
9 & \AppDevWhenI; after results, documentation only & One validation image is smaller than the frame and is upscaled, although the specification says that frames are never upscaled. It belongs to the frozen evaluation set and stays in both sets. & no number changed \\
\addlinespace[2pt]
8, add. & \AppDevWhenJ; after the enlarged run & Facts of the enlarged run and the platform comparison of Appendix~\ref{sec:app_sets}. One correction of the summary step: the Clopper--Pearson bound of the false-acceptance count used a fixed number of negatives and now uses the number of records; the summary of replicate~1 is identical before and after. & upper bound of the enlarged set \\
\bottomrule
\end{tabular}
\caption{Logged deviations from the frozen specification, in the names of this paper.
Times are in Pacific Standard Time; the first entry was logged on the day of the specification.}
\label{tab:app_dev}
\end{table}

\section{Replicate 1 and the enlarged set}
\label{sec:app_sets}
The specification names the \SetNValid{} DIV2K validation images as the evaluation set.
Replicate~1 ran on these images and covers every test.
Its tolerance sweeps used the first \OldHtN{} images (deviation~5 of Tab.~\ref{tab:app_dev}).
The enlarged set (deviation~8) adds the \SetNTrain{} DIV2K training images whose short side is at least \DesN{} pixels, all of which are downscaled.
One validation image is smaller and is upscaled (deviation~9).
The \SetNSkipped{} smaller training images are skipped, and the \DesNCalib{} calibration images stay excluded.
The enlarged set has \SetN{} images and uses all \SetNCoco{} COCO test images as negatives (replicate~1: a seeded draw of \OldSetNCoco).
It was computed on a second platform, on CPUs only, with newer versions of the numerical libraries.
The design, the parameters, the thresholds, and the pass rules are the same; only the sample size differs.

The enlarged run contains the \SetNValid{} validation images again.
For these images, the signing and verification records of all conditions and attacks are identical to replicate~1 in every field except the wall-time fields.
The marked images that were compared before the launch are byte-identical.
Values computed through the chirp-z transform, which are the mismatched-auditor sweeps of the key-sensitivity test, differ in late digits between the two platforms.
The transform rounds differently there, and the simulated camera noise follows the rounding.
No detection decision differs on these images.

The main text reports the enlarged set.
Table~\ref{tab:app_compare} lists the outcome numbers of both sets side by side, and the tables of Appendices~\ref{sec:app_rates} to~\ref{sec:app_key} give every rate, sweep point, and quality value for both sets.

\begin{table}[!htbp]
\centering\footnotesize
\begin{tabular}{@{}p{0.62\linewidth}rr@{}}
\toprule
quantity & replicate 1 & enlarged set \\
\midrule
images $n$ & \OldSetN{} & \SetN{} \\
unmarked COCO images & \OldSetNCoco{} & \SetNCoco{} \\
\addlinespace[2pt]
ideal-optics equivalence: largest relative field error & $\OldHzMaxRel{}$ & $\HzMaxRel{}$ \\
ideal-optics equivalence: images with identical 8-bit output & \OldHzIdentical{} of \OldHzN{} & \HzIdentical{} of \HzN{} \\
ideal-optics equivalence: outcome (field level) & \OldVerHz{} & \VerHz{} \\
\addlinespace[2pt]
optical fidelity: bit agreement, nominal optics under the public verifier (\%) & \OldHoBitAgree{} & \HoBitAgree{} \\
optical fidelity: raw-bit agreement of the optical with the public verifier (\%) & \OldHoOptDigAgree{} & \HoOptDigAgree{} \\
optical fidelity: pooled VERIFIED, nominal / digital / nominal under the optical verifier (\%) & \OldHoPoolOpt{} / \OldHoPoolDig{} / \OldHoPoolVopt{} & \HoPoolOpt{} / \HoPoolDig{} / \HoPoolVopt{} \\
optical fidelity: outcome & \OldVerHo{} & \VerHo{} \\
\addlinespace[2pt]
post-sensor robustness: lowest VERIFIED rate over the attacks (\%, rounded down) & \OldHtwoMin{} & \HtwoMin{} \\
post-sensor robustness: outcome & \OldVerHtwo{} & \VerHtwo{} \\
\addlinespace[2pt]
tolerance sweeps: images in the sweeps (descriptive) & \OldHtN{} & \HtN{} \\
tolerance sweeps: budget, modulator bits / shift (bins) / tilt (deg) / defocus (mm) & \OldBudBslm{} / \OldBudShift{} / \OldBudTilt{} / \OldBudDz{} & \BudBslm{} / \BudShift{} / \BudTilt{} / \BudDz{} \\
tolerance sweeps: budget, sensor full well (e$^-$) & $\OldBudSne{}$ & $\BudSne{}$ \\
\addlinespace[2pt]
false acceptance: VERIFIED among unmarked COCO, public / optical verifier & \OldHfCocoVd{} / \OldHfCocoVo{} of \OldHfCocoN{} & \HfCocoVd{} / \HfCocoVo{} of \HfCocoN{} \\
false acceptance: one-sided \CrCpLevel\% upper bound & $\OldHfCpUpper{}$ & $\HfCpUpper{}$ \\
false acceptance: VERIFIED among unmarked DIV2K frames & \OldHfDivVd{} of \OldHfDivN{} & \HfDivVd{} of \HfDivN{} \\
false acceptance: VERIFIED with a wrong public key & \OldHfWrongPk{} of \OldHfWrongPkN{} & \HfWrongPk{} of \HfWrongPkN{} \\
false acceptance: Layer-2 detections with a wrong mask seed & \OldHfWrongSeed{} of \OldHfWrongSeedN{} & \HfWrongSeed{} of \HfWrongSeedN{} \\
false acceptance: largest null $z_c^2$ (threshold \DesThr) & \OldHfNullMax{} & \HfNullMax{} \\
false acceptance: outcome & \OldVerHf{} & \VerHf{} \\
\addlinespace[2pt]
wrong variants: VERIFIED with wrong layout seed / wrong public key / dark-field signer & \OldHvWone{} / \OldHvWtwo{} / \OldHvWthree{} & \HvWone{} / \HvWtwo{} / \HvWthree{} \\
wrong variants: outcome & \OldVerHv{} & \VerHv{} \\
\addlinespace[2pt]
key sensitivity: matched key detected, auditor sweep (\%) & \OldHsMatchedDet{} & \HsMatchedDet{} \\
key sensitivity: detected at $+$\CrScaleTest\% / $-$\CrScaleTest\% scale / \CrShiftTest{} bin shift (\%) & \OldHsScalePlus{} / \OldHsScaleMinus{} / \OldHsShiftOne{} & \HsScalePlus{} / \HsScaleMinus{} / \HsShiftOne{} \\
key sensitivity: tolerance, relative scale (\%) / mask shift (bins) & \OldHsTolRel{} / \OldHsTolShift{} & \HsTolRel{} / \HsTolShift{} \\
key sensitivity: outcome & \OldVerHs{} & \VerHs{} \\
\addlinespace[2pt]
payload copy: copied payloads VERIFIED & \OldHeVerified{} of \OldHeN{} & \HeVerified{} of \HeN{} \\
payload copy: one-sided \CrCpLevel\% upper bound & \OldHeCpUpper{} & \HeCpUpper{} \\
payload copy: outcome & \OldVerHe{} & \VerHe{} \\
\addlinespace[2pt]
PSNR, fifth percentile, digital signer / nominal optics / wide-pupil optics (dB) & \OldQDigPfive{} / \OldQOptDPfive{} / \OldQOptWPfive{} & \QDigPfive{} / \QOptDPfive{} / \QOptWPfive{} \\
independent recount: values compared / mismatches & \OldRcCompared{} / \OldRcMismatch{} & \RcCompared{} / \RcMismatch{} \\
\bottomrule
\end{tabular}

\caption{Outcome numbers of replicate~1 (\OldSetN{} images) and of the enlarged set (\SetN{} images) for the criteria of Tab.~\ref{tab:hyp}, numerical simulation.
Same design, thresholds, and pass rules.}
\label{tab:app_compare}
\end{table}

\section{Optical-digital cross-verification}
\label{sec:app_cross}

A deployed system would sign and verify in both domains.
A server signs or checks many images in software, while a device in the field writes or reads the mark with light.
The two domains must therefore accept each other's images.
This section arranges the existing evaluation records as a signer-by-verifier matrix; no new simulation was run for it.

\paragraph{Records.}
The evaluation signed each of the \VtXN{} images of the enlarged set once with the digital signer and once with the nominal optics and the per-image mask, called the optical signer here.
Each marked image was passed through the \VtXNatt{} post-sensor attacks and then read twice: by the public verifier, which is software, and by the optical verifier, which reads the Fourier magnitudes with the nominal camera model.
This gives four cells with \VtXN{} images per cell and attack.
The optical verifier is simulated with input quantization, camera noise, and exposure bracketing; lens aberrations and misregistration of its camera are not modeled.

\begin{table}[!tb]
\centering\small
\begin{tabular}{@{}l@{\hspace{8pt}}r@{\hspace{4pt}}l@{\hspace{10pt}}r@{\hspace{4pt}}l@{\hspace{10pt}}r@{\hspace{4pt}}l@{\hspace{10pt}}r@{\hspace{4pt}}l@{}}
\toprule
signer & \multicolumn{4}{c}{digital signer} & \multicolumn{4}{c}{optical signer (nominal optics)} \\
\cmidrule(r){2-5}\cmidrule(l){6-9}
verifier & \multicolumn{2}{c}{public} & \multicolumn{2}{c}{optical} & \multicolumn{2}{c}{public} & \multicolumn{2}{c}{optical} \\
\midrule
none & \VtXkDpNone & [\VtXloDpNone, \VtXhiDpNone] & \VtXkDoNone & [\VtXloDoNone, \VtXhiDoNone] & \VtXkOpNone & [\VtXloOpNone, \VtXhiOpNone] & \VtXkOoNone & [\VtXloOoNone, \VtXhiOoNone] \\
JPEG 95 & \VtXkDpJpegA & [\VtXloDpJpegA, \VtXhiDpJpegA] & \VtXkDoJpegA & [\VtXloDoJpegA, \VtXhiDoJpegA] & \VtXkOpJpegA & [\VtXloOpJpegA, \VtXhiOpJpegA] & \VtXkOoJpegA & [\VtXloOoJpegA, \VtXhiOoJpegA] \\
JPEG 75 & \VtXkDpJpegB & [\VtXloDpJpegB, \VtXhiDpJpegB] & \VtXkDoJpegB & [\VtXloDoJpegB, \VtXhiDoJpegB] & \VtXkOpJpegB & [\VtXloOpJpegB, \VtXhiOpJpegB] & \VtXkOoJpegB & [\VtXloOoJpegB, \VtXhiOoJpegB] \\
JPEG 50 & \VtXkDpJpegC & [\VtXloDpJpegC, \VtXhiDpJpegC] & \VtXkDoJpegC & [\VtXloDoJpegC, \VtXhiDoJpegC] & \VtXkOpJpegC & [\VtXloOpJpegC, \VtXhiOpJpegC] & \VtXkOoJpegC & [\VtXloOoJpegC, \VtXhiOoJpegC] \\
WebP 75 & \VtXkDpWebp & [\VtXloDpWebp, \VtXhiDpWebp] & \VtXkDoWebp & [\VtXloDoWebp, \VtXhiDoWebp] & \VtXkOpWebp & [\VtXloOpWebp, \VtXhiOpWebp] & \VtXkOoWebp & [\VtXloOoWebp, \VtXhiOoWebp] \\
resize 0.5x & \VtXkDpScaleA & [\VtXloDpScaleA, \VtXhiDpScaleA] & \VtXkDoScaleA & [\VtXloDoScaleA, \VtXhiDoScaleA] & \VtXkOpScaleA & [\VtXloOpScaleA, \VtXhiOpScaleA] & \VtXkOoScaleA & [\VtXloOoScaleA, \VtXhiOoScaleA] \\
resize 0.75x & \VtXkDpScaleB & [\VtXloDpScaleB, \VtXhiDpScaleB] & \VtXkDoScaleB & [\VtXloDoScaleB, \VtXhiDoScaleB] & \VtXkOpScaleB & [\VtXloOpScaleB, \VtXhiOpScaleB] & \VtXkOoScaleB & [\VtXloOoScaleB, \VtXhiOoScaleB] \\
JPEG 75, 0.5x & \VtXkDpCombo & [\VtXloDpCombo, \VtXhiDpCombo] & \VtXkDoCombo & [\VtXloDoCombo, \VtXhiDoCombo] & \VtXkOpCombo & [\VtXloOpCombo, \VtXhiOpCombo] & \VtXkOoCombo & [\VtXloOoCombo, \VtXhiOoCombo] \\
\midrule
all \VtXNatt{} attacks & \VtXkDpAll & [\VtXloDpAll, \VtXhiDpAll] & \VtXkDoAll & [\VtXloDoAll, \VtXhiDoAll] & \VtXkOpAll & [\VtXloOpAll, \VtXhiOpAll] & \VtXkOoAll & [\VtXloOoAll, \VtXhiOoAll] \\
sum of \VtXNtrials & \VtXkDpPool & (\VtXrDpPool\%) & \VtXkDoPool & (\VtXrDoPool\%) & \VtXkOpPool & (\VtXrOpPool\%) & \VtXkOoPool & (\VtXrOoPool\%) \\
\bottomrule
\end{tabular}

\caption{Cross-verification in the numerical simulation: number of VERIFIED images of \VtXN{} per cell and attack, with the two-sided 95\% Clopper--Pearson interval of the rate in percent, rounded outward (100 without decimals is exact).
All \VtXNatt{} attacks: images that are VERIFIED after every attack.
The last row sums image-attack pairs, which are not independent, so it has no interval.}
\label{tab:app_cross}
\end{table}

\begin{figure}[!tb]
\centering
\includegraphics{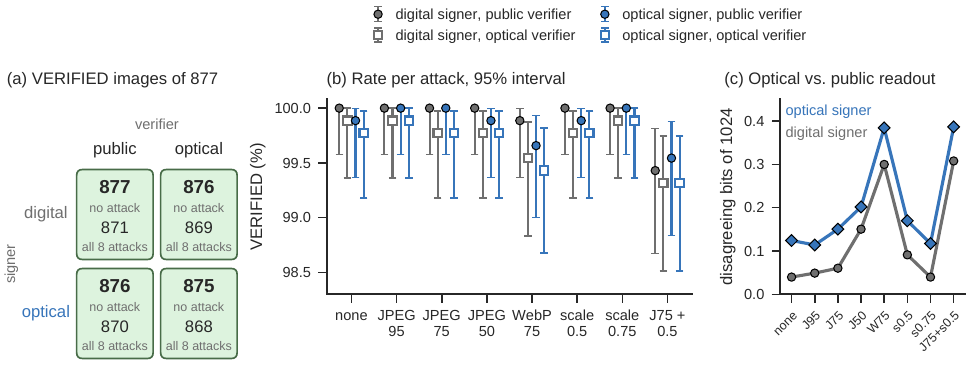}
\caption{Cross-verification in the numerical simulation.
(a)~VERIFIED images of \VtXN{} per signer and verifier, without attack and after every one of the \VtXNatt{} attacks.
(b)~VERIFIED rate per attack with 95\% Clopper--Pearson intervals.
(c)~Mean number of raw payload bits, of \VtXNbits, on which the optical and the public readout of the same image disagree.}
\label{fig:app_cross}
\end{figure}

\paragraph{Result.}
Table~\ref{tab:app_cross} and Fig.~\ref{fig:app_cross} give the matrix.
Every cell verifies between \VtXminK{} and \VtXmaxK{} of \VtXN{} images under every attack.
The lowest rate is \VtXminRate\%, after JPEG~75 followed by resizing, with a lower interval bound of \VtXminLo\%.
Under one attack the four cells differ by at most \VtXmaxGap{} percentage points.
Between \VtXminAllK{} and \VtXmaxAllK{} images per cell are VERIFIED after all \VtXNatt{} attacks.
The cross cells behave like the same-domain cells: an image signed digitally is read by the optical verifier as reliably as an optically signed one, and an optically signed image passes the public verifier as reliably as a digitally signed one, within these intervals.

\begin{table}[!htbp]
\centering\footnotesize\setlength{\tabcolsep}{4.2pt}
\begin{tabular}{@{}l@{\hspace{8pt}}rrrrrr@{\hspace{12pt}}rrrrrr@{}}
\toprule
 & \multicolumn{6}{c}{digital signer} & \multicolumn{6}{c}{optical signer (nominal optics)} \\
\cmidrule(r){2-7}\cmidrule(l){8-13}
 & \multicolumn{4}{c}{verdicts of the two verifiers} & \multicolumn{2}{c}{raw bits} & \multicolumn{4}{c}{verdicts of the two verifiers} & \multicolumn{2}{c}{raw bits} \\
attack & both & pub. & opt. & none & mean & max & both & pub. & opt. & none & mean & max \\
\midrule
none & \VtXcDBothNone & \VtXcDPubNone & \VtXcDOptNone & \VtXcDNeiNone & \VtXbDMeanNone & \VtXbDMaxNone & \VtXcOBothNone & \VtXcOPubNone & \VtXcOOptNone & \VtXcONeiNone & \VtXbOMeanNone & \VtXbOMaxNone \\
JPEG 95 & \VtXcDBothJpegA & \VtXcDPubJpegA & \VtXcDOptJpegA & \VtXcDNeiJpegA & \VtXbDMeanJpegA & \VtXbDMaxJpegA & \VtXcOBothJpegA & \VtXcOPubJpegA & \VtXcOOptJpegA & \VtXcONeiJpegA & \VtXbOMeanJpegA & \VtXbOMaxJpegA \\
JPEG 75 & \VtXcDBothJpegB & \VtXcDPubJpegB & \VtXcDOptJpegB & \VtXcDNeiJpegB & \VtXbDMeanJpegB & \VtXbDMaxJpegB & \VtXcOBothJpegB & \VtXcOPubJpegB & \VtXcOOptJpegB & \VtXcONeiJpegB & \VtXbOMeanJpegB & \VtXbOMaxJpegB \\
JPEG 50 & \VtXcDBothJpegC & \VtXcDPubJpegC & \VtXcDOptJpegC & \VtXcDNeiJpegC & \VtXbDMeanJpegC & \VtXbDMaxJpegC & \VtXcOBothJpegC & \VtXcOPubJpegC & \VtXcOOptJpegC & \VtXcONeiJpegC & \VtXbOMeanJpegC & \VtXbOMaxJpegC \\
WebP 75 & \VtXcDBothWebp & \VtXcDPubWebp & \VtXcDOptWebp & \VtXcDNeiWebp & \VtXbDMeanWebp & \VtXbDMaxWebp & \VtXcOBothWebp & \VtXcOPubWebp & \VtXcOOptWebp & \VtXcONeiWebp & \VtXbOMeanWebp & \VtXbOMaxWebp \\
resize 0.5x & \VtXcDBothScaleA & \VtXcDPubScaleA & \VtXcDOptScaleA & \VtXcDNeiScaleA & \VtXbDMeanScaleA & \VtXbDMaxScaleA & \VtXcOBothScaleA & \VtXcOPubScaleA & \VtXcOOptScaleA & \VtXcONeiScaleA & \VtXbOMeanScaleA & \VtXbOMaxScaleA \\
resize 0.75x & \VtXcDBothScaleB & \VtXcDPubScaleB & \VtXcDOptScaleB & \VtXcDNeiScaleB & \VtXbDMeanScaleB & \VtXbDMaxScaleB & \VtXcOBothScaleB & \VtXcOPubScaleB & \VtXcOOptScaleB & \VtXcONeiScaleB & \VtXbOMeanScaleB & \VtXbOMaxScaleB \\
JPEG 75, 0.5x & \VtXcDBothCombo & \VtXcDPubCombo & \VtXcDOptCombo & \VtXcDNeiCombo & \VtXbDMeanCombo & \VtXbDMaxCombo & \VtXcOBothCombo & \VtXcOPubCombo & \VtXcOOptCombo & \VtXcONeiCombo & \VtXbOMeanCombo & \VtXbOMaxCombo \\
\midrule
sum / all & \VtXcDBoth & \VtXcDPubOnly & \VtXcDOptOnly & \VtXcDNeither & \VtXbDMeanAll & \VtXbDMaxAll & \VtXcOBoth & \VtXcOPubOnly & \VtXcOOptOnly & \VtXcONeither & \VtXbOMeanAll & \VtXbOMaxAll \\
\bottomrule
\end{tabular}

\caption{Agreement of the two verifiers on the same marked and attacked image (\VtXN{} images per row).
Verdicts: VERIFIED by both, by the public verifier only, by the optical verifier only, or by neither.
Raw bits: number of the \VtXNbits{} payload bits on which the two readouts disagree, mean over images and largest value.}
\label{tab:app_agree}
\end{table}

\paragraph{Agreement of the two readouts.}
Table~\ref{tab:app_agree} compares the two verifiers on the same image.
Of the \VtXNtrials{} image-attack pairs per signer, they return the same verdict on \VtXcDSame{} for the digital signer and on \VtXcOSame{} for the optical signer.
All other pairs are accepted by the public verifier alone, \VtXcDPubOnly{} and \VtXcOPubOnly{} pairs.
No image is accepted by the optical verifier and rejected by the public verifier (\VtXoptOnly{} pairs), so the optical readout is the stricter one.
Before error correction, the two readouts differ on \VtXminDis{} to \VtXmaxDis{} of the \VtXNbits{} raw bits per image on average, an agreement of at least \VtXminAgree\%; the largest disagreement on a single image is \VtXmaxDisBits{} bits.
Every failure in the matrix is a NO PAYLOAD outcome of the Reed--Solomon decoder, \VtXnoPayload{} of \VtXallTrials{} verifications; none is NOT VERIFIED (\VtXnotVerified).
The failures come from \VtXfailUnion{} distinct images, of which \VtXfailOptOnly{} fail under the optical verifier only.

\paragraph{Limits.}
The matrix is a property of the numerical model, not of hardware.
It covers the public layer only: the digital signer writes no Layer~2, and Layer-2 detection for the optical signer is reported in Sec.~\ref{sec:keysens}.
The intervals are per cell and attack and carry no correction for the number of cells.

\section{Optical layouts}
\label{sec:app_layouts}

The architecture figure of the main text (Fig.~\ref{fig:pipeline}) is a schematic.
This section gives component-level layouts of the simulated unit, in the style of the optical setups that the review of Jiao \etal~\cite{jiao2019review} collects for optical image hiding.
Every layout describes the numerical model of Sec.~\ref{sec:optics}; none describes a built instrument.

\subsection{What the review offers and what we adopt}
\label{sec:app_review}

The review sorts optical watermarking by the optical system that encodes a hidden image and by the method that embeds the system output into a host.
Three of its practices are adopted here.
First, each scheme is shown as a layout of named components, with one drawing for writing and one for reading (its joint-transform-correlator figure); Fig.~\ref{fig:app_unfolded} does this for our signer, optical verifier, and correlator.
Second, the systems are compared in one table of properties; Tab.~\ref{tab:app_review} places our unit next to its seven systems.
Third, the review warns that double random phase encoding and cascaded phase masks are hard to realize, because random phase masks need precise alignment under coherent light, and that most such studies are numerical simulations only.
Our study is also a simulation, so Appendix~\ref{sec:app_budget} turns that warning into numbers: the alignment tolerances measured in the tolerance sweeps.
We do not adopt any of the reviewed encodings.
They hide an image under a symmetric key, whereas our public layer carries a signature that anyone can check.

\begin{table}[!tb]
\centering\footnotesize
\begin{tabular}{@{}p{0.185\linewidth}p{0.36\linewidth}p{0.40\linewidth}@{}}
\toprule
system in the review~\cite{jiao2019review} & defining elements, as the review describes them & relation to the layouts of this work \\
\midrule
double random phase encoding~\cite{refregier1995,kishk2002hiding} & two random phase masks, in the input plane and in the Fourier plane; the complex output is added to the host with a weight; hard to align, mostly simulated & one phase mask, in the Fourier plane of a reference arm; it writes a keyed carrier, not a hidden image; its alignment tolerance is in Tab.~\ref{tab:app_budget} \\
off-axis Fourier holography & hidden image behind a random phase mask, Fourier transformed, interferes with an off-axis reference wave; the hologram is added to the host & Layer~2 is also a Fourier hologram added by interference; the reference is in line: the host field itself, read by the square-root sensor \\
phase-shifting interferometry & two-arm interferometer with phase retarders; several interferograms per hidden image, each added to a host & same two-arm topology (Fig.~\ref{fig:app_unfolded}a); one exposure per write, no phase steps \\
cascaded phase-only masks & masks computed by iterative phase retrieval; the host stays unchanged; hard to align & not used \\
joint transform correlator & hidden image and key code side by side in the input plane; modulator and two lenses; easy to implement & reading uses a matched-filter correlator~\cite{vanderlugt1964} instead (Fig.~\ref{fig:app_unfolded}c); the paired drawings are adopted \\
ghost imaging & single-pixel detector and many illumination patterns; long acquisition & not used \\
ptychography & several diffraction patterns from overlapping probes; easy to implement, many exposures & not used \\
\midrule
embedding methods & weighted addition in the spatial or a transform domain; quantization of transform coefficients; adaptive embedding & Layer~1 is a quantization embedding in Fourier magnitudes, executed by the modulator; Layer~2 is a weighted addition, executed by interference \\
\bottomrule
\end{tabular}
\caption{The optical systems and embedding methods of the review, and how the layouts of this work relate to them.
The middle column paraphrases the review.}
\label{tab:app_review}
\end{table}

\subsection{Unfolded layouts}
\label{sec:app_unfolded}

\begin{figure}[!htbp]
\centering
\includegraphics{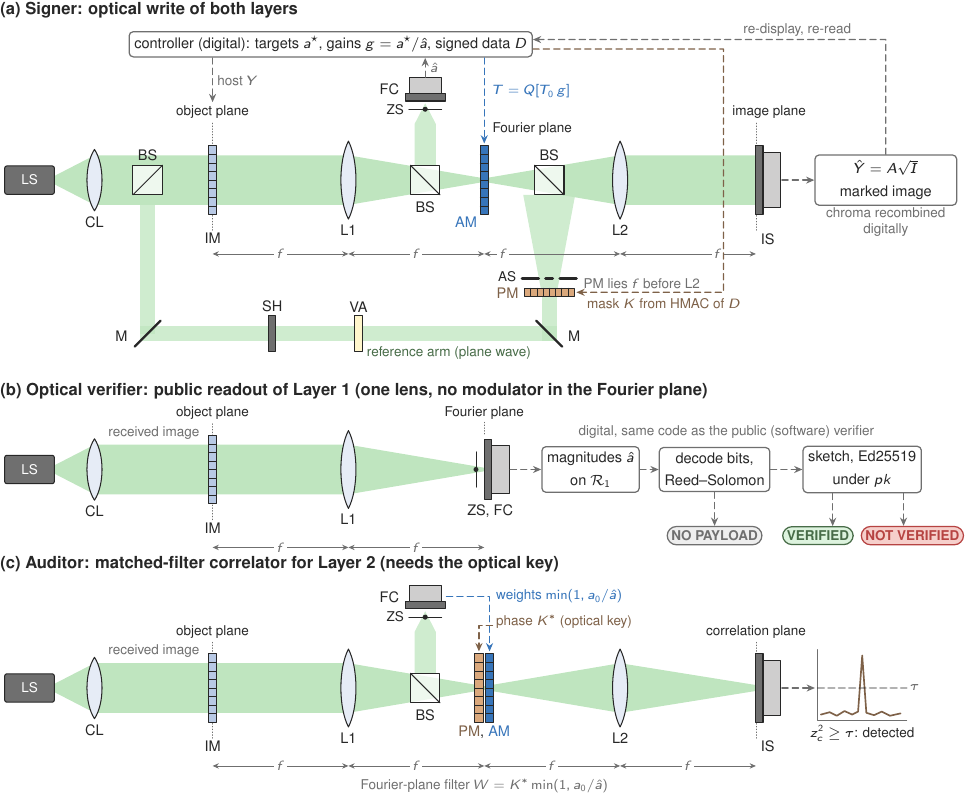}
\caption{Unfolded layouts of the simulated unit; schematic, not to scale, modulators drawn in transmission.
LS: laser source. CL: collimator. BS: beam splitter. IM: input modulator. L1, L2: Fourier lenses of focal length $f$. FC: Fourier camera. ZS: zero-order stop. AM: amplitude modulator. PM: phase modulator. AS: annular stop. SH: shutter. VA: variable attenuator. M: mirror. IS: image sensor.
(a)~The signer writes Layer~1 with AM and adds the Layer-2 carrier through the reference arm.
(b)~The optical verifier reads the Fourier magnitudes; decoding and the signature check are digital.
(c)~The correlator applies the filter $W$, which combines amplitude weights and the conjugate mask in one Fourier plane.
Dashed lines carry data.}
\label{fig:app_unfolded}
\end{figure}

Figure~\ref{fig:app_unfolded} unfolds the three configurations along the optical axis.
All three share the front half: a collimated beam of wavelength \VtLambda\,nm illuminates the input modulator, which displays the luminance on \VtN$\times$\VtN{} pixels of \VtObjPitch\,$\mu$m, an object \VtObjW\,mm wide.
The input modulator lies in the front focal plane of L1, so the back focal plane holds the Fourier transform without a quadratic phase~\cite{goodman1968fourier}.
A beam splitter taps this plane for the Fourier camera.

\paragraph{Signer.}
The amplitude modulator sits in the Fourier plane and multiplies each bin by the quantized gain $T$.
L2 transforms back, and the sensor in its back focal plane records the marked luminance.
The reference arm leaves the beam before the object, passes a shutter and an attenuator, and illuminates the phase modulator as a plane wave.
The phase modulator is one focal length in front of L2.
Seen from L2 through the last beam splitter it lies in the Fourier plane, so L2 turns the mask $K$ into the carrier $r$ at the sensor.
An annular stop limits the reference light to the ring $\mathcal{R}_2$.
The two arms meet in the last beam splitter and interfere on the sensor.
No off-axis angle and no phase stepping is needed, because the real, positive host field acts as the in-line reference of the mark and the square-root readout returns the field.
The attenuator sets the carrier amplitude $\beta$, which the design varies over a factor of \VtRefRange{} in amplitude (\VtRefRangeDb\,dB in intensity) from host to host.
The shutter blocks the reference when the signer re-reads its own output.

\paragraph{Optical verifier.}
The verifier is the front half alone: input modulator, L1, zero-order stop, and Fourier camera.
It needs no modulator in the Fourier plane and no second lens, because the public mark is carried by magnitudes.
The camera region covers the bins up to radius \VtBinsRoi.

\paragraph{Correlator.}
The auditor's unit is a full 4F system with the filter $W$ in the Fourier plane.
$W$ has an amplitude part, the whitening weights, and a phase part, the conjugate mask.
Figure~\ref{fig:app_unfolded}c draws them as two adjacent elements; a hardware version must image one panel onto the other or encode both on one panel.
The sensor in the correlation plane records $|C(\mathbf{x})|^2$.

\paragraph{Light and electronics.}
Table~\ref{tab:app_division} lists which step of signing and verification is done by light.
The optical path covers both transforms, the spectral read, the multiplication by the gains, the addition of the carrier, and detection.
The lattice arithmetic, the error-correcting codes, the content feature, and the signature stay electronic, as in the main text.

\begin{table}[!htbp]
\centering\footnotesize
\begin{tabular}{@{}p{0.36\linewidth}p{0.3\linewidth}p{0.28\linewidth}@{}}
\toprule
step & optical signer and optical verifier & digital signer and public verifier \\
\midrule
Fourier transform of the luminance & lens L1 & FFT \\
ring magnitudes $\hat a$ & Fourier camera, \VtFcNexp{} exposures & exact magnitudes \\
lattice targets $a^{\star}$ and gains $g$ & electronic & electronic \\
feature, sketch, signature, Reed--Solomon code & electronic & electronic \\
multiplication of the spectrum by $g$ & amplitude modulator & array multiplication \\
addition of the Layer-2 carrier & reference arm, interference & not written by the digital signer \\
inverse transform & lens L2 & inverse FFT \\
readout of the marked luminance & image sensor, square root & rounding to 8 bits \\
bit decisions, signature check & electronic, from camera magnitudes & electronic, from FFT magnitudes \\
\bottomrule
\end{tabular}
\caption{Division of work between light and electronics in the simulated unit, next to the digital pipeline.}
\label{tab:app_division}
\end{table}

\subsection{Fourier plane}
\label{sec:app_fplane}

\begin{figure}[!htbp]
\centering
\includegraphics{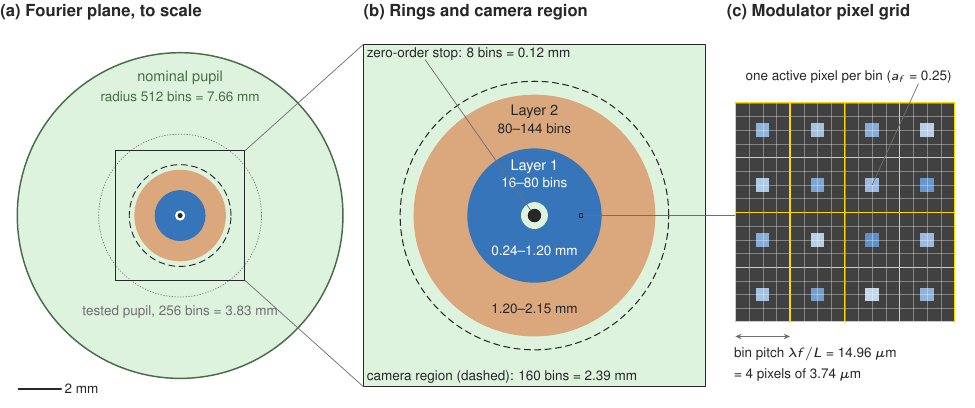}
\caption{Fourier plane of the simulated unit.
(a)~To scale: nominal pupil, a smaller tested pupil, the two rings, and the zero-order stop.
(b)~Layer~1 and Layer~2 rings and the region that the Fourier camera reads.
(c)~One bin spans \VtPxPerBin$\times$\VtPxPerBin{} modulator pixels, of which one is open; the grey levels are illustrative.}
\label{fig:app_fplane}
\end{figure}

Figure~\ref{fig:app_fplane} gives the Fourier plane in physical units.
One DFT bin has the pitch $\lambda f/L=\VtBinPitch\,\mu$m, which equals \VtPxPerBin{} modulator pixels of \VtModPitch\,$\mu$m and fixes the focal length at \VtFzero\,mm.
The public ring $\mathcal{R}_1$ extends from \VtRLoneIn{} to \VtRLoneOut\,mm from the axis and the keyed ring $\mathcal{R}_2$ from \VtRLoneOut{} to \VtRLtwoOut\,mm.
Both rings and the camera region, of diameter \VtRoiD\,mm, lie well inside the nominal pupil of diameter \VtPupilD\,mm.
The pupil is large because it must pass the whole sampled spectrum of the host image, not because of the mark.
The zero-order stop has a diameter of \VtStopD\,mm.
With a pixel aperture fraction of \VtAf, one modulator pixel of \VtActivePx\,$\mu$m is open in each bin.

\section{Reference hardware design: laboratory bench}
\label{sec:app_hardware}

\textbf{No hardware was built.}
Nothing in this section was printed, assembled, aligned, or measured, and no commercial part was bought or tested.
The section turns the simulated parameters into a bench layout, a parts list, a printable mount kit, and a tolerance budget, so that the simulated unit can be judged as an engineering object.
Part classes are examples.
Where a class does not meet the simulated design, Appendix~\ref{sec:app_gaps} says so, and the simulated design is kept.
No hardware was built for either design; Tab.~\ref{tab:app_simperf} gives a simulation of both designs under favorable published tolerances with calibration.
The CAD sources and STL files of this bench and of the portable unit of Appendix~\ref{sec:app_portable} are released at \href{https://github.com/GY19A/asymmetric-optical-image-watermarking}{\texttt{github.com/GY19A/asymmetric-optical-image-watermarking}} (design files only).

\subsection{Folded bench}
\label{sec:app_bench}

\begin{figure}[!htbp]
\centering
\begin{minipage}[c]{0.712\linewidth}
\includegraphics{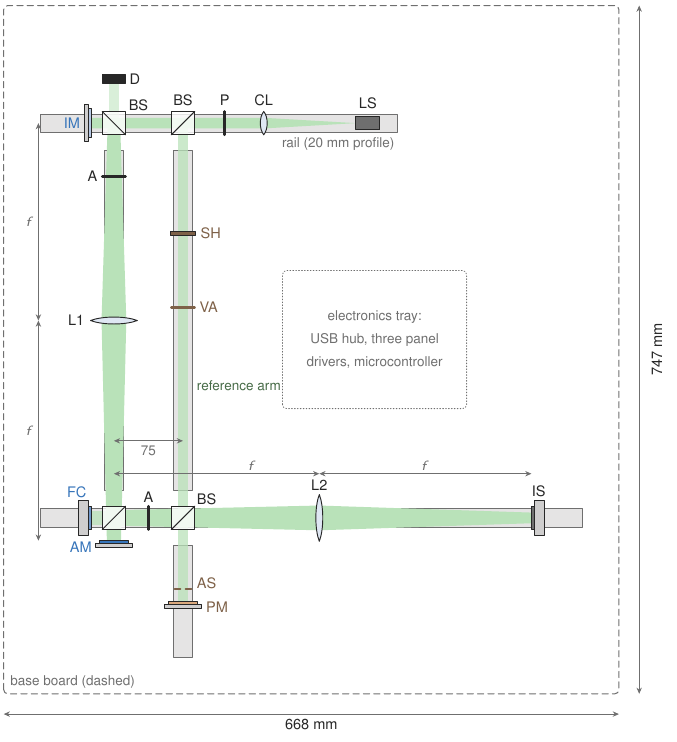}
\end{minipage}\hfill
\begin{minipage}[c]{0.275\linewidth}
\footnotesize\setlength{\tabcolsep}{4pt}
\begin{tabular}{@{}ll@{}}
\toprule
\multicolumn{2}{@{}l}{layout value (mm)} \\
\midrule
focal length $f$ & \VtFzero \\
cube edge & \VtCube \\
cube to panel & \VtGap \\
cube to L1 & \VtLegAone \\
combiner to L2 & \VtLegAthree \\
combiner to PM & \VtLegGthree \\
line spacing & \VtXthree \\
axis above rail & \VtAxisH \\
rails & \VtRailN, \VtRailS, \VtRailW, \\
 & \VtRailRef, \VtRailStub \\
base board & \VtFootW$\times$\VtFootD \\
arm mismatch & \VtArmMismatch \\
\bottomrule
\end{tabular}

\vspace{6pt}
\raggedright
P: polarizer. A: analyzer. D: beam dump. Other labels as in Fig.~\ref{fig:app_unfolded}.
The distances count each pass through a cube as its thickness divided by the assumed glass index \VtGlassN.
\end{minipage}
\caption{Top view of the folded bench of the reference design, to scale; beams show the required clear apertures.
Reflective panels sit behind beam-splitter cubes.
The same bench without AM, PM, L2, and IS is the optical verifier.
Design only.}
\label{fig:app_bench}
\end{figure}

Small-pitch modulators are reflective, so the bench of Fig.~\ref{fig:app_bench} folds each panel behind a cube beam splitter.
The unfolded path, the focal distances, and the conjugate planes are those of Fig.~\ref{fig:app_unfolded}a.
The beam enters along the north rail, is split into the signal and the reference line, and meets the input modulator behind the first cube.
The returning light is reflected south through L1 to the Fourier cube.
The part reflected there on the first pass reaches the Fourier camera, the transmitted part reaches the amplitude modulator, and the light returning from it is reflected east through the combiner cube and L2 to the image sensor.
The reference line runs south at a spacing of \VtXthree\,mm, passes the combiner cube, and is reflected by the phase modulator back into it and on to L2.
Each cube pass is counted with its air-equivalent thickness, so that the input modulator, both Fourier-plane panels, and the Fourier camera sit at one focal length from their lens.
The required beam diameter is largest at the lenses, \VtCaLens\,mm, against a clear aperture of \VtCaLensHave\,mm, and tightest at the combiner cube, \VtCaComb\,mm against \VtCaCubeHave\,mm.
The base board measures \VtFootW$\times$\VtFootD\,mm, with room for the electronics next to the beams.

\subsection{Parts}
\label{sec:app_parts}

\begin{table}[!htb]
\centering\footnotesize
\begin{tabular}{@{}p{0.15\linewidth}p{0.295\linewidth}p{0.50\linewidth}@{}}
\toprule
element & part class (example) & required by the simulated design \\
\midrule
laser source & single-frequency laser, fiber coupled & \VtLambda\,nm; coherence length well above the arm mismatch of \VtArmMismatch\,mm; scale match to an auditor unit within Tab.~\ref{tab:app_budget} \\
collimator & achromat, focal length \VtFcol\,mm & uniform beam over the object diagonal of \VtObjDiag\,mm \\
polarizer, analyzers & \VtOptD\,mm film or glass polarizers & amplitude operation of the panels (polarization is not modeled) \\
beam splitters & four non-polarizing cubes, \VtCube\,mm & clear aperture of at least \VtCaComb\,mm at the combiner \\
input modulator & reflective liquid-crystal-on-silicon panel, amplitude mode, \VtObjPitch\,$\mu$m class & \VtN$\times$\VtN{} pixels of \VtObjPitch\,$\mu$m (\VtObjW\,mm), \VtBIn{} amplitude bits \\
lenses L1, L2 & achromats, focal length trimmed & $f=\VtFzero$\,mm, matched pair, clear aperture of at least \VtCaLens\,mm \\
amplitude modulator & reflective liquid-crystal-on-silicon panel, \VtModPitch\,$\mu$m class & pitch \VtModPitch\,$\mu$m (\VtPxPerBin{} pixels per bin), \VtBSlm{} amplitude bits, one open pixel per bin; pupil of \VtPupilPx{} pixels (\VtPupilD\,mm) in both axes \\
phase modulator & reflective phase-only panel, \VtModPitch\,$\mu$m class & pitch \VtModPitch\,$\mu$m, \VtBPhi{} phase bits over $2\pi$; ring $\mathcal{R}_2$ needs \VtLtwoPx{} pixels (\VtRLtwoOut\,mm radius) \\
Fourier camera & monochrome USB board camera & samples at the bin centers, pitch \VtBinPitch\,$\mu$m, over \VtRoiD\,mm; $\VtFcNe$\,e$^-$ full well, \VtFcRead\,e$^-$ read noise, \VtFcBits{} bits, \VtFcNexp{} exposures in steps of \VtFcGain; stop of \VtStopD\,mm \\
image sensor & monochrome USB board camera & \VtN$\times$\VtN{} samples of \VtObjPitch\,$\mu$m (\VtObjW\,mm square); $\VtSNe$\,e$^-$ effective full well from \VtSFrames{} frames of $\VtSWell$\,e$^-$, \VtSRead\,e$^-$ read noise, \VtSBits{} bits \\
shutter, attenuator & servo flag; variable neutral-density filter & reference amplitude over a range of \VtRefRange{} (\VtRefRangeDb\,dB); mean reference intensity between $\VtRefRelLo$ and $\VtRefRelHi$ of the sensor full scale \\
rails, base & five \VtRail\,mm slotted profiles; flat board & rail lengths \VtRailN, \VtRailS, \VtRailW, \VtRailRef, and \VtRailStub\,mm; board \VtFootW$\times$\VtFootD\,mm \\
\bottomrule
\end{tabular}
\caption{Commercial parts of the reference design by class, with the values that the simulated design requires of them.
No part was bought or tested.}
\label{tab:app_bom}
\end{table}

Table~\ref{tab:app_bom} lists the commercial parts by class.
The right column repeats the simulated configuration of Tab.~\ref{tab:optics} in the units of a parts list; it adds no new parameter.

\paragraph{Light budget.}
The unit is not light efficient, and the following estimate is arithmetic on the design values, not a measurement.
One open pixel per bin sends $a_f^{4}=1/\VtThrCombInv$ of the intensity into the image order, the four ideal cubes pass $1/\VtThrCubeInv$, and the modulator bias $T_0$ passes about a quarter.
The signal arm then delivers at most one part in $\VtThrInv$ of the light at the object to the sensor.
Filling every pixel of one frame to the full well takes \VtThrFrameNj\,nJ at the sensor, hence at least \VtThrFrameMj\,mJ at the object, or \VtThrPower\,mW for an exposure of \VtThrExp\,s.
Polarizer losses, panel reflectivity, quantum efficiency, and beam overfill come on top.
A fully open pixel aperture would lower the bound to \VtThrPowerFull\,mW; the tolerance sweeps tested that aperture (Tab.~\ref{tab:app_budget}).
At such powers the beam must be enclosed, so the housing is light tight and its lid switch disables the laser.

\subsection{Printed mount kit}
\label{sec:app_kit}

\begin{figure}[!htbp]
\centering
\includegraphics{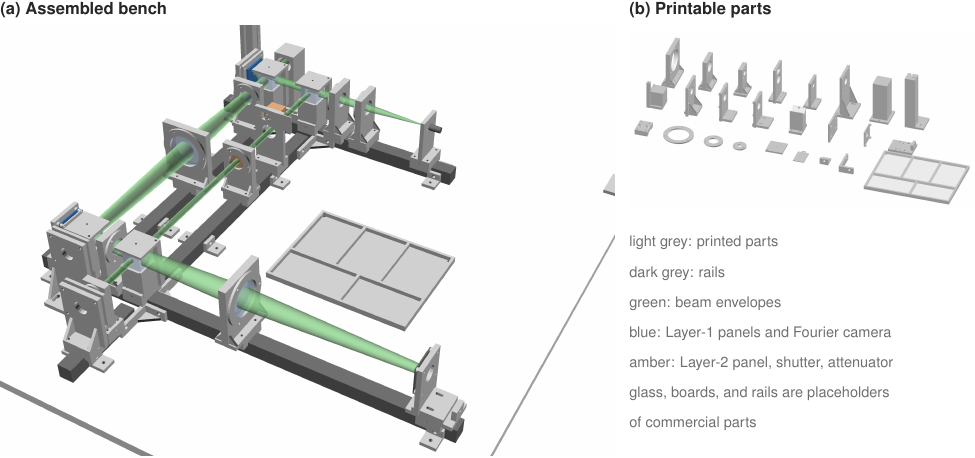}
\caption{Printed mount kit of the reference design, rendered from the parametric sources.
(a)~The bench with its \VtCadInstances{} printed parts; the housing sheets are omitted.
(b)~The \VtCadParts{} printable parts.
Design only: nothing was printed.}
\label{fig:app_kit}
\end{figure}

The kit holds every commercial part of Tab.~\ref{tab:app_bom} on a base board with \VtCadRails{} slotted rails (Fig.~\ref{fig:app_kit}).
It has \VtCadParts{} printable parts; the bench uses \VtCadUsed{} of them in \VtCadInstances{} instances, with a printed volume of \VtCadVolume\,cm$^3$ (Tab.~\ref{tab:app_printed}).
Every holder bolts to a rail carrier, a saddle of \VtCarL$\times$\VtCarW\,mm that is fixed to the rail with two screws.
The optical axis runs \VtAxisH\,mm above the rails.
Round optics rest in a seat under a retaining ring, and each cube rests against a locating corner under a clamp.
Each panel board is screwed to a plate that tilts on three spring-loaded screws; slots allow a lateral shift of $\pm$\VtPanShift\,mm, and the fine registration is done on the panel itself (Appendix~\ref{sec:app_budget}).
The image sensor rides on a slide with $\pm$\VtSlideTravel\,mm of travel along the beam.
The amplitude modulator lies beside its rail and stands on the base board, as does the beam dump.
Corner posts carry the sheets of the housing, and a tray holds the electronics.

The sources are parametric and read every position from one layout file, which is generated from the simulated parameters.
Three checks were run on the model.
Every exported mesh is a closed solid that fits a build volume of \VtCadBuildX$\times$\VtCadBuildX$\times$\VtCadBuildZ\,mm.
The beam envelopes of Fig.~\ref{fig:app_bench}, enlarged by \VtCadMargin\,mm in diameter, intersect no printed part.
The union of all printed instances has the same volume as the sum of the parts, to within \VtCadUnionDiff\,cm$^3$, so no two parts overlap.
These checks say nothing about stiffness, thermal drift, or print tolerances, which were not analyzed.

\begin{table}[!htbp]
\centering\footnotesize
\begin{tabular}{@{}lrlrl@{}}
\toprule
part & qty. & box (mm) & cm$^3$ & function \\
\midrule
rail carrier & 18 & 30$\times$40$\times$12 & 11.0 & saddle on a rail; carries one holder \\
lens holder 50 & 2 & 30$\times$71$\times$87 & 51.1 & Fourier lenses L1, L2 \\
retaining ring 50 & 2 & 69$\times$69$\times$3 & 5.7 & holds a lens in its seat \\
lens holder 25 & 5 & 30$\times$45$\times$75 & 34.9 & collimator, polarizer, analyzers, attenuator \\
retaining ring 25 & 5 & 43$\times$43$\times$3 & 3.2 & holds a small optic in its seat \\
lens holder 12 & 1 & 30$\times$40$\times$68 & 26.8 & annular stop in front of the phase modulator \\
retaining ring 12 & 1 & 31$\times$31$\times$3 & 1.9 & holds the stop window \\
cube table & 4 & 37$\times$41$\times$66 & 45.4 & seat with locating corner for a beam-splitter cube \\
cube clamp & 4 & 40$\times$40$\times$4 & 6.2 & top clamp of a cube \\
panel frame rail & 2 & 30$\times$56$\times$76 & 30.9 & fixed frame of a panel mount, on a rail \\
panel frame board & 1 & 44$\times$60$\times$104 & 49.0 & same, standing on the base board (amplitude modulator) \\
panel tilt plate & 3 & 4$\times$56$\times$48 & 10.4 & tilt plate that carries a panel board \\
camera mount fixed & 1 & 31$\times$46$\times$75 & 24.7 & Fourier camera board \\
stop cap & 1 & 5$\times$39$\times$39 & 2.8 & holds the zero-order stop in front of the Fourier camera \\
camera slide base & 1 & 56$\times$40$\times$20 & 14.4 & base of the focus slide of the image sensor \\
camera slide carriage & 1 & 38$\times$46$\times$69 & 24.7 & carriage with the image sensor board \\
fiber holder & 1 & 30$\times$40$\times$68 & 21.4 & fiber adapter at the collimator focus \\
laser vblock & 0 & 30$\times$40$\times$54 & 38.1 & alternative: clamp for a laser module barrel \\
shutter bracket & 1 & 30$\times$46$\times$88 & 25.7 & servo bracket of the reference shutter \\
shutter flag & 1 & 26$\times$43$\times$2 & 1.6 & flag of the shutter \\
beam dump & 1 & 44$\times$50$\times$97 & 105.7 & trap for the unused port of the input cube \\
rail foot & 18 & 20$\times$22$\times$15 & 3.0 & fixes a rail to the base board \\
corner bracket & 5 & 34$\times$54$\times$15 & 6.8 & joins two rails at a right angle \\
enclosure post & 4 & 44$\times$44$\times$130 & 73.5 & corner post for the housing sheets \\
electronics tray & 1 & 200$\times$150$\times$8 & 86.3 & bays for hub, three panel drivers, microcontroller \\
\bottomrule
\end{tabular}

\caption{Printable parts of the mount kit: quantity in the bench, bounding box, and solid volume of the exported mesh.
Design only.}
\label{tab:app_printed}
\end{table}

\subsection{USB peripheral}
\label{sec:app_usb}

\begin{figure}[!htbp]
\centering
\includegraphics{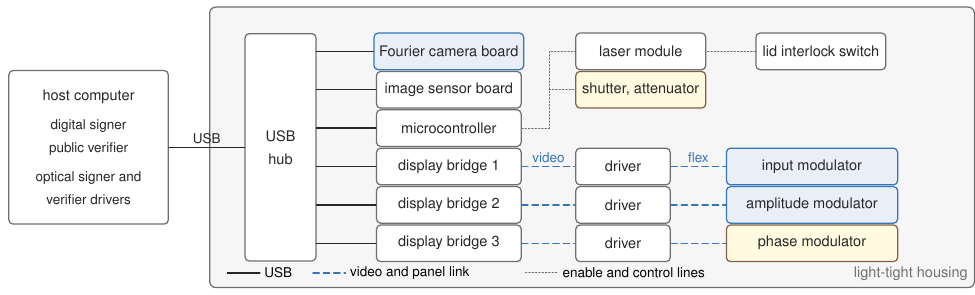}
\caption{Electronics of the reference design as one USB peripheral.
Blue: parts that serve Layer~1. Amber: parts that serve Layer~2. Design only.}
\label{fig:app_usb}
\end{figure}

Figure~\ref{fig:app_usb} shows the electronics.
The housing exposes one USB connection and a power input.
An internal hub serves the two camera boards, three display bridges, and a microcontroller.
Each bridge feeds the driver board of one panel; the microcontroller switches the laser, the reference shutter, and the attenuator, and reads the lid switch.
The host computer runs the same software for both domains: it computes targets, gains, codes, and signatures, and it either transforms the image with an FFT or sends it through the unit.
The bridges must deliver uncompressed, pixel-exact frames, because a single grey level on a panel is a modulator code.
A bridge that compresses video is excluded, and a read-back test of displayed patterns belongs to commissioning.

\subsection{Tolerance and alignment budget}
\label{sec:app_budget}

\begin{table}[!htbp]
\centering\footnotesize
\begin{tabular}{@{}p{0.17\linewidth}p{0.065\linewidth}p{0.17\linewidth}p{0.15\linewidth}p{0.34\linewidth}@{}}
\toprule
quantity & nominal & last tested value that passes & next tested value & consequence for the bench \\
\midrule
modulator shift (bins) & 0 & \VtBudShift: \VtBudVnShift{} / \VtBudVjShift & \VtNxtShift: \VtNxtVnShift{} / \VtNxtVjShift & \VtShiftBudUm\,$\mu$m passes, \VtShiftFailUm\,$\mu$m fails; the pattern is registered in steps of one pixel (\VtPixShift{} bin), residual at most \VtResidShift{} bin \\
modulator tilt (degrees) & 0 & \VtBudTilt: \VtBudVnTilt{} / \VtBudVjTilt & \VtNxtTilt: \VtNxtVnTilt{} / \VtNxtVjTilt & foreshortening of \VtTiltFore\% passes, \VtTiltForeFail\% fails; three-screw tilt plate \\
sensor defocus (mm) & 0 & \VtBudDz: \VtBudVnDz{} / \VtBudVjDz & \VtNxtDz: \VtNxtVnDz{} / \VtNxtVjDz & quality sets the limit: median PSNR \VtPsDzA\,dB at \VtDzA\,mm, \VtPsDzB\,dB at \VtDzB\,mm; focus slide \\
amplitude modulator bits & \VtBSlm & \VtBudBslm: \VtBudVnBslm{} / \VtBudVjBslm & \VtNxtBslm: \VtNxtVnBslm{} / \VtNxtVjBslm & a panel with \VtBSlm{} bits has \VtBslmMargin{} bits of margin \\
pixel aperture fraction & \VtAf & \VtBudAf: \VtBudVnAf{} / \VtBudVjAf & none fails & a fully open pixel is tolerated; Layer~2 is then detected on \VtLttAfFull\% \\
sensor full well (e$^-$) & $\VtSNe$ & \VtBudSne: \VtBudVnSne{} / \VtBudVjSne & \VtNxtSne: \VtNxtVnSne{} / \VtNxtVjSne & median PSNR \VtPsSneA\,dB at \VtSneA{} and \VtPsSneB\,dB at \VtSneB \\
sensor bits & \VtSBits & \VtBudSbits: \VtBudVnSbits{} / \VtBudVjSbits & none fails & median PSNR \VtPsSbitsA\,dB at \VtSbitsA{} bits, \VtPsSbitsB\,dB at \VtSbitsB{} bits \\
Fourier camera, exposures $\times$ bits & \VtFcNexp$\times$\VtFcBits & 2$\times$12: 99.2 / 99.3; 3$\times$8: 99.8 / 99.8; 3$\times$10: 99.8 / 99.8 & 1$\times$12: 19.5 / 18.9 & exposure bracketing is required; the bit depth is not critical \\
pupil radius (bins) & \VtBinsPupil & \VtBudPupil: \VtBudVnPupil{} / \VtBudVjPupil & none fails & quality sets the limit: median PSNR \VtPsPupilSmall\,dB at \VtPupilSmall{} bins \\
arm coherence & \VtGamma & \VtBudGamma: \VtBudVnGamma{} / \VtBudVjGamma & none fails & Layer~2 is detected on \VtLttGamA\% at \VtGamA, \VtLttGamB\% at \VtGamB, \VtLttGamC\% at \VtGamC \\
\midrule
\multicolumn{5}{@{}l}{auditor unit against signer unit: Layer-2 detection rate in percent, \VtKeyN{} unattacked images} \\
\midrule
scale $\lambda f$ (\%) & 0 & \VtKeyScaleOk: \VtKeyScaleOkDetP{} / \VtKeyScaleOkDetM & \VtKeyScaleBad: \VtKeyScaleBadDetP{} / \VtKeyScaleBadDetM & \VtKeyScaleOkNm\,nm of wavelength or \VtKeyScaleOkFmm\,mm of focal length pass; the median statistic crosses the threshold at \VtKeyScaleTol\% \\
mask shift (bins) & 0 & \VtKeyShiftOk: \VtKeyShiftOkDet & \VtKeyShiftBad: \VtKeyShiftBadDet & \VtKeyShiftOkUm\,$\mu$m passes; the median crosses at \VtKeyShiftTol{} bins (\VtKeyShiftTolUm\,$\mu$m) \\
\bottomrule
\end{tabular}

\caption{Tolerance and alignment budget from the one-at-a-time sweeps of the numerical simulation (\VtBudN{} images, nominal optics, public verifier).
Upper block: VERIFIED rate in percent without attack / after JPEG~75; a value passes if both rates are at least \VtBudMin\%.
Lower block: the two signs of the scale mismatch.
Combined errors are not part of these sweeps; the simulation of Appendix~\ref{sec:app_simperf} draws them jointly.}
\label{tab:app_budget}
\end{table}

Table~\ref{tab:app_budget} derives the alignment budget from the tolerance and key-sensitivity sweeps of the evaluation.
The sweeps change one parameter at a time, so the table bounds each error alone.

\paragraph{What needs care.}
Three quantities need alignment at the micrometer or sub-percent level.
The first is the lateral registration of the amplitude modulator to the spectrum.
Verification survives \VtShiftBudUm\,$\mu$m and fails at \VtShiftFailUm\,$\mu$m.
No printed part positions a panel that well, and none has to: the pattern is shifted on the panel in whole pixels of \VtModPitch\,$\mu$m, which leaves at most \VtResidShiftUm\,$\mu$m.
The tested shifts on either side of this residual, \VtBracketLo{} and \VtBracketHi{} bins, keep the VERIFIED rate at \VtVnBracketLo\% and \VtVnBracketHi\% without attack.
The second is the scale $\lambda f$, which must map bin $\mathbf{u}$ to its \VtPxPerBin{} pixels.
A foreshortening of \VtTiltFore\% along one axis passes, which corresponds to \VtTiltEdgeBins{} bins at the outer edge of $\mathcal{R}_1$; an isotropic scale error of the signer was not swept.
Between a signer and an auditor unit, a scale difference of \VtKeyScaleOk\% is tolerated, that is \VtKeyScaleOkNm\,nm of wavelength or \VtKeyScaleOkFmm\,mm of focal length.
The simulated focal length is no catalogue value, so the reference design trims the focal length of a two-lens combination and checks it with gratings displayed on the input modulator.
The third is the exposure bracketing of the Fourier camera: a single exposure fails.

\paragraph{What does not.}
Tilt of the modulator, defocus of the sensor, the bit depths, and the coherence of the two arms have wide margins for verification.
Defocus and a small pupil cost image quality long before they cost verification, because the public mark lives in magnitudes.
Coherence matters only for Layer~2.
With an arm mismatch of \VtArmMismatch\,mm, a source with a short coherence length would lose the keyed layer and leave the public layer intact.

\paragraph{Arm phase.}
The keyed layer has one requirement that the sweeps cover only indirectly.
For the real signal field of the model, a static phase offset $\varphi$ between the two arms acts on the sensor exactly like a coherence lowered by the factor $\cos\varphi$.
The tested coherence of \VtGamA{} therefore stands for an offset of \VtPhaseA$^\circ$, or \VtPhasePathA\,nm of path, and the tested coherence of \VtGamB{} for \VtPhaseB$^\circ$.
Layer~2 thus needs interferometric stability of the arm difference, which a printed bench does not provide by itself.
The phase modulator could re-zero the offset before each write; this is not designed.
Layer~1 does not depend on the arm phase.

\paragraph{Alignment sequence.}
The intended sequence was not executed.
(1)~Collimate: minimize the zero-order spot on the Fourier camera with a blank input.
(2)~Set the Fourier plane: slide the camera to the smallest spot; set the amplitude modulator to the same reduced distance.
(3)~Set the scale: display gratings of \VtBinsLoneIn, \VtBinsLoneOut, and \VtBinsLtwoOut{} cycles; trim the focal length until their orders fall on the bins of the same index.
(4)~Register the modulator: attenuate one known bin and shift the pattern in whole pixels until the response at the sensor peaks.
(5)~Focus the sensor with a displayed checkerboard, to within \VtDzA\,mm.
(6)~Open the reference arm, set the attenuator, and check the fringe contrast.

\subsection{Where commercial part classes fall short of the simulated design}
\label{sec:app_gaps}

The simulated design is kept in every case below; the gaps are open issues of the hardware design.
(i)~\emph{Pupil.} The nominal pupil needs \VtPupilPx{} modulator pixels in both axes. The reflective panel class at \VtModPitch\,$\mu$m offers \VtPanelPxW$\times$\VtPanelPxH{} pixels, that is $\pm$\VtPanelHalfW{} by $\pm$\VtPanelHalfH{} bins, and covers \VtPanelFrac\% of the pupil disc. Both rings fit. A rectangular aperture was not simulated; the circular pupil of \VtPupilSmall{} bins keeps verification at \VtVnnPupilSmall\% and lowers the median PSNR from \VtNomPsnr{} to \VtPsPupilSmall\,dB.
(ii)~\emph{Input bit depth.} The design displays the host with \VtBIn{} bits. The amplitude panel class at \VtObjPitch\,$\mu$m is addressed with \VtInBitsClass{} bits, and this parameter was not swept.
(iii)~\emph{Dark sub-pixels.} A phase-only panel cannot switch fifteen of sixteen pixels dark, so the reference arm needs an aperture mask or an off-pixel pattern. The fully open aperture is the tested alternative.
(iv)~\emph{Amplitude operation.} A liquid-crystal panel modulates amplitude between polarizers with a coupled phase and finite contrast; neither is modeled.
(v)~\emph{Sampling.} The model takes point samples at the bin centers and \VtObjPitch\,$\mu$m samples at the sensor. Real pixels integrate, and their grids must be registered; this is not part of the model (the simulation of Appendix~\ref{sec:app_simperf} adds the registration of the Fourier camera).
(vi)~\emph{Arm phase.} The model adds the carrier in phase with the host field; holding that phase on a printed bench is unsolved (Appendix~\ref{sec:app_budget}).
(vii)~\emph{Correlator.} The mount kit covers the signer and the optical verifier. The correlator of Fig.~\ref{fig:app_unfolded}c is an optical layout only.

\section{Reference hardware design: portable unit}
\label{sec:app_portable}

\begin{figure}[!htbp]
\centering
\begin{minipage}[c]{0.56\linewidth}
\includegraphics[width=\linewidth]{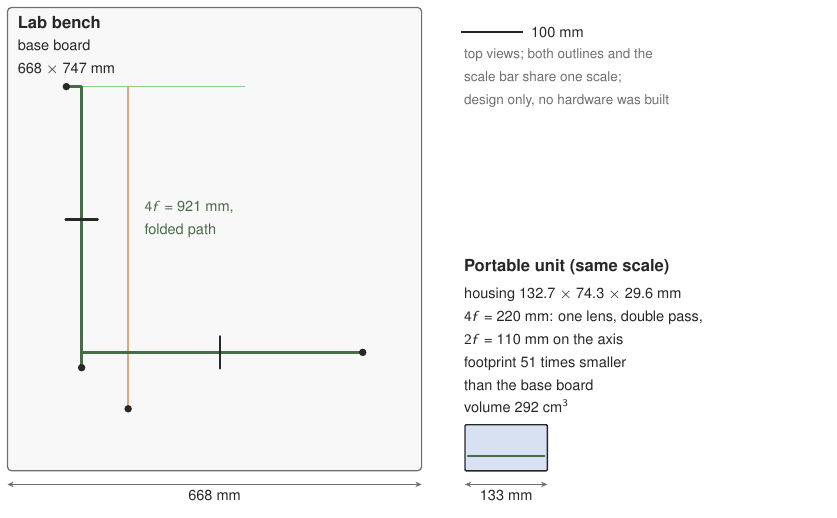}
\end{minipage}\hfill
\begin{minipage}[c]{0.38\linewidth}
\caption{The base board of the laboratory bench and the outline of the portable unit at the same scale (design only).}
\label{fig:app_size}
\end{minipage}
\end{figure}

\textbf{No hardware was built.}
The bench is an instrument for a laboratory.
This section asks how small the same scheme can be made and designs a unit of the size of a power bank, as a portable, open-hardware option next to the bench.
It gives the design: its CAD sources and STL files are released with those of the bench (Appendix~\ref{sec:app_hardware}), and Tab.~\ref{tab:app_simperf} gives a simulation of the unit under favorable published tolerances with calibration.
The scheme is unchanged: the signed payload in the Fourier magnitudes (Layer~1), the keyed carrier (Layer~2), the frame of \VtN$\times$\VtN{} pixels, both rings, and the Reed--Solomon and BCH layout are those of the main text.
Only the optics is scaled.
Figure~\ref{fig:app_size} compares its footprint with the base board of the bench.

\subsection{Optical parameters}
\label{sec:app_popt}

\paragraph{Scaling rule.}
One bin of the Fourier plane has the pitch $p=\lambda f/L$, so the focal length follows from the object width, the bin pitch, and the wavelength: $f=p\,L/\lambda$.
The bench uses an object pitch of \VtObjPitch\,$\mu$m and bins of \VtPxPerBin{} modulator pixels, which gives \VbLabF\,mm and a path of $4f=\VbLabFourF$\,mm.
Table~\ref{tab:app_trade} lists what a smaller object pitch, fewer pixels per bin, and other diode wavelengths give.
Three constraints select the design.
The frame must keep \VtN{} pixels, because the frame rule, the rings, and the payload of \VbPBitsNeeded{} ring bins of \VbPBitsHave{} are defined on it.
A bin of one pixel cannot be registered better than half a bin, which is beyond the budget of \VtBudShift{} bin.
A bin of four pixels puts the nominal pupil of \VtPupilPx{} pixels off the short side of the panel class (Appendix~\ref{sec:app_gaps}) and needs $2f$ above \VbEnvLen\,mm at every listed wavelength.

\begin{table}[!htbp]
\centering\footnotesize
\begin{tabular}{@{}rrrrrrrll@{}}
\toprule
$N$ & object pitch & pixels & $\lambda$ & $f$ & $2f$ & pupil & pupil on & admissible \\
 & ($\mu$m) & per bin & (nm) & (mm) & (mm) & (mm) & the panel & \\
\midrule
1024 & 3.74 & 1 & 405 & 35.4 & 71 & 3.83 & yes & no: registration \\
1024 & 3.74 & 1 & 520 & 27.5 & 55 & 3.83 & yes & no: registration \\
1024 & 3.74 & 1 & 638 & 22.5 & 45 & 3.83 & yes & no: registration \\
1024 & 3.74 & 2 & 405 & 70.7 & 141 & 7.66 & yes & no: $2f$ too long \\
\textbf{1024} & \textbf{3.74} & \textbf{2} & \textbf{520} & \textbf{55.1} & \textbf{110} & \textbf{7.66} & \textbf{yes} & \textbf{yes (chosen)} \\
1024 & 3.74 & 2 & 638 & 44.9 & 90 & 7.66 & yes & yes \\
1024 & 3.74 & 4 & 405 & 141.5 & 283 & 15.32 & no & no: pupil off panel \\
1024 & 3.74 & 4 & 520 & 110.2 & 220 & 15.32 & no & no: pupil off panel \\
1024 & 3.74 & 4 & 638 & 89.8 & 180 & 15.32 & no & no: pupil off panel \\
1024 & 8.00 & 1 & 520 & 58.9 & 118 & 3.83 & yes & no: registration \\
1024 & 8.00 & 2 & 520 & 117.8 & 236 & 7.66 & yes & no: $2f$ too long \\
1024 & 8.00 & 4 & 520 & 235.7 & 471 & 15.32 & no & no: pupil off panel \\
\bottomrule
\end{tabular}

\caption{Scaling the optics: focal length $f=p\,L/\lambda$ for an object of $N$ pixels (width $L$) and bins of whole modulator pixels of \VtModPitch\,$\mu$m (bin pitch $p$).
A row is admissible if it keeps $N$, has the nominal pupil on a panel of the \VbPanelPx{} class in both axes, leaves a registration residual of at most \VbPRegRes{} bin, and fits a single-lens double pass ($2f$ plus panels and walls) into \VbEnvLen\,mm.
Design arithmetic.}
\label{tab:app_trade}
\end{table}

\begin{figure}[!tb]
\centering
\includegraphics{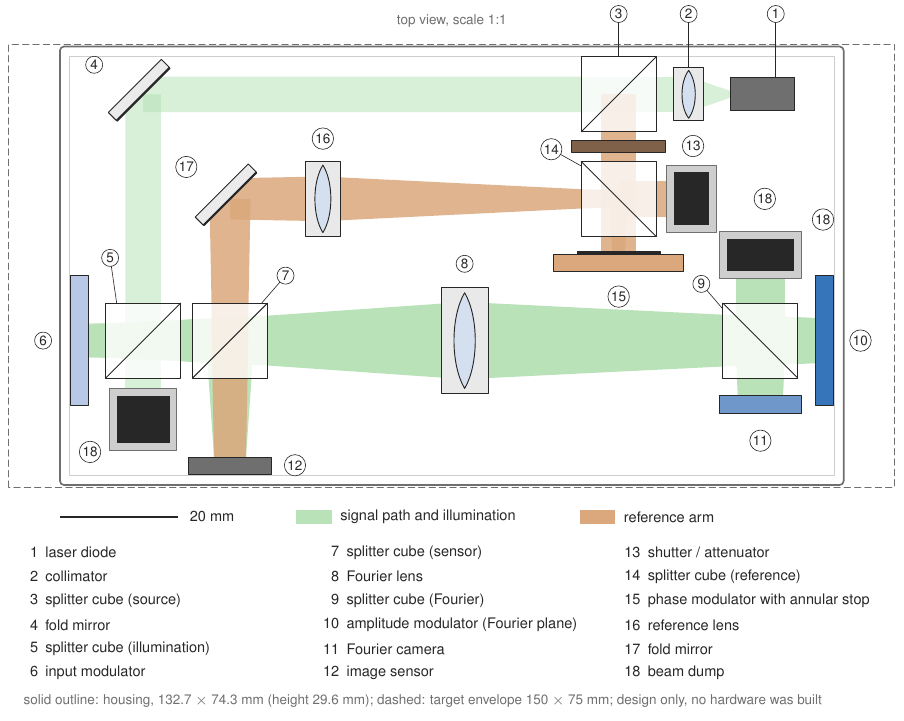}
\caption{Optical layout of the portable unit, top view, to scale (design only).
The signal path runs from the input modulator through the Fourier lens to the amplitude modulator in the Fourier plane and back through the same lens to the image sensor; the reference arm adds the Layer-2 carrier at the sensor cube.}
\label{fig:app_playout}
\end{figure}

\paragraph{Chosen design.}
The portable unit displays the object on a panel of \VbPObjPitch\,$\mu$m pitch, so the object is \VbPObjW\,mm wide, and uses bins of \VbPPxPerBin{} pixels, \VbPBin\,$\mu$m.
At \VbPLambda\,nm, a single-mode diode wavelength next to the \VtLambda\,nm of the bench, the focal length is \VbPF\,mm.
The amplitude modulator in the Fourier plane is reflective, so one lens serves as both Fourier lenses: the light passes it, is modulated, and passes it again.
The optical path is then $2f=\VbPTwoF$\,mm long instead of $4f$.
The pupil of \VbPupilBins{} bins is \VbPPupil\,mm wide, \VbPPupilPx{} pixels, and fits the panel in both axes, which the bench design does not achieve.
The rings shrink with the bin: Layer~2 ends at a radius of \VbPRingLL\,mm, and the zero-order stop is \VbPStop\,mm wide.
Three properties are less favorable than on the bench.
One open pixel per bin is now half of the bin (aperture fraction \VbPAf{} instead of \VtAf; Tab.~\ref{tab:app_budget} covers this value for the bench parameters only).
The registration residual doubles to \VbPRegRes{} bin.
The beams are faster and the field angle larger (half field \VbPHalfField$^\circ$), which raises the lens aberrations.
A wavelength of \VbPAltLambda\,nm would shorten the unit further ($f=\VbPAltF$\,mm) and is the alternative of Tab.~\ref{tab:app_trade}.
The estimated light throughput of the signal arm is \VbPThroughGain{} times that of the bench, because the pixel aperture is larger; the estimate ignores the losses of the panels and polarizers as before.
The image sensor samples the output at the object pitch of \VbPObjPitch\,$\mu$m; with a well of $10^4$ electrons per pixel, the simulated effective well needs \VbPFramesWell{} averaged frames.

\subsection{Layout and enclosure}
\label{sec:app_pcad}

\begin{figure}[!htbp]
\centering
\includegraphics{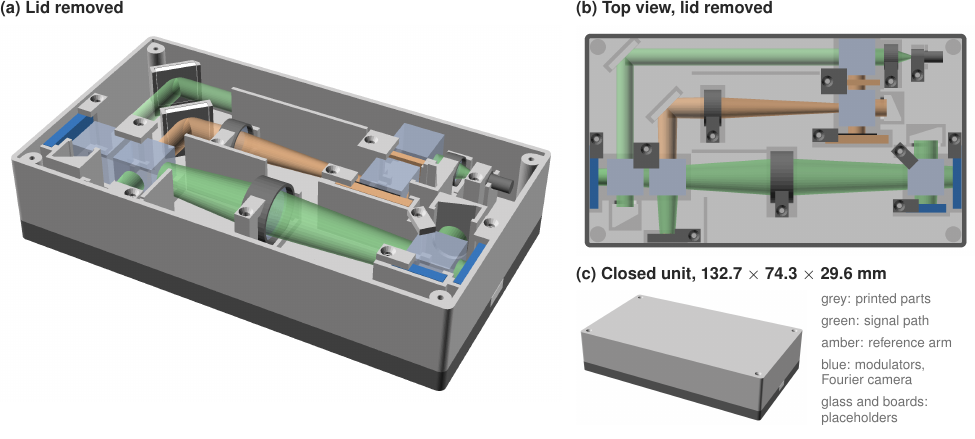}
\caption{Printable enclosure of the portable unit, rendered from the parametric model (not printed, not built).}
\label{fig:app_pcad}
\end{figure}

\paragraph{Layout.}
Figure~\ref{fig:app_playout} shows the folded path.
On the main axis lie the input modulator, the illumination cube, the sensor cube, the Fourier lens, the Fourier cube, and the amplitude modulator.
Collimated light of the laser diode enters through the illumination cube and is reflected onto the input modulator.
On the way to the Fourier plane, the Fourier cube sends half of the light to the Fourier camera behind its zero-order stop; this arm alone is the optical verifier, so the signer and the optical verifier share one unit.
The light that returns from the amplitude modulator passes the lens a second time and is reflected by the sensor cube onto the image sensor.
The reference arm runs parallel to the main axis: a second cube, the phase modulator with its annular stop, a reference lens of the same focal length, and a fold mirror that sends the carrier through the sensor cube onto the image sensor.
Every modulator and both sensors sit at one focal length from their lens when each pass through a cube is counted with its air-equivalent thickness; the parametric layout satisfies the six conjugate conditions, keeps a clearance of at least \VbPClear\,mm between all bodies, and leaves the beam paths free.
The lenses are placed as thin lenses at the centers of their cells.
A two-element lens of this focal length is \VbLensPLen\,mm long with focal distances of \VbLensPFfd{} and \VbLensPBfd\,mm from its outer vertices, so a detailed design moves the elements on the main axis by a few millimeters, within the spare length.
The correlator of the auditor needs amplitude and phase in one Fourier plane and is not laid out, as for the bench.

\paragraph{Enclosure.}
The printed enclosure has three main parts: a tray for the electronics board with a USB-C opening, a chassis whose deck carries pockets and fences for the cubes, seats for the lens cells, panels and sensors, beam dumps, and light baffles, and a lid (Fig.~\ref{fig:app_pcad}).
Its outer size is \VbPOutL$\times$\VbPOutW$\times$\VbPOutH\,mm, inside the target of \VbEnvTarget\,mm in all three axes, a volume of \VbPVol\,cm$^3$.
The length is set by $2f$ plus three cube passes, two panels, and two walls and leaves \VbPMarginL\,mm; the width is set by the back focus of the image sensor, the offset of the reference line, and the source line and leaves \VbPMarginW\,mm; the height is set by the lens cell and the board bay and leaves \VbPMarginH\,mm.
All \VbPNParts{} printed part types export as closed meshes without warnings, fit the print bed, and intersect no beam.
The outlines of the panels, cameras, cells, and the diode are assumed; no supplier part was selected.

\begin{table}[!htbp]
\centering\footnotesize
\begin{tabular}{@{}p{0.47\linewidth}p{0.235\linewidth}p{0.235\linewidth}@{}}
\toprule
 & laboratory bench & portable unit \\
\midrule
wavelength (nm) & 532 & 520 \\
object: pixels, pitch ($\mu$m), width (mm) & 1024, 8.0, 8.19 & 1024, 3.74, 3.83 \\
bin pitch ($\mu$m), modulator pixels per bin & 14.96, 4 & 7.48, 2 \\
pixel aperture fraction & 0.25 & 0.5 \\
focal length (mm) & 230.36 & 55.09 \\
optical path, object to sensor (mm) & 921 (two lenses) & 110 (one lens, two passes) \\
nominal pupil: diameter (mm), modulator pixels & 15.32, 4096 & 7.66, 2048 \\
pupil on a 4160 $\times$ 2464 panel in both axes & no (72\% of the disc) & yes \\
pattern registration residual (bin; $\mu$m) & 0.125; 1.87 & 0.25; 1.87 \\
depth of focus at the sensor (mm) & 0.241 & 0.054 \\
size & base board 668 $\times$ 747 mm & 132.7 $\times$ 74.3 $\times$ 29.6 mm \\
printed part types; printed volume (cm$^3$) & 25; 1569 & 7; 80 \\
\bottomrule
\end{tabular}

\caption{The two designs side by side.
Design values; no hardware was built for either.}
\label{tab:app_devices}
\end{table}

\section{Simulated performance of the reference designs}
\label{sec:app_simperf}

The results of the main text use the ideal thin-lens model of Sec.~\ref{sec:optics}.
Table~\ref{tab:app_simperf} adds a simulation of the two reference designs themselves.
It replaces the ideal lenses by the aberrations of each design's lens prescription, draws assembly and thermal tolerances for every unit, and adds a calibration in software. The Fourier-camera read is registered to the frame, the scale of the Fourier plane is estimated from a displayed grating, the object-plane map of the lens is stored and compensated, and each exposure is the sum of \SpFrames{} frames. In the portable unit, the modulator patterns are also registered to within $\pm$\SpTolShift{} bin.
Each tolerance value is taken at the favorable end of the range reported in published measurements, published simulations, or vendor data sheets; none was measured on our hardware.
Those that matter most are the residual of the registered camera read, \SpTolCamReg{} bin per axis (one standard deviation, \SpTolCamRegPx{} camera pixel), the pattern registration of the modulators, $\pm$\SpTolShift{} bin, and the scale of the Fourier plane, $\pm$\SpTolScalePct\% plus a thermal term for a temperature excursion of $\pm$\SpTolDTLab\,K on the bench and $\pm$\SpTolDTPort\,K in the portable unit.
In the rows with assembly tolerances, the residuals of the camera registration and of the scale estimate are drawn at these accuracies, and the object-plane map is taken as known.
The software of the public verifier is unchanged, and no hardware was built.

\begin{table}[!htb]
\centering\footnotesize
\begin{tabular}{@{}llcc@{}}
\toprule
design & configuration & public verifier & optical verifier \\
\midrule
laboratory bench & lens design + calibration & \SpLabLensPubK{} (\SpLabLensPubPct) & \SpLabLensOptK{} (\SpLabLensOptPct) \\
laboratory bench & lens design + assembly tolerances + calibration & \SpLabTolPubK{} (\SpLabTolPubPct) & \SpLabTolOptK{} (\SpLabTolOptPct) \\
portable unit & lens design + assembly tolerances + calibration & \SpPortTolPubK{} (\SpPortTolPubPct) & \SpPortTolOptK{} (\SpPortTolOptPct) \\
\bottomrule
\end{tabular}

\caption{Simulated performance of the reference designs: number of marked images that are VERIFIED at the least favorable of the \SpNSettings{} post-sensor settings (no attack, compression, downscaling), with the percentage in parentheses; $n=\SpN$ images per row.
Simulation with tolerances at the favorable end of published ranges and software calibration; no hardware was built.}
\label{tab:app_simperf}
\end{table}

With a single frame per exposure, the same calibrated optical verifier accepted 0 of \SpNegN{} unmarked images for each design under these tolerances, and 0 of \SpNegWideN{} unmarked images, a larger set that contains \SpNegOverlap{} of them, at a wider tolerance set.

\section{Conditions, attacks, and image quality}
\label{sec:app_rates}
Table~\ref{tab:app_rates} lists, for every signer condition and every post-sensor attack, the VERIFIED rate under the public verifier and under the optical verifier, the mean raw bit error rate, the Layer-2 detection rate, and the median Layer-2 statistic, for the enlarged set and for replicate~1.
Within a set, all conditions use the same images.
The wide-pupil optics, with a pupil radius of \DesPupilWide{} bins, is the exploratory condition of deviation~3 (Tab.~\ref{tab:app_dev}) and supports no pre-registered criterion.
Table~\ref{tab:app_quality} lists the image quality per condition.

\begin{table}[!htb]
\centering\footnotesize
\begin{tabular}{lrrrrr}
\toprule
signer & PSNR p5 (dB) & PSNR median (dB) & PSNR min (dB) & SSIM p5 & binding loop converged (\%) \\
\midrule
\multicolumn{6}{@{}l}{\emph{enlarged set, \SetN{} images}} \\
digital signer & \QDigPfive & \QDigMed & \QDigMin & \QDigSsimPfive & \QDigConv \\
ideal optics & \QIdealPfive & \QIdealMed & \QIdealMin & \QIdealSsimPfive & \QIdealConv \\
nominal optics & \QOptDPfive & \QOptDMed & \QOptDMin & \QOptDSsimPfive & \QOptDConv \\
nominal optics, static mask & \QOptSPfive & \QOptSMed & \QOptSMin & \QOptSSsimPfive & \QOptSConv \\
wide-pupil optics & \QOptWPfive & \QOptWMed & \QOptWMin & \QOptWSsimPfive & \QOptWConv \\
\midrule
\multicolumn{6}{@{}l}{\emph{replicate 1, \OldSetN{} images}} \\
digital signer & \OldQDigPfive & \OldQDigMed & \OldQDigMin & \OldQDigSsimPfive & \OldQDigConv \\
ideal optics & \OldQIdealPfive & \OldQIdealMed & \OldQIdealMin & \OldQIdealSsimPfive & \OldQIdealConv \\
nominal optics & \OldQOptDPfive & \OldQOptDMed & \OldQOptDMin & \OldQOptDSsimPfive & \OldQOptDConv \\
nominal optics, static mask & \OldQOptSPfive & \OldQOptSMed & \OldQOptSMin & \OldQOptSSsimPfive & \OldQOptSConv \\
wide-pupil optics & \OldQOptWPfive & \OldQOptWMed & \OldQOptWMin & \OldQOptWSsimPfive & \OldQOptWConv \\
\bottomrule
\end{tabular}

\caption{Image quality of the marked images against the hosts (PSNR and SSIM; p5: fifth percentile), and share of the images on which the content-binding loop converged.}
\label{tab:app_quality}
\end{table}

\begin{table}[!t]
\centering\footnotesize\setlength{\tabcolsep}{3.6pt}
\begin{tabular}{@{}ll@{\hspace{12pt}}rrrrr@{\hspace{16pt}}rrrrr@{}}
\toprule
 & & \multicolumn{5}{c}{enlarged set (\SetN{} images)} & \multicolumn{5}{c}{replicate 1 (\OldSetN{} images)} \\
\cmidrule(lr){3-7}\cmidrule(l){8-12}
 & & \multicolumn{2}{c}{VERIFIED (\%)} & & & & \multicolumn{2}{c}{VERIFIED (\%)} & & & \\
\cmidrule(lr){3-4}\cmidrule(lr){8-9}
signer & attack & public & optical & raw BER & Layer 2 (\%) & median $z_c$ & public & optical & raw BER & Layer 2 (\%) & median $z_c$ \\
\midrule
digital & none & \RDigVdNone & \RDigVoNone & $\RDigBerNone$ & -- & -- & \OldRDigVdNone & \OldRDigVoNone & $\OldRDigBerNone$ & -- & -- \\
 & JPEG 95 & \RDigVdJpegA & \RDigVoJpegA & $\RDigBerJpegA$ & -- & -- & \OldRDigVdJpegA & \OldRDigVoJpegA & $\OldRDigBerJpegA$ & -- & -- \\
 & JPEG 75 & \RDigVdJpegB & \RDigVoJpegB & $\RDigBerJpegB$ & -- & -- & \OldRDigVdJpegB & \OldRDigVoJpegB & $\OldRDigBerJpegB$ & -- & -- \\
 & JPEG 50 & \RDigVdJpegC & \RDigVoJpegC & $\RDigBerJpegC$ & -- & -- & \OldRDigVdJpegC & \OldRDigVoJpegC & $\OldRDigBerJpegC$ & -- & -- \\
 & WebP 75 & \RDigVdWebp & \RDigVoWebp & $\RDigBerWebp$ & -- & -- & \OldRDigVdWebp & \OldRDigVoWebp & $\OldRDigBerWebp$ & -- & -- \\
 & resize 0.5x & \RDigVdScaleA & \RDigVoScaleA & $\RDigBerScaleA$ & -- & -- & \OldRDigVdScaleA & \OldRDigVoScaleA & $\OldRDigBerScaleA$ & -- & -- \\
 & resize 0.75x & \RDigVdScaleB & \RDigVoScaleB & $\RDigBerScaleB$ & -- & -- & \OldRDigVdScaleB & \OldRDigVoScaleB & $\OldRDigBerScaleB$ & -- & -- \\
 & JPEG 75, 0.5x & \RDigVdCombo & \RDigVoCombo & $\RDigBerCombo$ & -- & -- & \OldRDigVdCombo & \OldRDigVoCombo & $\OldRDigBerCombo$ & -- & -- \\
\midrule
ideal & none & \RIdealVdNone & \RIdealVoNone & $\RIdealBerNone$ & \RIdealLtNone & \RIdealZcNone & \OldRIdealVdNone & \OldRIdealVoNone & $\OldRIdealBerNone$ & \OldRIdealLtNone & \OldRIdealZcNone \\
 & JPEG 95 & \RIdealVdJpegA & \RIdealVoJpegA & $\RIdealBerJpegA$ & \RIdealLtJpegA & \RIdealZcJpegA & \OldRIdealVdJpegA & \OldRIdealVoJpegA & $\OldRIdealBerJpegA$ & \OldRIdealLtJpegA & \OldRIdealZcJpegA \\
 & JPEG 75 & \RIdealVdJpegB & \RIdealVoJpegB & $\RIdealBerJpegB$ & \RIdealLtJpegB & \RIdealZcJpegB & \OldRIdealVdJpegB & \OldRIdealVoJpegB & $\OldRIdealBerJpegB$ & \OldRIdealLtJpegB & \OldRIdealZcJpegB \\
 & JPEG 50 & \RIdealVdJpegC & \RIdealVoJpegC & $\RIdealBerJpegC$ & \RIdealLtJpegC & \RIdealZcJpegC & \OldRIdealVdJpegC & \OldRIdealVoJpegC & $\OldRIdealBerJpegC$ & \OldRIdealLtJpegC & \OldRIdealZcJpegC \\
 & WebP 75 & \RIdealVdWebp & \RIdealVoWebp & $\RIdealBerWebp$ & \RIdealLtWebp & \RIdealZcWebp & \OldRIdealVdWebp & \OldRIdealVoWebp & $\OldRIdealBerWebp$ & \OldRIdealLtWebp & \OldRIdealZcWebp \\
 & resize 0.5x & \RIdealVdScaleA & \RIdealVoScaleA & $\RIdealBerScaleA$ & \RIdealLtScaleA & \RIdealZcScaleA & \OldRIdealVdScaleA & \OldRIdealVoScaleA & $\OldRIdealBerScaleA$ & \OldRIdealLtScaleA & \OldRIdealZcScaleA \\
 & resize 0.75x & \RIdealVdScaleB & \RIdealVoScaleB & $\RIdealBerScaleB$ & \RIdealLtScaleB & \RIdealZcScaleB & \OldRIdealVdScaleB & \OldRIdealVoScaleB & $\OldRIdealBerScaleB$ & \OldRIdealLtScaleB & \OldRIdealZcScaleB \\
 & JPEG 75, 0.5x & \RIdealVdCombo & \RIdealVoCombo & $\RIdealBerCombo$ & \RIdealLtCombo & \RIdealZcCombo & \OldRIdealVdCombo & \OldRIdealVoCombo & $\OldRIdealBerCombo$ & \OldRIdealLtCombo & \OldRIdealZcCombo \\
\midrule
nominal & none & \ROptDVdNone & \ROptDVoNone & $\ROptDBerNone$ & \ROptDLtNone & \ROptDZcNone & \OldROptDVdNone & \OldROptDVoNone & $\OldROptDBerNone$ & \OldROptDLtNone & \OldROptDZcNone \\
 & JPEG 95 & \ROptDVdJpegA & \ROptDVoJpegA & $\ROptDBerJpegA$ & \ROptDLtJpegA & \ROptDZcJpegA & \OldROptDVdJpegA & \OldROptDVoJpegA & $\OldROptDBerJpegA$ & \OldROptDLtJpegA & \OldROptDZcJpegA \\
 & JPEG 75 & \ROptDVdJpegB & \ROptDVoJpegB & $\ROptDBerJpegB$ & \ROptDLtJpegB & \ROptDZcJpegB & \OldROptDVdJpegB & \OldROptDVoJpegB & $\OldROptDBerJpegB$ & \OldROptDLtJpegB & \OldROptDZcJpegB \\
 & JPEG 50 & \ROptDVdJpegC & \ROptDVoJpegC & $\ROptDBerJpegC$ & \ROptDLtJpegC & \ROptDZcJpegC & \OldROptDVdJpegC & \OldROptDVoJpegC & $\OldROptDBerJpegC$ & \OldROptDLtJpegC & \OldROptDZcJpegC \\
 & WebP 75 & \ROptDVdWebp & \ROptDVoWebp & $\ROptDBerWebp$ & \ROptDLtWebp & \ROptDZcWebp & \OldROptDVdWebp & \OldROptDVoWebp & $\OldROptDBerWebp$ & \OldROptDLtWebp & \OldROptDZcWebp \\
 & resize 0.5x & \ROptDVdScaleA & \ROptDVoScaleA & $\ROptDBerScaleA$ & \ROptDLtScaleA & \ROptDZcScaleA & \OldROptDVdScaleA & \OldROptDVoScaleA & $\OldROptDBerScaleA$ & \OldROptDLtScaleA & \OldROptDZcScaleA \\
 & resize 0.75x & \ROptDVdScaleB & \ROptDVoScaleB & $\ROptDBerScaleB$ & \ROptDLtScaleB & \ROptDZcScaleB & \OldROptDVdScaleB & \OldROptDVoScaleB & $\OldROptDBerScaleB$ & \OldROptDLtScaleB & \OldROptDZcScaleB \\
 & JPEG 75, 0.5x & \ROptDVdCombo & \ROptDVoCombo & $\ROptDBerCombo$ & \ROptDLtCombo & \ROptDZcCombo & \OldROptDVdCombo & \OldROptDVoCombo & $\OldROptDBerCombo$ & \OldROptDLtCombo & \OldROptDZcCombo \\
\midrule
static & none & \ROptSVdNone & \ROptSVoNone & $\ROptSBerNone$ & \ROptSLtNone & \ROptSZcNone & \OldROptSVdNone & \OldROptSVoNone & $\OldROptSBerNone$ & \OldROptSLtNone & \OldROptSZcNone \\
 & JPEG 95 & \ROptSVdJpegA & \ROptSVoJpegA & $\ROptSBerJpegA$ & \ROptSLtJpegA & \ROptSZcJpegA & \OldROptSVdJpegA & \OldROptSVoJpegA & $\OldROptSBerJpegA$ & \OldROptSLtJpegA & \OldROptSZcJpegA \\
 & JPEG 75 & \ROptSVdJpegB & \ROptSVoJpegB & $\ROptSBerJpegB$ & \ROptSLtJpegB & \ROptSZcJpegB & \OldROptSVdJpegB & \OldROptSVoJpegB & $\OldROptSBerJpegB$ & \OldROptSLtJpegB & \OldROptSZcJpegB \\
 & JPEG 50 & \ROptSVdJpegC & \ROptSVoJpegC & $\ROptSBerJpegC$ & \ROptSLtJpegC & \ROptSZcJpegC & \OldROptSVdJpegC & \OldROptSVoJpegC & $\OldROptSBerJpegC$ & \OldROptSLtJpegC & \OldROptSZcJpegC \\
 & WebP 75 & \ROptSVdWebp & \ROptSVoWebp & $\ROptSBerWebp$ & \ROptSLtWebp & \ROptSZcWebp & \OldROptSVdWebp & \OldROptSVoWebp & $\OldROptSBerWebp$ & \OldROptSLtWebp & \OldROptSZcWebp \\
 & resize 0.5x & \ROptSVdScaleA & \ROptSVoScaleA & $\ROptSBerScaleA$ & \ROptSLtScaleA & \ROptSZcScaleA & \OldROptSVdScaleA & \OldROptSVoScaleA & $\OldROptSBerScaleA$ & \OldROptSLtScaleA & \OldROptSZcScaleA \\
 & resize 0.75x & \ROptSVdScaleB & \ROptSVoScaleB & $\ROptSBerScaleB$ & \ROptSLtScaleB & \ROptSZcScaleB & \OldROptSVdScaleB & \OldROptSVoScaleB & $\OldROptSBerScaleB$ & \OldROptSLtScaleB & \OldROptSZcScaleB \\
 & JPEG 75, 0.5x & \ROptSVdCombo & \ROptSVoCombo & $\ROptSBerCombo$ & \ROptSLtCombo & \ROptSZcCombo & \OldROptSVdCombo & \OldROptSVoCombo & $\OldROptSBerCombo$ & \OldROptSLtCombo & \OldROptSZcCombo \\
\midrule
wide pupil & none & \ROptWVdNone & \ROptWVoNone & $\ROptWBerNone$ & \ROptWLtNone & \ROptWZcNone & \OldROptWVdNone & \OldROptWVoNone & $\OldROptWBerNone$ & \OldROptWLtNone & \OldROptWZcNone \\
 & JPEG 95 & \ROptWVdJpegA & \ROptWVoJpegA & $\ROptWBerJpegA$ & \ROptWLtJpegA & \ROptWZcJpegA & \OldROptWVdJpegA & \OldROptWVoJpegA & $\OldROptWBerJpegA$ & \OldROptWLtJpegA & \OldROptWZcJpegA \\
 & JPEG 75 & \ROptWVdJpegB & \ROptWVoJpegB & $\ROptWBerJpegB$ & \ROptWLtJpegB & \ROptWZcJpegB & \OldROptWVdJpegB & \OldROptWVoJpegB & $\OldROptWBerJpegB$ & \OldROptWLtJpegB & \OldROptWZcJpegB \\
 & JPEG 50 & \ROptWVdJpegC & \ROptWVoJpegC & $\ROptWBerJpegC$ & \ROptWLtJpegC & \ROptWZcJpegC & \OldROptWVdJpegC & \OldROptWVoJpegC & $\OldROptWBerJpegC$ & \OldROptWLtJpegC & \OldROptWZcJpegC \\
 & WebP 75 & \ROptWVdWebp & \ROptWVoWebp & $\ROptWBerWebp$ & \ROptWLtWebp & \ROptWZcWebp & \OldROptWVdWebp & \OldROptWVoWebp & $\OldROptWBerWebp$ & \OldROptWLtWebp & \OldROptWZcWebp \\
 & resize 0.5x & \ROptWVdScaleA & \ROptWVoScaleA & $\ROptWBerScaleA$ & \ROptWLtScaleA & \ROptWZcScaleA & \OldROptWVdScaleA & \OldROptWVoScaleA & $\OldROptWBerScaleA$ & \OldROptWLtScaleA & \OldROptWZcScaleA \\
 & resize 0.75x & \ROptWVdScaleB & \ROptWVoScaleB & $\ROptWBerScaleB$ & \ROptWLtScaleB & \ROptWZcScaleB & \OldROptWVdScaleB & \OldROptWVoScaleB & $\OldROptWBerScaleB$ & \OldROptWLtScaleB & \OldROptWZcScaleB \\
 & JPEG 75, 0.5x & \ROptWVdCombo & \ROptWVoCombo & $\ROptWBerCombo$ & \ROptWLtCombo & \ROptWZcCombo & \OldROptWVdCombo & \OldROptWVoCombo & $\OldROptWBerCombo$ & \OldROptWLtCombo & \OldROptWZcCombo \\
\bottomrule
\end{tabular}

\caption{All conditions and attacks in the numerical simulation, for the enlarged set and for replicate~1.
Signers: digital signer; ideal optics; nominal optics with the per-image mask; nominal optics with the static mask; wide-pupil optics.
Rates in percent. The last column of each block is $z_c$, not $z_c^2$; detection requires $z_c^2\ge\DesThr$.}
\label{tab:app_rates}
\end{table}

\section{Tolerance sweeps}
\label{sec:app_sweeps}
Table~\ref{tab:app_sweeps} lists every sweep point for both sets.
Each point changes one parameter of the nominal configuration.
The sweeps ran on \HtN{} evaluation images in the enlarged set and on the first \OldHtN{} images in replicate~1 (deviation~5 of Tab.~\ref{tab:app_dev}).
The pupil radius of \DesPupilWide{} bins is the exploratory point of deviation~3.
Appendix~\ref{sec:app_budget} turns these sweeps into an alignment budget for the reference hardware design.

\begin{table}[!t]
\centering\scriptsize\setlength{\tabcolsep}{3.8pt}
\begin{tabular}{@{}p{0.175\linewidth}l@{\hspace{8pt}}rrrrr@{\hspace{12pt}}rrrrr@{}}
\toprule
 & & \multicolumn{5}{c}{enlarged set (\HtN{} images)} & \multicolumn{5}{c}{replicate 1 (\OldHtN{} images)} \\
\cmidrule(lr){3-7}\cmidrule(l){8-12}
 & & \multicolumn{2}{c}{VERIFIED (\%)} & & & & \multicolumn{2}{c}{VERIFIED (\%)} & & & \\
\cmidrule(lr){3-4}\cmidrule(lr){8-9}
parameter & value & none & JPEG 75 & raw BER & Layer 2 (\%) & PSNR (dB) & none & JPEG 75 & raw BER & Layer 2 (\%) & PSNR (dB) \\
\midrule
modulator amplitude bits & 3 & \HtVnBslmH & \HtVjBslmH & $\HtBerBslmH$ & \HtLtBslmH & \HtPsBslmH & \OldHtVnBslmH & \OldHtVjBslmH & $\OldHtBerBslmH$ & \OldHtLtBslmH & \OldHtPsBslmH \\
 & 4 & \HtVnBslmF & \HtVjBslmF & $\HtBerBslmF$ & \HtLtBslmF & \HtPsBslmF & \OldHtVnBslmF & \OldHtVjBslmF & $\OldHtBerBslmF$ & \OldHtLtBslmF & \OldHtPsBslmF \\
 & 5 & \HtVnBslmV & \HtVjBslmV & $\HtBerBslmV$ & \HtLtBslmV & \HtPsBslmV & \OldHtVnBslmV & \OldHtVjBslmV & $\OldHtBerBslmV$ & \OldHtLtBslmV & \OldHtPsBslmV \\
 & 6 & \HtVnBslmS & \HtVjBslmS & $\HtBerBslmS$ & \HtLtBslmS & \HtPsBslmS & \OldHtVnBslmS & \OldHtVjBslmS & $\OldHtBerBslmS$ & \OldHtLtBslmS & \OldHtPsBslmS \\
 & 7 & \HtVnBslmN & \HtVjBslmN & $\HtBerBslmN$ & \HtLtBslmN & \HtPsBslmN & \OldHtVnBslmN & \OldHtVjBslmN & $\OldHtBerBslmN$ & \OldHtLtBslmN & \OldHtPsBslmN \\
 & 10 & \HtVnBslmOZ & \HtVjBslmOZ & $\HtBerBslmOZ$ & \HtLtBslmOZ & \HtPsBslmOZ & \OldHtVnBslmOZ & \OldHtVjBslmOZ & $\OldHtBerBslmOZ$ & \OldHtLtBslmOZ & \OldHtPsBslmOZ \\
\midrule
sensor full well (e$^-$) & 100.0 & \HtVnSneOZZpZ & \HtVjSneOZZpZ & $\HtBerSneOZZpZ$ & \HtLtSneOZZpZ & \HtPsSneOZZpZ & \OldHtVnSneOZZpZ & \OldHtVjSneOZZpZ & $\OldHtBerSneOZZpZ$ & \OldHtLtSneOZZpZ & \OldHtPsSneOZZpZ \\
 & 300.0 & \HtVnSneHZZpZ & \HtVjSneHZZpZ & $\HtBerSneHZZpZ$ & \HtLtSneHZZpZ & \HtPsSneHZZpZ & \OldHtVnSneHZZpZ & \OldHtVjSneHZZpZ & $\OldHtBerSneHZZpZ$ & \OldHtLtSneHZZpZ & \OldHtPsSneHZZpZ \\
 & 1000.0 & \HtVnSneOZZZpZ & \HtVjSneOZZZpZ & $\HtBerSneOZZZpZ$ & \HtLtSneOZZZpZ & \HtPsSneOZZZpZ & \OldHtVnSneOZZZpZ & \OldHtVjSneOZZZpZ & $\OldHtBerSneOZZZpZ$ & \OldHtLtSneOZZZpZ & \OldHtPsSneOZZZpZ \\
 & 3000.0 & \HtVnSneHZZZpZ & \HtVjSneHZZZpZ & $\HtBerSneHZZZpZ$ & \HtLtSneHZZZpZ & \HtPsSneHZZZpZ & \OldHtVnSneHZZZpZ & \OldHtVjSneHZZZpZ & $\OldHtBerSneHZZZpZ$ & \OldHtLtSneHZZZpZ & \OldHtPsSneHZZZpZ \\
 & 10000.0 & \HtVnSneOZZZZpZ & \HtVjSneOZZZZpZ & $\HtBerSneOZZZZpZ$ & \HtLtSneOZZZZpZ & \HtPsSneOZZZZpZ & \OldHtVnSneOZZZZpZ & \OldHtVjSneOZZZZpZ & $\OldHtBerSneOZZZZpZ$ & \OldHtLtSneOZZZZpZ & \OldHtPsSneOZZZZpZ \\
 & 30000.0 & \HtVnSneHZZZZpZ & \HtVjSneHZZZZpZ & $\HtBerSneHZZZZpZ$ & \HtLtSneHZZZZpZ & \HtPsSneHZZZZpZ & \OldHtVnSneHZZZZpZ & \OldHtVjSneHZZZZpZ & $\OldHtBerSneHZZZZpZ$ & \OldHtLtSneHZZZZpZ & \OldHtPsSneHZZZZpZ \\
\midrule
Fourier camera (exposures x bits) & 2$\times$12 & \HtVnFcamTxOT & \HtVjFcamTxOT & $\HtBerFcamTxOT$ & \HtLtFcamTxOT & \HtPsFcamTxOT & \OldHtVnFcamTxOT & \OldHtVjFcamTxOT & $\OldHtBerFcamTxOT$ & \OldHtLtFcamTxOT & \OldHtPsFcamTxOT \\
 & 3$\times$8 & \HtVnFcamHxE & \HtVjFcamHxE & $\HtBerFcamHxE$ & \HtLtFcamHxE & \HtPsFcamHxE & \OldHtVnFcamHxE & \OldHtVjFcamHxE & $\OldHtBerFcamHxE$ & \OldHtLtFcamHxE & \OldHtPsFcamHxE \\
 & 3$\times$10 & \HtVnFcamHxOZ & \HtVjFcamHxOZ & $\HtBerFcamHxOZ$ & \HtLtFcamHxOZ & \HtPsFcamHxOZ & \OldHtVnFcamHxOZ & \OldHtVjFcamHxOZ & $\OldHtBerFcamHxOZ$ & \OldHtLtFcamHxOZ & \OldHtPsFcamHxOZ \\
 & 1$\times$12 & \HtVnFcamOxOT & \HtVjFcamOxOT & $\HtBerFcamOxOT$ & \HtLtFcamOxOT & \HtPsFcamOxOT & \OldHtVnFcamOxOT & \OldHtVjFcamOxOT & $\OldHtBerFcamOxOT$ & \OldHtLtFcamOxOT & \OldHtPsFcamOxOT \\
\midrule
pixel aperture fraction & 0.0 & \HtVnAfZpZ & \HtVjAfZpZ & $\HtBerAfZpZ$ & \HtLtAfZpZ & \HtPsAfZpZ & \OldHtVnAfZpZ & \OldHtVjAfZpZ & $\OldHtBerAfZpZ$ & \OldHtLtAfZpZ & \OldHtPsAfZpZ \\
 & 0.125 & \HtVnAfZpOTV & \HtVjAfZpOTV & $\HtBerAfZpOTV$ & \HtLtAfZpOTV & \HtPsAfZpOTV & \OldHtVnAfZpOTV & \OldHtVjAfZpOTV & $\OldHtBerAfZpOTV$ & \OldHtLtAfZpOTV & \OldHtPsAfZpOTV \\
 & 0.5 & \HtVnAfZpV & \HtVjAfZpV & $\HtBerAfZpV$ & \HtLtAfZpV & \HtPsAfZpV & \OldHtVnAfZpV & \OldHtVjAfZpV & $\OldHtBerAfZpV$ & \OldHtLtAfZpV & \OldHtPsAfZpV \\
 & 0.75 & \HtVnAfZpNV & \HtVjAfZpNV & $\HtBerAfZpNV$ & \HtLtAfZpNV & \HtPsAfZpNV & \OldHtVnAfZpNV & \OldHtVjAfZpNV & $\OldHtBerAfZpNV$ & \OldHtLtAfZpNV & \OldHtPsAfZpNV \\
 & 1.0 & \HtVnAfOpZ & \HtVjAfOpZ & $\HtBerAfOpZ$ & \HtLtAfOpZ & \HtPsAfOpZ & \OldHtVnAfOpZ & \OldHtVjAfOpZ & $\OldHtBerAfOpZ$ & \OldHtLtAfOpZ & \OldHtPsAfOpZ \\
\midrule
pupil radius (bins) & 96.0 & \HtVnPupilISpZ & \HtVjPupilISpZ & $\HtBerPupilISpZ$ & \HtLtPupilISpZ & \HtPsPupilISpZ & \OldHtVnPupilISpZ & \OldHtVjPupilISpZ & $\OldHtBerPupilISpZ$ & \OldHtLtPupilISpZ & \OldHtPsPupilISpZ \\
 & 128.0 & \HtVnPupilOTEpZ & \HtVjPupilOTEpZ & $\HtBerPupilOTEpZ$ & \HtLtPupilOTEpZ & \HtPsPupilOTEpZ & \OldHtVnPupilOTEpZ & \OldHtVjPupilOTEpZ & $\OldHtBerPupilOTEpZ$ & \OldHtLtPupilOTEpZ & \OldHtPsPupilOTEpZ \\
 & 160.0 & \HtVnPupilOSZpZ & \HtVjPupilOSZpZ & $\HtBerPupilOSZpZ$ & \HtLtPupilOSZpZ & \HtPsPupilOSZpZ & \OldHtVnPupilOSZpZ & \OldHtVjPupilOSZpZ & $\OldHtBerPupilOSZpZ$ & \OldHtLtPupilOSZpZ & \OldHtPsPupilOSZpZ \\
 & 256.0 & \HtVnPupilTVSpZ & \HtVjPupilTVSpZ & $\HtBerPupilTVSpZ$ & \HtLtPupilTVSpZ & \HtPsPupilTVSpZ & \OldHtVnPupilTVSpZ & \OldHtVjPupilTVSpZ & $\OldHtBerPupilTVSpZ$ & \OldHtLtPupilTVSpZ & \OldHtPsPupilTVSpZ \\
 & 724.0 & \HtVnPupilNTFpZ & \HtVjPupilNTFpZ & $\HtBerPupilNTFpZ$ & \HtLtPupilNTFpZ & \HtPsPupilNTFpZ & \OldHtVnPupilNTFpZ & \OldHtVjPupilNTFpZ & $\OldHtBerPupilNTFpZ$ & \OldHtLtPupilNTFpZ & \OldHtPsPupilNTFpZ \\
\midrule
defocus (mm) & 0.1 & \HtVnDzZpO & \HtVjDzZpO & $\HtBerDzZpO$ & \HtLtDzZpO & \HtPsDzZpO & \OldHtVnDzZpO & \OldHtVjDzZpO & $\OldHtBerDzZpO$ & \OldHtLtDzZpO & \OldHtPsDzZpO \\
 & 0.25 & \HtVnDzZpTV & \HtVjDzZpTV & $\HtBerDzZpTV$ & \HtLtDzZpTV & \HtPsDzZpTV & \OldHtVnDzZpTV & \OldHtVjDzZpTV & $\OldHtBerDzZpTV$ & \OldHtLtDzZpTV & \OldHtPsDzZpTV \\
 & 0.5 & \HtVnDzZpV & \HtVjDzZpV & $\HtBerDzZpV$ & \HtLtDzZpV & \HtPsDzZpV & \OldHtVnDzZpV & \OldHtVjDzZpV & $\OldHtBerDzZpV$ & \OldHtLtDzZpV & \OldHtPsDzZpV \\
 & 1.0 & \HtVnDzOpZ & \HtVjDzOpZ & $\HtBerDzOpZ$ & \HtLtDzOpZ & \HtPsDzOpZ & \OldHtVnDzOpZ & \OldHtVjDzOpZ & $\OldHtBerDzOpZ$ & \OldHtLtDzOpZ & \OldHtPsDzOpZ \\
 & 2.0 & \HtVnDzTpZ & \HtVjDzTpZ & $\HtBerDzTpZ$ & \HtLtDzTpZ & \HtPsDzTpZ & \OldHtVnDzTpZ & \OldHtVjDzTpZ & $\OldHtBerDzTpZ$ & \OldHtLtDzTpZ & \OldHtPsDzTpZ \\
 & 4.0 & \HtVnDzFpZ & \HtVjDzFpZ & $\HtBerDzFpZ$ & \HtLtDzFpZ & \HtPsDzFpZ & \OldHtVnDzFpZ & \OldHtVjDzFpZ & $\OldHtBerDzFpZ$ & \OldHtLtDzFpZ & \OldHtPsDzFpZ \\
 & 8.0 & \HtVnDzEpZ & \HtVjDzEpZ & $\HtBerDzEpZ$ & \HtLtDzEpZ & \HtPsDzEpZ & \OldHtVnDzEpZ & \OldHtVjDzEpZ & $\OldHtBerDzEpZ$ & \OldHtLtDzEpZ & \OldHtPsDzEpZ \\
\midrule
modulator tilt (deg) & 1.0 & \HtVnTiltOpZ & \HtVjTiltOpZ & $\HtBerTiltOpZ$ & \HtLtTiltOpZ & \HtPsTiltOpZ & \OldHtVnTiltOpZ & \OldHtVjTiltOpZ & $\OldHtBerTiltOpZ$ & \OldHtLtTiltOpZ & \OldHtPsTiltOpZ \\
 & 2.0 & \HtVnTiltTpZ & \HtVjTiltTpZ & $\HtBerTiltTpZ$ & \HtLtTiltTpZ & \HtPsTiltTpZ & \OldHtVnTiltTpZ & \OldHtVjTiltTpZ & $\OldHtBerTiltTpZ$ & \OldHtLtTiltTpZ & \OldHtPsTiltTpZ \\
 & 5.0 & \HtVnTiltVpZ & \HtVjTiltVpZ & $\HtBerTiltVpZ$ & \HtLtTiltVpZ & \HtPsTiltVpZ & \OldHtVnTiltVpZ & \OldHtVjTiltVpZ & $\OldHtBerTiltVpZ$ & \OldHtLtTiltVpZ & \OldHtPsTiltVpZ \\
 & 10.0 & \HtVnTiltOZpZ & \HtVjTiltOZpZ & $\HtBerTiltOZpZ$ & \HtLtTiltOZpZ & \HtPsTiltOZpZ & \OldHtVnTiltOZpZ & \OldHtVjTiltOZpZ & $\OldHtBerTiltOZpZ$ & \OldHtLtTiltOZpZ & \OldHtPsTiltOZpZ \\
\midrule
modulator shift (bins) & 0.02 & \HtVnShiftZpZT & \HtVjShiftZpZT & $\HtBerShiftZpZT$ & \HtLtShiftZpZT & \HtPsShiftZpZT & \OldHtVnShiftZpZT & \OldHtVjShiftZpZT & $\OldHtBerShiftZpZT$ & \OldHtLtShiftZpZT & \OldHtPsShiftZpZT \\
 & 0.05 & \HtVnShiftZpZV & \HtVjShiftZpZV & $\HtBerShiftZpZV$ & \HtLtShiftZpZV & \HtPsShiftZpZV & \OldHtVnShiftZpZV & \OldHtVjShiftZpZV & $\OldHtBerShiftZpZV$ & \OldHtLtShiftZpZV & \OldHtPsShiftZpZV \\
 & 0.1 & \HtVnShiftZpO & \HtVjShiftZpO & $\HtBerShiftZpO$ & \HtLtShiftZpO & \HtPsShiftZpO & \OldHtVnShiftZpO & \OldHtVjShiftZpO & $\OldHtBerShiftZpO$ & \OldHtLtShiftZpO & \OldHtPsShiftZpO \\
 & 0.2 & \HtVnShiftZpT & \HtVjShiftZpT & $\HtBerShiftZpT$ & \HtLtShiftZpT & \HtPsShiftZpT & \OldHtVnShiftZpT & \OldHtVjShiftZpT & $\OldHtBerShiftZpT$ & \OldHtLtShiftZpT & \OldHtPsShiftZpT \\
 & 0.35 & \HtVnShiftZpHV & \HtVjShiftZpHV & $\HtBerShiftZpHV$ & \HtLtShiftZpHV & \HtPsShiftZpHV & \OldHtVnShiftZpHV & \OldHtVjShiftZpHV & $\OldHtBerShiftZpHV$ & \OldHtLtShiftZpHV & \OldHtPsShiftZpHV \\
 & 0.5 & \HtVnShiftZpV & \HtVjShiftZpV & $\HtBerShiftZpV$ & \HtLtShiftZpV & \HtPsShiftZpV & \OldHtVnShiftZpV & \OldHtVjShiftZpV & $\OldHtBerShiftZpV$ & \OldHtLtShiftZpV & \OldHtPsShiftZpV \\
 & 1.0 & \HtVnShiftOpZ & \HtVjShiftOpZ & $\HtBerShiftOpZ$ & \HtLtShiftOpZ & \HtPsShiftOpZ & \OldHtVnShiftOpZ & \OldHtVjShiftOpZ & $\OldHtBerShiftOpZ$ & \OldHtLtShiftOpZ & \OldHtPsShiftOpZ \\
\midrule
sensor ADC bits & 6 & \HtVnSbitsS & \HtVjSbitsS & $\HtBerSbitsS$ & \HtLtSbitsS & \HtPsSbitsS & \OldHtVnSbitsS & \OldHtVjSbitsS & $\OldHtBerSbitsS$ & \OldHtLtSbitsS & \OldHtPsSbitsS \\
 & 8 & \HtVnSbitsE & \HtVjSbitsE & $\HtBerSbitsE$ & \HtLtSbitsE & \HtPsSbitsE & \OldHtVnSbitsE & \OldHtVjSbitsE & $\OldHtBerSbitsE$ & \OldHtLtSbitsE & \OldHtPsSbitsE \\
 & 10 & \HtVnSbitsOZ & \HtVjSbitsOZ & $\HtBerSbitsOZ$ & \HtLtSbitsOZ & \HtPsSbitsOZ & \OldHtVnSbitsOZ & \OldHtVjSbitsOZ & $\OldHtBerSbitsOZ$ & \OldHtLtSbitsOZ & \OldHtPsSbitsOZ \\
\midrule
arm coherence & 0.25 & \HtVnGammaZpTV & \HtVjGammaZpTV & $\HtBerGammaZpTV$ & \HtLtGammaZpTV & \HtPsGammaZpTV & \OldHtVnGammaZpTV & \OldHtVjGammaZpTV & $\OldHtBerGammaZpTV$ & \OldHtLtGammaZpTV & \OldHtPsGammaZpTV \\
 & 0.5 & \HtVnGammaZpV & \HtVjGammaZpV & $\HtBerGammaZpV$ & \HtLtGammaZpV & \HtPsGammaZpV & \OldHtVnGammaZpV & \OldHtVjGammaZpV & $\OldHtBerGammaZpV$ & \OldHtLtGammaZpV & \OldHtPsGammaZpV \\
 & 0.75 & \HtVnGammaZpNV & \HtVjGammaZpNV & $\HtBerGammaZpNV$ & \HtLtGammaZpNV & \HtPsGammaZpNV & \OldHtVnGammaZpNV & \OldHtVjGammaZpNV & $\OldHtBerGammaZpNV$ & \OldHtLtGammaZpNV & \OldHtPsGammaZpNV \\
\bottomrule
\end{tabular}

\caption{One-at-a-time tolerance sweeps in the numerical simulation (public verifier), for the enlarged set and for replicate~1.
VERIFIED: rate without attack and after JPEG~75. Layer~2: detection rate without attack. PSNR: median.}
\label{tab:app_sweeps}
\end{table}

\section{Key sensitivity of the optical carrier}
\label{sec:app_key}
Table~\ref{tab:keysweep} lists the median statistic and the detection rate of an auditor unit whose optical configuration differs from the signer in one parameter, for the enlarged set and for replicate~1.
On the enlarged set, the matched median is \HsMatchedMed{} (quartiles \HsMatchedQlo{} and \HsMatchedQhi).
The tolerance of a parameter is the mismatch at which the median statistic crosses the threshold \DesThr.
It is \HsTolRel\% in relative scale and \HsTolShift{} bins in mask position.

\begin{table}[!htb]
\centering\footnotesize
\begin{minipage}[t]{0.48\linewidth}
\centering
\begin{tabular}{llrr}
\toprule
mismatch & value & median $z_c^2$ & detected (\%) \\
\midrule
relative scale $\delta$ (fraction) & +0.0001 & \HcScPZpZZZOMed & \HcScPZpZZZODet \\
 & -0.0001 & \HcScMZpZZZOMed & \HcScMZpZZZODet \\
 & +0.0002 & \HcScPZpZZZTMed & \HcScPZpZZZTDet \\
 & -0.0002 & \HcScMZpZZZTMed & \HcScMZpZZZTDet \\
 & +0.0005 & \HcScPZpZZZVMed & \HcScPZpZZZVDet \\
 & -0.0005 & \HcScMZpZZZVMed & \HcScMZpZZZVDet \\
 & +0.001 & \HcScPZpZZOMed & \HcScPZpZZODet \\
 & -0.001 & \HcScMZpZZOMed & \HcScMZpZZODet \\
 & +0.002 & \HcScPZpZZTMed & \HcScPZpZZTDet \\
 & -0.002 & \HcScMZpZZTMed & \HcScMZpZZTDet \\
 & +0.003 & \HcScPZpZZHMed & \HcScPZpZZHDet \\
 & -0.003 & \HcScMZpZZHMed & \HcScMZpZZHDet \\
 & +0.005 & \HcScPZpZZVMed & \HcScPZpZZVDet \\
 & -0.005 & \HcScMZpZZVMed & \HcScMZpZZVDet \\
 & +0.01 & \HcScPZpZOMed & \HcScPZpZODet \\
 & -0.01 & \HcScMZpZOMed & \HcScMZpZODet \\
 & +0.02 & \HcScPZpZTMed & \HcScPZpZTDet \\
 & -0.02 & \HcScMZpZTMed & \HcScMZpZTDet \\
\midrule
mask shift (bins) & 0.1 & \HcShZpOMed & \HcShZpODet \\
 & 0.25 & \HcShZpTVMed & \HcShZpTVDet \\
 & 0.5 & \HcShZpVMed & \HcShZpVDet \\
 & 0.75 & \HcShZpNVMed & \HcShZpNVDet \\
 & 1 & \HcShOMed & \HcShODet \\
 & 2 & \HcShTMed & \HcShTDet \\
\midrule
auditor phase bits & 1 & \HcPbOMed & \HcPbODet \\
 & 2 & \HcPbTMed & \HcPbTDet \\
 & 3 & \HcPbHMed & \HcPbHDet \\
 & 4 & \HcPbFMed & \HcPbFDet \\
\midrule
source width FWHM (nm) & 0.5 & \HcSwZpVMed & \HcSwZpVDet \\
 & 1 & \HcSwOMed & \HcSwODet \\
 & 2 & \HcSwTMed & \HcSwTDet \\
 & 5 & \HcSwVMed & \HcSwVDet \\
 & 10 & \HcSwOZMed & \HcSwOZDet \\
\bottomrule
\end{tabular}
\\[3pt]
(a) enlarged set, \DesNEval{} images
\end{minipage}\hfill
\begin{minipage}[t]{0.48\linewidth}
\centering
\begin{tabular}{llrr}
\toprule
mismatch & value & median $z_c^2$ & detected (\%) \\
\midrule
relative scale $\delta$ (fraction) & +0.0001 & \OldHcScPZpZZZOMed & \OldHcScPZpZZZODet \\
 & -0.0001 & \OldHcScMZpZZZOMed & \OldHcScMZpZZZODet \\
 & +0.0002 & \OldHcScPZpZZZTMed & \OldHcScPZpZZZTDet \\
 & -0.0002 & \OldHcScMZpZZZTMed & \OldHcScMZpZZZTDet \\
 & +0.0005 & \OldHcScPZpZZZVMed & \OldHcScPZpZZZVDet \\
 & -0.0005 & \OldHcScMZpZZZVMed & \OldHcScMZpZZZVDet \\
 & +0.001 & \OldHcScPZpZZOMed & \OldHcScPZpZZODet \\
 & -0.001 & \OldHcScMZpZZOMed & \OldHcScMZpZZODet \\
 & +0.002 & \OldHcScPZpZZTMed & \OldHcScPZpZZTDet \\
 & -0.002 & \OldHcScMZpZZTMed & \OldHcScMZpZZTDet \\
 & +0.003 & \OldHcScPZpZZHMed & \OldHcScPZpZZHDet \\
 & -0.003 & \OldHcScMZpZZHMed & \OldHcScMZpZZHDet \\
 & +0.005 & \OldHcScPZpZZVMed & \OldHcScPZpZZVDet \\
 & -0.005 & \OldHcScMZpZZVMed & \OldHcScMZpZZVDet \\
 & +0.01 & \OldHcScPZpZOMed & \OldHcScPZpZODet \\
 & -0.01 & \OldHcScMZpZOMed & \OldHcScMZpZODet \\
 & +0.02 & \OldHcScPZpZTMed & \OldHcScPZpZTDet \\
 & -0.02 & \OldHcScMZpZTMed & \OldHcScMZpZTDet \\
\midrule
mask shift (bins) & 0.1 & \OldHcShZpOMed & \OldHcShZpODet \\
 & 0.25 & \OldHcShZpTVMed & \OldHcShZpTVDet \\
 & 0.5 & \OldHcShZpVMed & \OldHcShZpVDet \\
 & 0.75 & \OldHcShZpNVMed & \OldHcShZpNVDet \\
 & 1 & \OldHcShOMed & \OldHcShODet \\
 & 2 & \OldHcShTMed & \OldHcShTDet \\
\midrule
auditor phase bits & 1 & \OldHcPbOMed & \OldHcPbODet \\
 & 2 & \OldHcPbTMed & \OldHcPbTDet \\
 & 3 & \OldHcPbHMed & \OldHcPbHDet \\
 & 4 & \OldHcPbFMed & \OldHcPbFDet \\
\midrule
source width FWHM (nm) & 0.5 & \OldHcSwZpVMed & \OldHcSwZpVDet \\
 & 1 & \OldHcSwOMed & \OldHcSwODet \\
 & 2 & \OldHcSwTMed & \OldHcSwTDet \\
 & 5 & \OldHcSwVMed & \OldHcSwVDet \\
 & 10 & \OldHcSwOZMed & \OldHcSwOZDet \\
\bottomrule
\end{tabular}
\\[3pt]
(b) replicate~1, \OldSetN{} images
\end{minipage}
\caption{Key sensitivity sweep in the numerical simulation (unattacked images, nominal optics, per-image mask): median Layer-2 statistic and detection rate of an auditor unit that differs from the signer in one parameter.
The source-width rows report the incoherent average over the spectrum.}
\label{tab:keysweep}
\end{table}

\end{document}